\documentclass[review]{elsarticle}

\usepackage[colorlinks,citecolor=blue,linktoc=all,linkcolor=cyan]{hyperref}
\usepackage{graphicx}

\usepackage[T1]{fontenc}
\usepackage{dsfont}               
\usepackage{mathrsfs}             
\usepackage{slashed}              
\usepackage{amsmath}
\usepackage{amssymb}
\usepackage{amsbsy}
\usepackage{amsfonts}
\usepackage{xcolor}

\numberwithin{equation}{section}
\numberwithin{table}{section}
\numberwithin{figure}{section}

\journal{Progress in Particle and Nuclear Physics}

\newcommand{\STr}{\mathrm{STr}}
\newcommand{\Tr}{\text{Tr}}
\newcommand{\ii}{\text{i}}
\newcommand{\tave}[1]{\langle\!\langle{#1}\rangle\!\rangle}
\DeclareMathOperator{\im}{Im}

\DeclareMathOperator{\tr}{tr}

\graphicspath{{figures/}}

\usepackage{titlesec}
\usepackage{sectsty}
\titleformat{\section}{\normalfont\Large\bfseries}{\thesection}{1em}{}
\titleformat{\subsection}{\normalfont\large\bfseries}{\thesubsection}{1em}{}
\titleformat{\subsubsection}{\normalfont\normalsize\bfseries}{\thesubsubsection}{1em}{}

\begin{document}
	
\begin{frontmatter}
	
	\title{Electromagnetic Probes: Theory and Experiment}
	
	\author[JLU,HFHF]{Ralf-Arno Tripolt\corref{cor}}
		\ead{Ralf-Arno.Tripolt@theo.physik.uni-giessen.de}
	\author[RU]{Frank Geurts}
	\cortext[cor]{Corresponding author}

    \address[JLU]{Institut f\"{u}r Theoretische Physik, Justus-Liebig-Universit\"{a}t, Heinrich-Buff-Ring 16, 35392 Giessen, Germany}
	\address[HFHF]{Helmholtz Research Academy Hesse for FAIR (HFHF), Campus Giessen, 35392 Giessen, Germany}
	\address[RU]{Department of Physics \& Astronomy, Rice University, Houston TX 77005, USA}
	
	\begin{abstract}
		We review the current state of research on electromagnetic probes in the context of heavy-ion collisions. The focus is on thermal photons and dileptons which provide unique insights into the properties of the created hot and dense matter. This review is intended to provide an introductory overview of the topic as well as a discussion of recent theoretical and experimental results. In particular, we discuss the role of vector-meson spectral functions in the calculation of photon and dilepton rates and present recent results obtained from different frameworks. Furthermore, we will highlight the special role of photons and dileptons to provide information on observables such as the temperature, the lifetime, the polarization and the electrical conductivity of the produced medium as well as their use to learn about chiral symmetry restoration and phase transitions.
	\end{abstract}
	
	
\end{frontmatter}

\newpage

\thispagestyle{empty}
\tableofcontents

\newpage

\section{Introduction}
\label{sec:introduction}

The investigation of matter under extreme conditions in temperature and density as prevailing in the aftermath of the Big Bang is one of the central aims of present theoretical as well as experimental research efforts in high-energy particle physics. Nowadays, such an extreme state of matter may occur naturally in the core of compact stellar objects like neutron stars or during neutron star merger events. In a laboratory environment, extreme temperatures and densities can be created in relativistic collisions of heavy particles, see for example \cite{Friman:2011zz, Braun-Munzinger:2015hba, Muller:2021ygo, Gelis:2021zmx, ParticleDataGroup:2020ssz} for reviews. Such heavy-ion collisions are furthermore the only means by which bulk properties of a non-Abelian gauge theory, such as Quantum Chromodynamics (QCD), can be assessed experimentally. Heavy-ion collision experiments are currently performed at the Large Hadron Collider (LHC) at CERN, the Relativistic Heavy Ion Collider (RHIC) at BNL, the Schwer-Ionen-Synchrotron (SIS) at GSI and planned at future facilities such as the Facility for Antiproton and Ion Research (FAIR), the Nuclotron-based Ion Collider fAcility (NICA), the High Intensity heavy ion Accelerator Facility (HIAF), and the heavy-ion program at the Japan Proton Accelerator Complex (J-PARC). 

Electromagnetic (EM) probes, i.e. photons and dileptons, have proven to be exceptionally versatile and useful probes to study the properties of the hot and dense medium created in such collisions, see for example \cite{Stankus:2005eq,Rapp:2009yu,Rapp:2011is,David:2019wpt2, Salabura:2020tou} for reviews. This is due to the fact that they don't (directly) interact `strongly' with the surrounding medium, i.e.~not via the strong interaction as described by QCD, but predominantly via the electromagnetic interaction as described by Quantum Electrodynamics (QED). Since the electromagnetic interaction is considerably weaker than the strong interaction, as for example evident by comparing the EM coupling strength $\alpha_{\text{EM}}\approx 1/137$ with the strong coupling $\alpha_s$ which is of the order $\mathcal{O}(10^{-1})-\mathcal{O}(1)$, photons and dileptons have a mean-free path that is larger than the extent of the created fireball. They can thus traverse the medium almost undisturbed and carry information from their production point to the detector. The smallness of the EM coupling, however, also entails that photons and dileptons are produced very rarely compared to strongly-interacting particles such as pions. For example, the decay of the $\rho(770)$ vector meson into dileptons, i.e.~into an electron-positron pair or into a muon-antimuon pair, is suppressed by a factor of $\sim 5\cdot 10^{-5}$ as compared to the decay into pions, see for example the corresponding experimental branching ratios \cite{ParticleDataGroup:2020ssz}.

Another important feature of photons and dileptons is that they are produced at all stages of the collision process. In principle, they can thus be used to obtain information on all phases of the fireball evolution, from initial hard scattering processes over the pre-equilibrium phase and the Quark-Gluon Plasma (QGP) phase to the hadron gas phase. This information is, however, convoluted with the space-time evolution of the medium which makes extracting information on a particular phase, such as the thermally-equilibrated QGP or the hadron-gas phase, very challenging. A good theoretical understanding of the underlying dilepton production rates within the various phases as well as of the space-time evolution of the collision process is therefore imperative for a robust interpretation of photon and dilepton spectra.

\begin{figure}[t]
	\centering\includegraphics[width=0.55\textwidth]{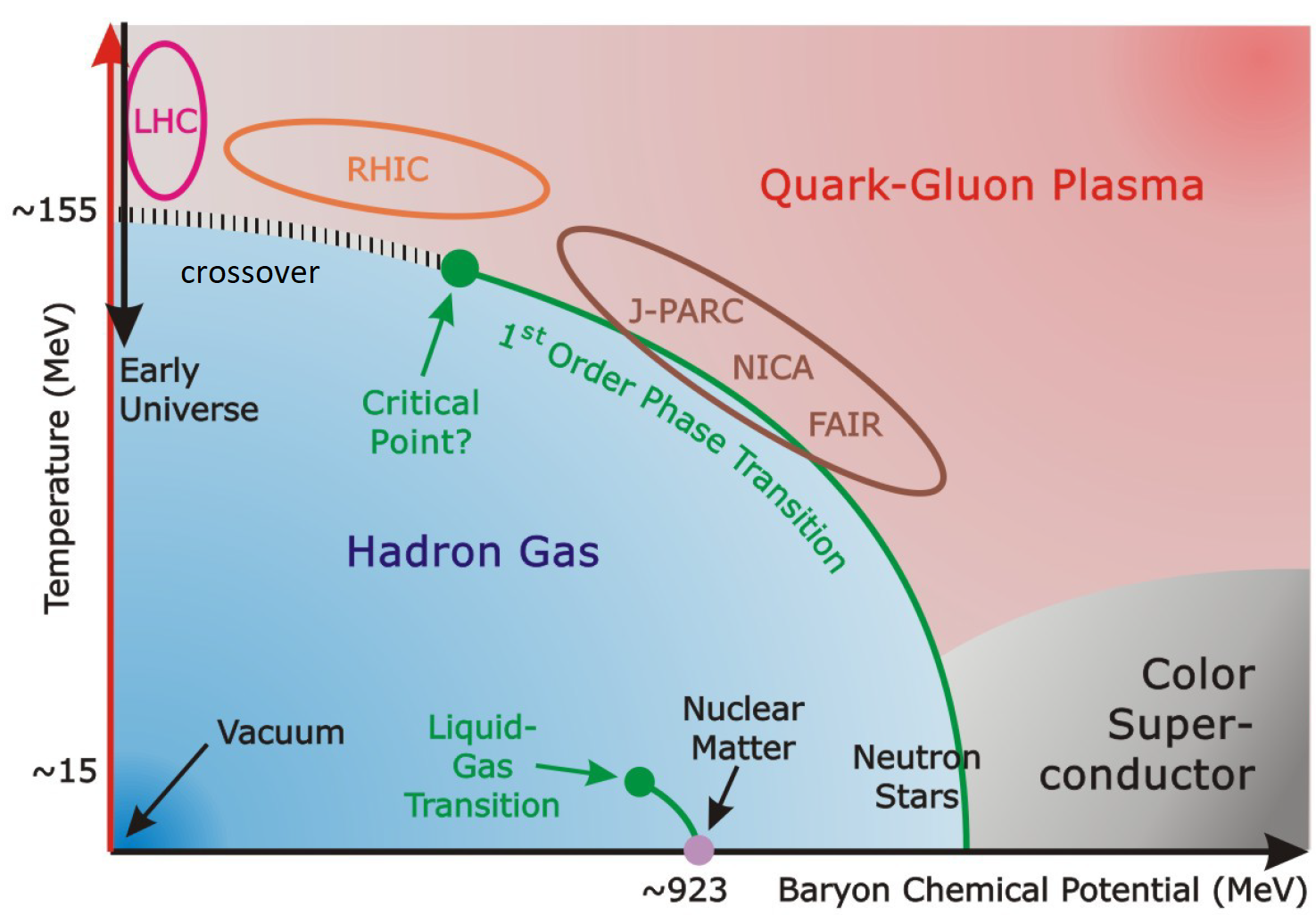}
	\caption{Sketch of the QCD phase diagram. Figure adapted from \cite{CRC:TR:211}.}
	\label{fig:QCD_phase_diagram}
\end{figure}

In this review, we will focus in particular on the theoretical description and experimental results concerning the soft thermal radiation from the QGP and the hadron gas phase. Those regimes are of particular interest since they correspond to the extreme state of matter that filled our Universe shortly after the Big Bang and since they allow to study fundamental properties of QCD such as confinement and chiral symmetry breaking. Color confinement, which describes the fact that no color-neutral objects have been observed in an isolated state, is expected to disappear at high enough temperatures and/or densities. Chiral symmetry, on the other hand, is a symmetry of the QCD Lagrangian for massless quarks that is spontaneously broken in the vacuum, i.e.~at zero temperature and density, but eventually gets restored at high temperatures and/or densities. Mapping out the corresponding QCD phase diagram is one of the central goals in high-energy physics, see also Fig.~\ref{fig:QCD_phase_diagram} which shows an illustration of the QCD phase diagram as well as of the approximate regimes\footnote{
We note that the experimental regimes shown in Fig.~\ref{fig:QCD_phase_diagram} are merely for illustrative purposes and should of course also extend into the hadron gas phase.
} 
where heavy-ion collision experiments can be used for its investigation, see for example \cite{Stephanov:2006zvm,Fukushima:2010bq} for reviews. EM probes, and in particular dileptons, can indeed be useful to learn about certain aspects of the QCD phase diagram such as the location of first-order phase transitions and the conjectured critical endpoint.

A special role in the theoretical description of thermal photon and dilepton spectra is played by the light vector mesons. This is due to the fact that vector mesons carry the same quantum numbers as the photon and can therefore directly transform into a real or virtual photon, where the virtual photon can subsequently decay into a lepton-antilepton pair, i.e.~a dilepton. The light vector mesons $\rho(770)$, $\omega(782)$, and $\phi(1020)$, therefore, act as an intermediary between the hadronic strong-interaction regime and the emitted electromagnetic particles. In fact, the resulting thermal EM spectra can almost exclusively be described by the decay of light vector mesons, with the largest contribution stemming from the $\rho(770)$ vector meson. This phenomenological result is known as Vector Meson Dominance (VMD) and will be discussed in more detail in the following.

One of the main challenges for a realistic description of thermal photon and dilepton rates is therefore the computation of in-medium vector-meson spectral functions. This can be achieved using different frameworks such as hadronic many-body theory (HMBT), QCD sum rules, the Massive Yang Mills (MYM) framework, or the Functional Renormalization Group (FRG) approach, see Secs.~\ref{sec:vector_mesons_in_medium} and \ref{sec:aFRG}. In particular, the Rapp-Wambach spectral functions as obtained from hadronic many-body theory have proven to be very successful for the description of experimental data and are still widely used. More recently, also in-medium spectral functions from the FRG have become available for different effective theories. The FRG approach, for example, allows to take the effects of fluctuations into account and to incorporate important aspects of chiral symmetry and its breaking pattern. In particular, recent results concerning the in-medium spectral function of the $\rho(770)$ vector meson and of its chiral partner, the $a_1(1260)$ axial-vector meson, will be discussed in this review.

\begin{figure}[t]
	\centering\includegraphics[width=0.5\textwidth]{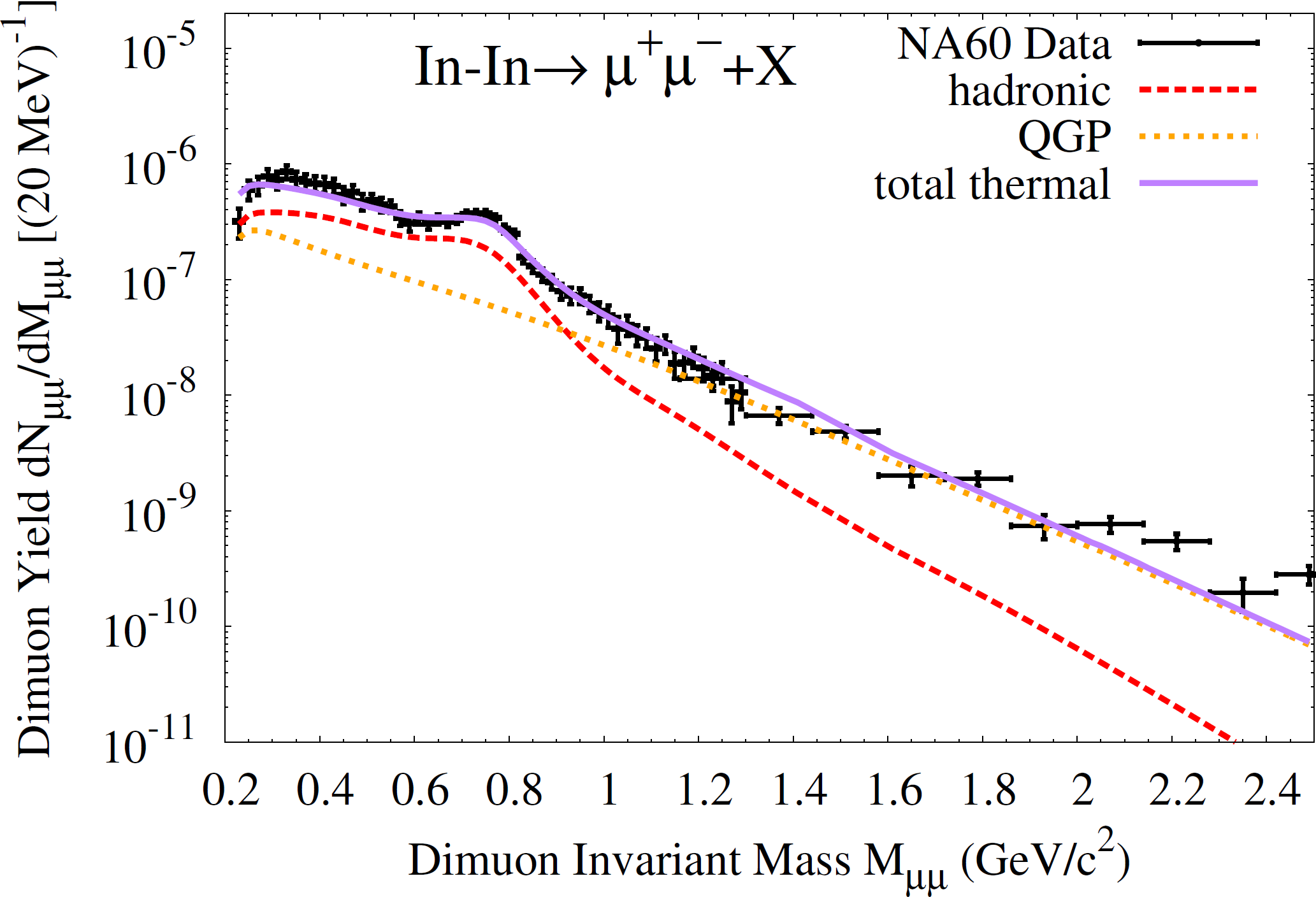}
	\caption{Excess dimuon invariant-mass spectrum as measured in In-In collisions at $\sqrt{s_{NN}}=17.3$~GeV by the NA60 collaboration at the SPS \cite{Specht:2010xu,NA60:2008ctj} together with theoretical results based on hadronic many-body theory \cite{Rapp:2014hha}. Figure adapted from \cite{Rapp:2014hha}.}
	\label{fig:NA60_Rapp}
\end{figure}

The resulting thermal photon and dilepton rates can then be combined with suitable descriptions of the space-time evolution of the heavy-ion collision process in order to obtain the measured spectra, see Fig.~\ref{fig:NA60_Rapp} for an example of a dilepton spectrum measured with high precision and the excellent agreement with theoretical predictions. For central heavy-ion collisions at high collision energies, ideal or viscous relativistic hydrodynamics is frequently used to describe the dynamics of the produced medium, while at lower collision energies transport descriptions, sometimes in combination with a coarse-graining procedure, have proven to be successful. The obtained results on spectra and other observables can then be used to learn about the properties of the produced medium and thus about the properties of hot and dense strong-interaction matter in general. In particular, we will discuss the connection to the temperature and the lifetime of the produced medium, the degree of collectivity, the underlying  spectral functions and chiral symmetry, phase transitions, changes in degrees of freedom, and to transport coefficients such as the electrical conductivity.

We close this introduction by giving an overview of the structure of the remaining parts of this review. In Sec.~\ref{sec:experimental_aspects} we discuss general aspects of experimental photon and dilepton measurements which includes a presentation of the heavy-ion collision process and an introduction to electromagnetic spectroscopy. In Sec.~\ref{sec:theoretical_aspects} we discuss general theoretical aspects of electromagnetic probes. This includes a short introduction to QCD, its phase diagram and symmetries, as well as a discussion on the theoretical computation of photon and dilepton rates, the EM spectral function and the idea of VMD. In Sec.~\ref{sec:vector_mesons_in_medium} we give a more detailed account of the theoretical approaches to describe vector mesons in a thermal medium which include low-density expansions and chiral mixing, lattice QCD, chiral and QCD sum rules, as well as Massive Yang Mills and hadronic many-body theory. In Sec.~\ref{sec:aFRG} we focus on a more recent theoretical framework for the computation of in-medium spectral functions, i.e.~the analytically-continued FRG (aFRG) method, and present results on vector and axial-vector mesons as obtained for nuclear matter. In Sec.~\ref{sec:thermal_photon_and_dilepton_rates} we discuss different theoretical results obtained for thermal photon and dilepton rates while experimental results on photon and dilepton spectra in heavy-ion collisions are discussed in Secs.~\ref{sec:photons_in_HICs} and \ref{sec:dileptons_in_HICs}, respectively. In the latter sections, a particular emphasis is on the interpretation of photon and dilepton spectra and what kind of information one can extract from them, such as on the temperature or the lifetime of the produced fireball. Finally, in Sec.~\ref{sec:conclusions} we conclude and provide an outlook on the future of electromagnetic probes in heavy-ion collisions.

\clearpage 
\section{Experimental aspects of photon and dilepton measurements}
\label{sec:experimental_aspects}

\subsection{Heavy-ion collision process}
\vspace{2mm}

\begin{figure}[b!]
	\centering\includegraphics[width=0.6\textwidth]{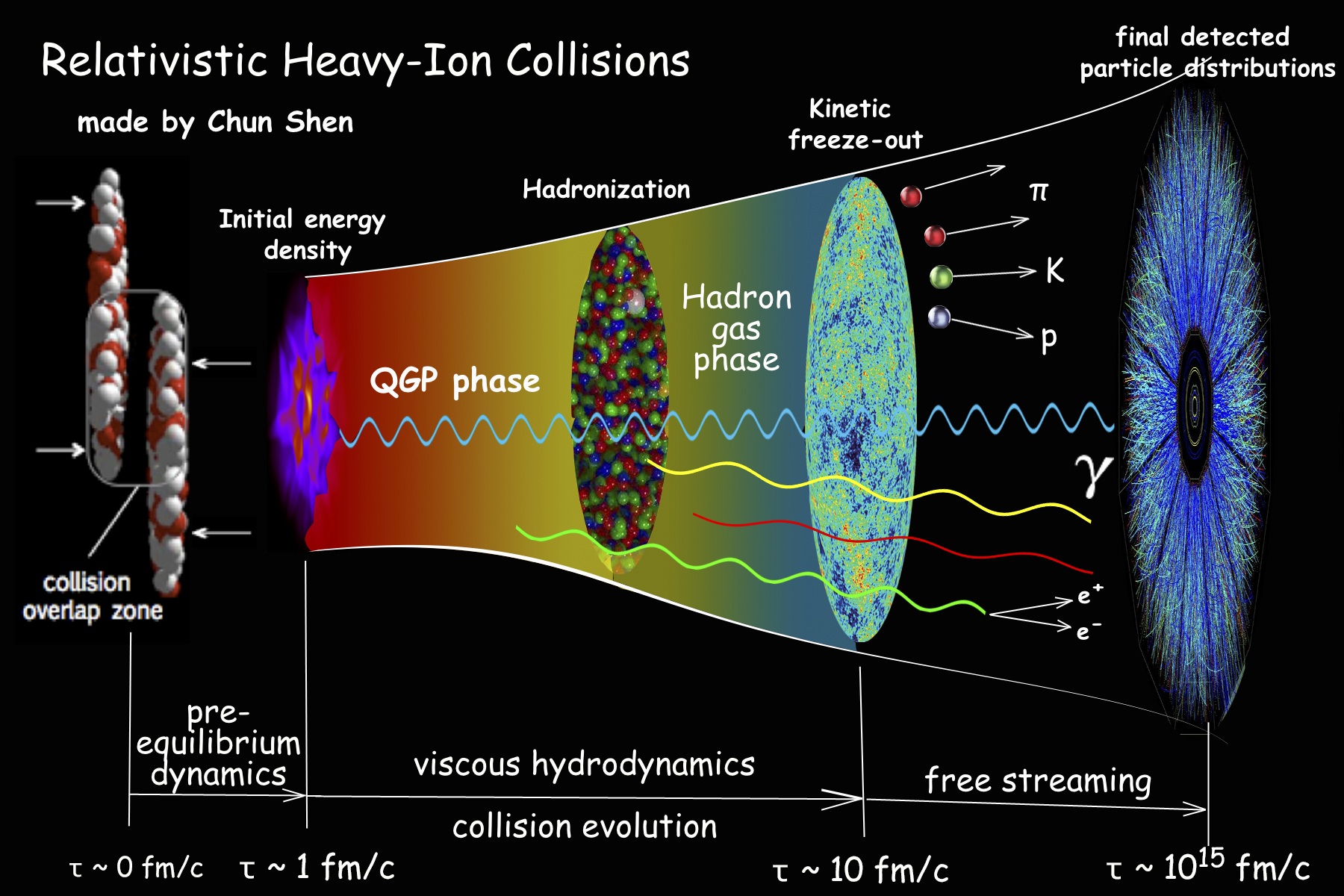}
	\caption{Sketch of the space-time evolution of a relativistic heavy-ion collision.  Figure created by Chun Shen \cite{ChunShenPlot}.}
	\label{fig:HIC}
\end{figure}

Collisions between heavy ions provide an ideal environment to study nuclear matter under extreme conditions. By preparing ion beams with different species and energies one can tune the initial conditions. Most accelerator facilities are able to accommodate a range of ion species and beam energies. This has allowed several experimental programs most notably at the RHIC and SPS accelerators to take full advantage of these capabilities. In what follows we will briefly discuss the main stages of the collision system, see also Fig.~\ref{fig:HIC}, with a particular focus on the production of electromagnetic radiation in the (ultra-)relativistic regime.

\paragraph{Initial State}
Despite all efforts to focus beams and optimize targets, collisions between nuclei are highly stochastic. The initial stage of nucleus-nucleus collisions is determined by the energy and geometrical overlap of the respective projectiles. The geometrical overlap between the colliding nuclei can be quantified in terms of the impact parameter $b$. This parameter is defined as  the distance of closest approach between the centers of the colliding nuclei, orthogonal to the beam direction. The centrality of a collision can be categorized in a range from head-on (central) to more glancing (peripheral) geometries corresponding to values of $0<b<(R_b + R_t)$, where $R_{b,t}$ are the radii of the two colliding species. Many of the soft observables scale with the number of participating nucleons. In contrast, hard processes typically scale with the number of binary collisions although such scaling can be more complex due to, e.g., shadowing effects. The experimental implementations of centrality determination are highly specific to each experiment and regularly involve the measurements of global event characteristics such as a count of the total number of particles, calorimetry, or a combination thereof. Experimental measurements often are cited in terms of a centrality fraction of its total cross-section, e.g. top-10\% most central, or 0-80\% minimum bias collisions. While the matching of these experimental measures to ranges of impact parameters through Glauber model calculation \cite{Glauber:1955qq,Franco:1965wi,Miller:2007ri} is a robust and well-established method in A$+$A to compare collision centralities between experiments, in very asymmetric collisions scattering yields can be modified as is discussed in e.g. \cite{David:2017vvn}. As was pointed out in \cite{Armesto:2015kwa}, in $p+$A collisions this leads to centrality biases in jet and nuclear modification factor measurements which need to be corrected for.

In addition to the size of the geometrical overlap, the centrality in heavy-ion collisions will also determine the shape of the overlap. Heavy-ion experiments going back as early as the Plastic Ball experiment at BEVALAC \cite{Gustafsson:1984ka} have shown evidence of the development of collective behavior in the angular distributions of particles. While the physical mechanism and its impact on the angular distributions strongly depend on the initial energy of the collisions, the effect nonetheless manifests itself as a preferential flow of particles  with respect to an event plane spanned between the impact parameter and the beam direction. Especially in non-central collisions at ultra-relativistic energies, the almond-shaped overlap region sets an initial anisotropy that is observed to result in a pressure-gradient-driven enhancement of in-plane yields when compared to yields that are perpendicular to the event plane. More careful studies of these flow patterns in terms of Fourier decomposition of the angular distributions with respect to the event plane allow for a better interpretation of the various components and at what stage of the evolution they play a role. For example, although the spatial anisotropy of the collision is determined by the geometrical overlap, the subsequent elliptic flow, or $v_2$ component, in ultra-relativistic collisions will arise as the system develops a certain degree of collectivity. On the other hand, the development of a directed flow $v_1$ component is closely tied to the initial compressibility of the colliding nuclei and a measure of its rapid change as a function of rapidity as an indication of the order of a phase transition to a quark-gluon plasma. Still, higher-order components are related to the lumpy structure of the initial collision, $v_3$, or otherwise sensitive to fluctuations in its initial geometry.

\begin{figure}
    \centering
    \includegraphics[width=0.5\textwidth]{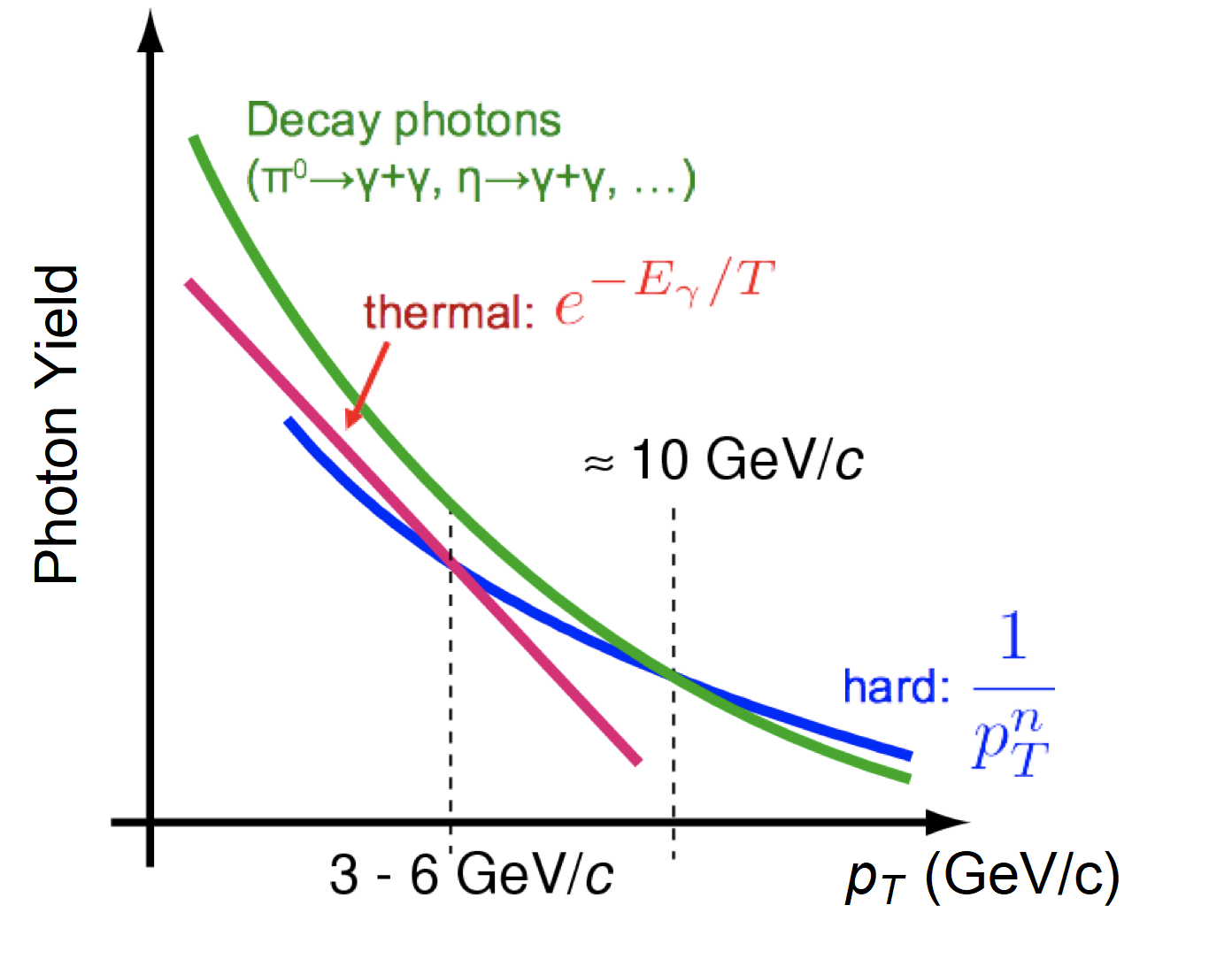}
    \includegraphics[width=0.4\textwidth]{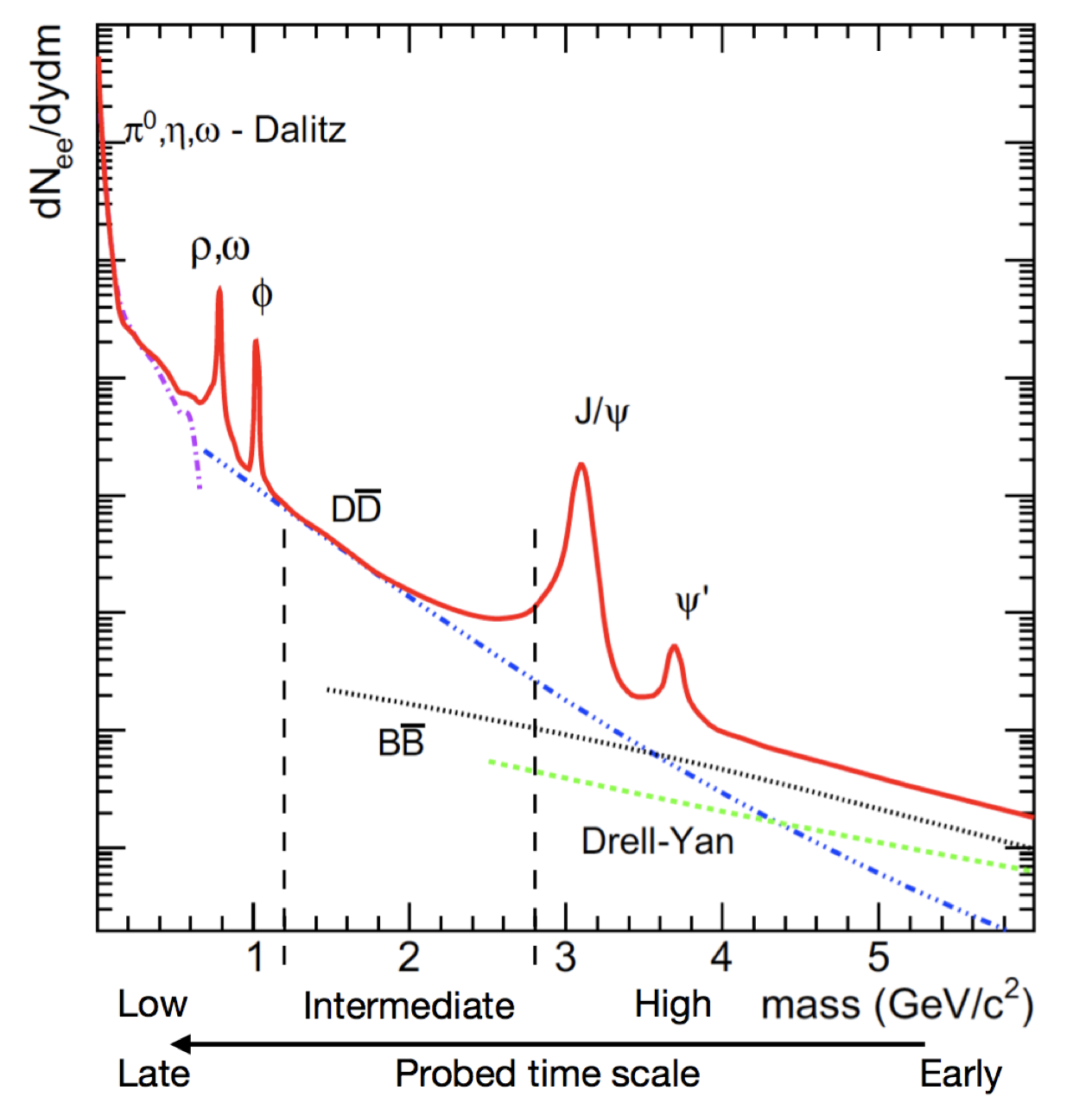}
    \caption{Left: Sketch of (logarithmic) photon yields vs.~transverse momentum from different sources to visualize their respective slopes. Figure created by K.~Reygers and adapted. Right: Sketch of a typical dilepton invariant mass spectrum from p+p collisions at 200 GeV. Figure adapted from \cite{Drees:2009xy}.}
    \label{fig:emsources}
\end{figure}

Where most observables that help probe the initial conditions are ultimately affected by its strongly-interacting constituents of the evolving system, electromagnetic probes by their very nature can escape the medium unscathed. The dominant physics mechanisms behind electromagnetic radiation from the initial state on the whole change from nucleon-nucleon bremsstrahlung at low energies, to parton-parton interactions such as Drell-Yan and heavy-flavor production at ultrarelativistic processes. These sources dominate at the higher dilepton invariant mass ranges, see also Fig.~\ref{fig:emsources}. Hard, i.e.~highly energetic, photons and weak bosons can serve as valuable references for energy-loss measurements using $\gamma$ or $Z$-boson-jet correlations. 

\paragraph{Hot and Dense Matter}
The high energy densities that are reached in the initial stages of the collisions set the stage for the formation of a partonic system which will rapidly expand and locally equilibrate. The build-up of anisotropic flow will be imprinted on the partons and hadrons. Throughout the evolution of the system, photons are created by Compton, annihilation, and bremsstrahlung processes with partonic and hadronic degrees of freedom. The energy spectra of the photons that leave the system will reflect the conditions, such as the temperature, prevalent at that time. The measured spectra, however, will reflect the convolution of the system's evolution which includes an adiabatic cool-down and change in the degrees of freedom of the dominant production processes. Thermal photons dominate the spectra for momenta between, approximately, 2 and 3 GeV$/c$ or for virtual photon masses between 1 and 3 GeV/$c^2$. Effective temperature measurements based on the final, time-integrated, momentum spectra are blue-shifted because of the build-up of radial flow of the expanding system. With 
\begin{align}
T_\mathrm{eff} = \sqrt{\frac{1+\beta_r}{1-\beta_r}} T 
\end{align}
such a blue-shift can be substantial and amount to a factor of 2 at, e.g.,  RHIC energies with a radial flow of $\beta_r\approx 0.6c$, see for example \cite{Shen:2013vja}.
As will be further elaborated in Sect.\ \ref{sec:photon_and_dilepton_rates}, the thermal production rates of photons and dileptons are closely connected. Dileptons play an important role as they can also give access to the time-like kinematic range of EM spectral functions. With invariant masses in the intermediate range, i.e. between the $\phi$ and $J/\psi$ meson,  dileptons can most notably provide for blue-shift free measurement of the thermal radiation. At lower invariant masses, this picture becomes more obscured as the leptonic decay modes of several hadrons will dominate. The additional kinematic `knob' that the dilepton invariant mass brings in addition to its momentum will also allow for experimental access to measurements of, for example, collectivity \cite{NA60:2008ctj,Chatterjee:2007xk,Adamczyk:2014lpa}, the lifetime of the system \cite{Rapp:2014hha,Heinz:1990jw,Adamczyk:2015mmx}, and possibly of transport coefficients such as the electrical conductivity \cite{Atchison:2017bpl}.

\paragraph{Freeze-Out} As the collision system continues to expand and the energy density and temperatures drop below their respective critical values, the color degrees of freedom, partons, are confined again to color-singlet states, hadrons. Within such a hot and dense hadron gas, inelastic collisions can still take place, and change the individual particle numbers. Eventually, the temperature decreases to a point where inelastic interactions are no longer possible and the particle ratios are fixed. This is called chemical freeze-out as it fixes the basic chemical composition of the system. The hot hadron gas will continue to expand and interact until the average interaction length between the hadrons becomes comparable to the size of the system. At that point, the hadrons decouple and the system undergoes a kinetic freeze-out and the momenta of the hadrons will reflect a blue-shifted freeze-out temperature. Electromagnetic radiation not only includes the earlier-mentioned thermal photons but also contributions from hadron decays. Especially light hadrons, such as the $\pi^0$ and $\eta$ mesons overwhelmingly dominate at low energies, while electromagnetic radiation from vector mesons is clearly visible at higher energies near their respective masses, see also Fig.~\ref{fig:emsources}. Not affected by a predominantly strongly interacting medium, dileptons furthermore provide for a unique opportunity to study the effects of chiral symmetry restoration on hadrons, such as the $\rho$ meson and its chiral partner, the $a_1$. Observations of a substantially modified  $\rho$ spectral function may soon see complementary measurements of a modified $a_1$ at masses just above that of the $\phi$ meson \cite{Scomparin:2022bsb}.


\subsection{Electromagnetic Spectroscopy}\label{sec:em-spectroscopy}
\vspace{2mm}

Measurements of electromagnetic probes from heavy-ion collisions can conveniently be divided into two categories: the direct measurement of photons and the indirect measurement of virtual photons in terms of lepton pairs. The former relies on electromagnetic calorimetry and was pioneered in experiments such as the SPS WA80 and WA98 experiments \cite{WA80:1990teq, WA98:2000vxl}. It offers the opportunity to directly measure photons through the detection of electromagnetic showers but typically at the cost of high energy thresholds. On the other hand, dilepton measurements provide means to reconstruct virtual photons and can be used to access low-energy photon signals. Its main challenge will be in careful control of a wide range of background sources. These can be substantial due to the large combinatorial backgrounds from uncorrelated pairs and impurities in the lepton identification. With high multiplicities in heavy-ion collisions, such combinatorial backgrounds lead to signal-to-background ratios of less than 1\%. Consequently, a precise and accurate determination of such backgrounds is paramount. In what follows we first discuss the aspects of lepton identification, followed by the determination of the combinatorial background. 

\paragraph{Particle Identification}
For the identification of electrons and positrons, several types of detectors are used. 
 Typically, the reconstruction of dileptons and single electrons and muons for that matter involves a combination of two or more detectors. The information from these combined detectors will have to establish the particle identification with high purity and provide high precision momentum information. Experimental strategies that address such demands require the combination of trackers with excellent momentum resolution and detectors that can suppress the often very large backgrounds from other particles.
The main challenge in establishing a pure electron sample will be contamination by charged pions.

Several experiments, including CERES, HADES, and PHENIX \cite{CERES:1995rik, HADES:2011nqx, PHENIX:2009gyd}, use Ring-Imaging Cherenkov (RICH) detectors with very high $\gamma$ thresholds ranging between 22 to 35 which, for the latter, effectively suppress pions below momenta of 4.65~GeV/$c$ in contrast to a lower cut-off for electrons at about 20~MeV/$c$. RICH detectors operated in that mode are usually referred to as Hadron Blind Detectors (HBD).  For each of these experiments, tracking detectors serve two main purposes: provide entry and/or exit information for the RICH detector and -in combination with an externally applied magnetic field- measure the lepton momentum and charge. Differences between the experiments are predominantly in the tracking systems and the addition of time-of-flight and pre-shower capabilities to further improve electron identification. Especially in high-multiplicity environments, RICH-only electron identification may suffer from pions whose trajectory may be more likely to overlap with the real electron's ring \cite{PHENIX:2009gyd}.

Time Projection Chambers (TPC) provide for excellent momentum resolutions and e.g.~were utilized in an upgrade of the CERES experiment \cite{CERES:2008asn}.  An additional advantage of the TPC is that at very low momenta, its energy-loss measurements come with sufficient separation between electrons and pions to allow for high-purity lepton identification. Yet, with increasing momentum the electron identification will be contaminated by the pions and kaons. Experiments, such as STAR and ALICE \cite{STAR:2012dzw, ALICE:2020umb} explicitly rely on the inclusion of time-of-flight information to help suppress slower hadrons and thus substantially improve the purity of the energy-loss-based electron identification. However, toward higher momenta, and finite time resolutions, the advantage of the combination of energy-loss and time-of-flight measurements runs out. Not unlike RICH detectors, Transition Radiation Detectors (TRDs) can push the pion-electron separation to higher momenta \cite{ALICE:2017ymw} and when combined with energy-loss-based methods bridge electron identification to the use of other technologies. More recent detector designs such as the CBM detector at GSI expect pion suppression up to $10^3$ and when combined with a TRD of the order of $10^4$, see e.g.~\cite{Lebedev:2010zzb}. Finally, electromagnetic calorimeters combined with externally provided momentum information can push electron identification to much higher momenta, but also require more energetic electrons for the electromagnetic cascades to develop.

Where electron identification suffers from conversions in detector material, for muons, this is not the case thus eliminating a substantial source of background contributions stemming from such conversions. Using muons instead of electrons also eliminates the substantial physical background from $\pi^0$ decays. On the other hand, the purity of muons will be affected by the weak decay of charged pions which yield a substantial `fake' rate of muons. Time-of-flight techniques will only allow access to very low momenta as there is only little difference between muon and pion masses. An effective approach to reducing the contamination from pions is the strategic placement of hadron absorbers. Placing absorbers close to the interaction point will be most effective as it will reduce hadron contamination and the `fake' muon rates. This strategy is employed in experiments such as NA60 \cite{NA60:2008dcb} and the muon arms of the PHENIX and ALICE detectors \cite{PHENIX:2003yhi, ALICE:2008ngc}. For forward measurements in such collider experiments, hadron absorbers can be mounted close to the beamline. However, for mid-rapidity measurements in multi-purpose experiments such as STAR and CMS \cite{Yang:2014xta, CMS:2008xjf}, the absorbers need to be positioned outside time-of-flight walls or electromagnetic calorimeters. In addition to dedicated absorbers, detector layouts often take advantage of the return yokes of their respective magnet systems.

The passage of muons through an absorber involves multiple Rutherford scatterings with its nuclei. While the individual scatterings will only have little effect on the muon's trajectory, many collisions in thick absorbers do lead to substantial deviations. Multiple scattering is independent of the particle's momentum. Consequently, the resolution of momentum measurements based on trackers located behind the absorber will be especially limited at lower momenta where the tracker resolution does not yet dominate. For precise measurements, it is therefore important to match the tracking information from muon detectors behind the absorbers with tracking information from trackers that have been positioned in front of the absorbers. In all, muon identification benefits from higher lab momenta and is thus more feasible in fixed-target experiments and forward instrumentation in collider experiments.

With the different phase-space considerations for muon and electron identification in mind, the reconstruction of virtual photons from dileptons is similar and provides us with both a measure of the momentum and invariant mass of the virtual photon with $\vec{p}_{ll} = \vec{p}_+ + \vec{p}_-$ and 
\begin{align}
 M_{ll} = \sqrt{m^2_- + m^2_+ + 2(E_+ E_- - \vec{p}_+\cdot\vec{p}_-)}
\end{align}
where $m_\pm$, $E_\pm$, and $\vec{p}_\pm$ denote the mass, energy, and momentum of the respective (anti-)leptons. For the case of electrons, $m_e$ typically is much smaller than the invariant dielectron mass, leading to the approximation $M_{ee} = \sqrt{2p_+ p_-(1-\cos\theta)}$. Throughout this paper we follow a naming convention: the low-mass range (LMR) which runs up to $M_{ll}<1.1$~GeV/$c^2$ and is mostly dominated by vector meson decays, followed by the intermediate mass range (IMR) which runs up to 2.9~GeV/$c^2$ marked by the start of the high mass range (HMR) and its charmonium contributions, see also Fig.~\ref{fig:emsources}.

\begin{figure}[t]
	\centering\includegraphics[width=0.44\textwidth]{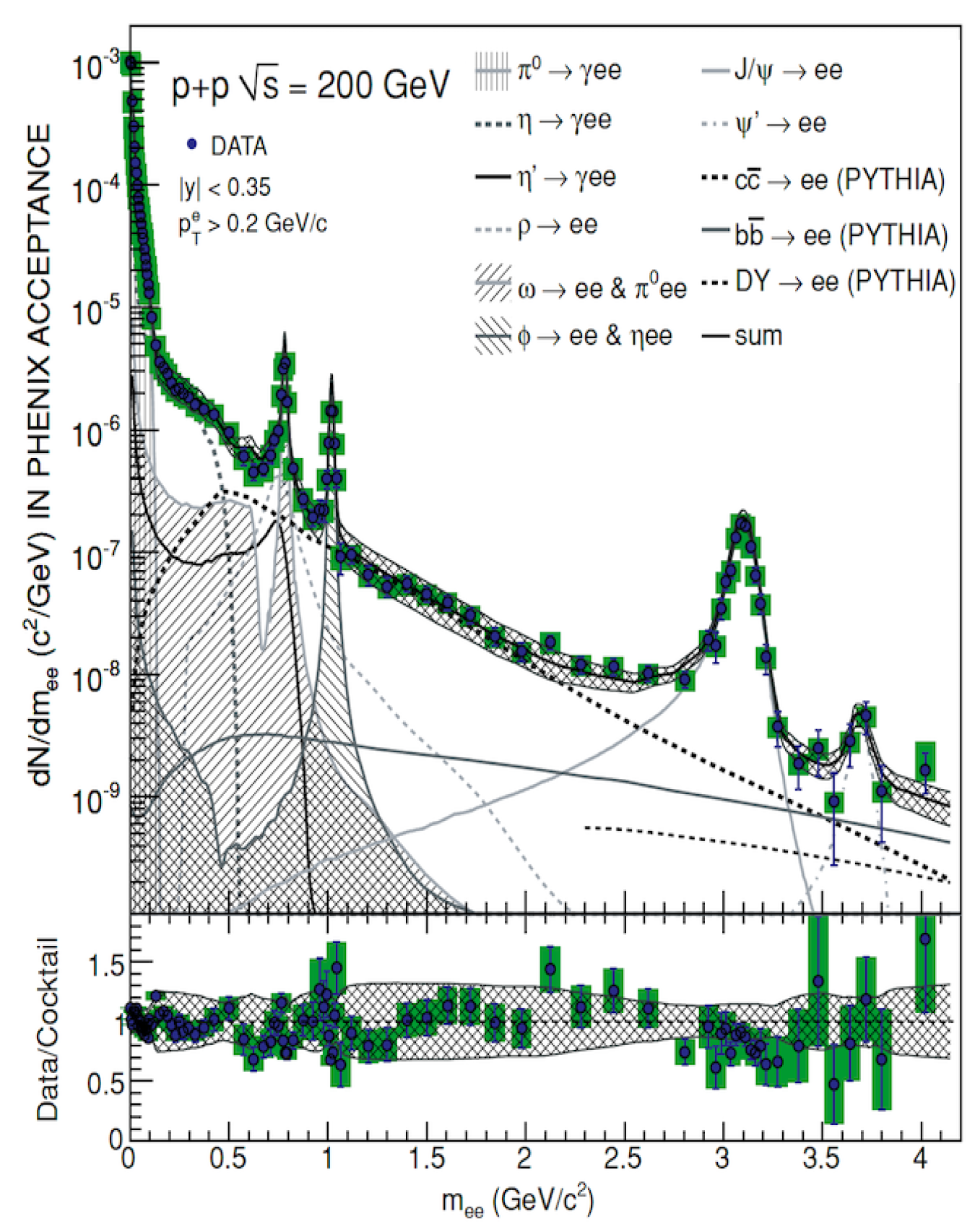}
    \includegraphics[width=0.55\textwidth]{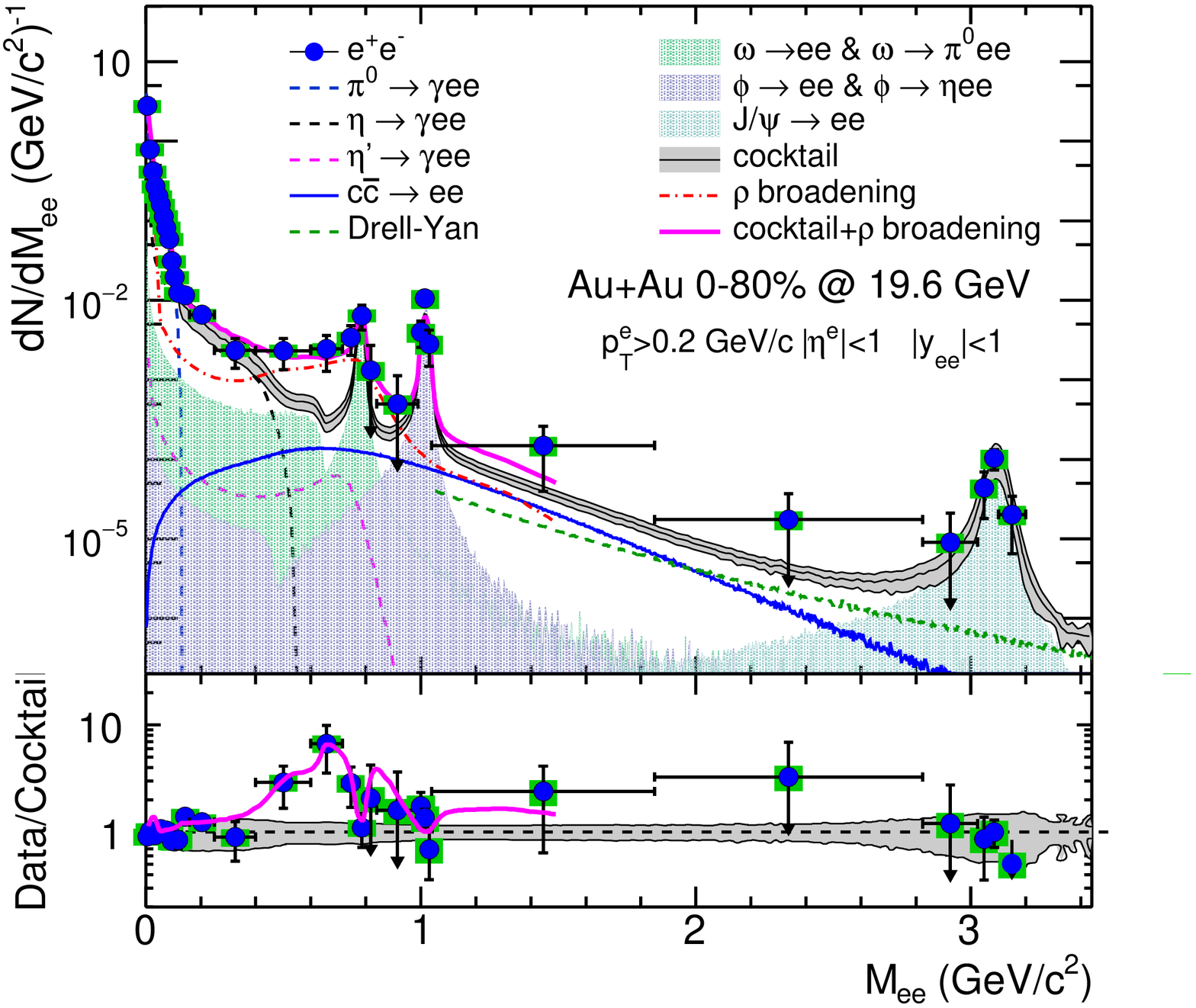}
    \caption{Inclusive invariant mass distributions of dielectron pairs. Left panel: measured by PHENIX in p-p collisions at $\sqrt{s_{\text{NN}}}=200$~GeV compared to cocktail sources originating from the decay of light hadrons ($\pi^0$, $\eta$, $\eta'$), vector meson decay ($\rho$, $\omega$, $\phi$, $J/\psi$, $\psi'$), correlated charm ($c\bar{c}$) and bottom ($b\bar{b}$) decays, and Drell-Yan scattering (DY). Figure adapted from \cite{PHENIX:2009gyd}. Right panel: measured by STAR in Au$+$Au collisions at $\sqrt{s_{\text{NN}}}=19.6$~GeV \cite{Adamczyk:2015mmx} compared to known cocktail sources that exclude the $\rho$ meson. Here, the $\rho$ meson spectral function is based on a theoretical calculation \cite{vanHees:2006ng}.}
	\label{fig:cocktail}
\end{figure}

\paragraph{Background Rejection}
In addition to a background that arises from impurities in the identification of photons or leptons, a large low-momentum source of dilepton background involves conversion pairs from interactions with detector support materials. Rejection techniques, based on a combination of small pair momentum and small opening angles have been very effective in removing such contributions, albeit at the price of a reduced phase space. Such loss of phase space can only be recouped by careful use of ultra-low mass materials such as the proposed ALICE ITS3 silicon tracker \cite{Suljic:2021hfl}.

A major challenge for lepton pair invariant mass spectroscopy is the large combinatorial background that follows from the many random combinations between unlike sign leptons on top of which the physical signals of interest sits. Not only do wrongly identified leptons contribute to this \cite{NA60:2008dcb}, but so do the many uncorrelated pairs that are reconstructed from leptons that did originate from the original virtual photon. Such uncorrelated pairs scale quadratically with event multiplicity and can quickly overwhelm the signal of interest resulting in signal-to-background ratios that can be at the per-mille level in heavy-ion collisions at the LHC. Even small statistical fluctuations in the background will therefore have large effects on the signal extraction and the determination of this background requires high precision. The combinatorial background strongly depends on the measured multiplicity, but is also highly sensitive to the detector's acceptance, which can be different on an event-by-event basis in collider experiments as different collision vertices result in variations of the acceptance for each event.  As described in \cite{Crochet:2001qd}, basically two approaches exist to determine and subtract the combinatorial background: a like-sign technique and a mixed-event technique. Using like-sign lepton pairs from the same collision event is an effective way to accurately describe the background but suffers from the limited statistical precision of a single event. The mixed-event method \cite{PHENIX:2009gyd, STAR:2015tnn, ALICE:2018fvj} uses unlike-sign pairs formed from multiple events. These events, however, must resemble the original event in terms of the event multiplicity and geometric acceptance. The latter is ensured by considering events grouped in terms of collision vertex location (in collider experiments) and the angle of the 2$^\mathrm{nd}$-order event plane. This technique can improve the precision of the combinatorial background  by including many more similar events.

However, both techniques come with limitations in the accuracy with which the background can be described. On the one hand, the mixed-event methodology cannot account for physical background effects from jet correlations at predominantly higher invariant masses and from correlated cross-pair backgrounds at low invariant masses where two or more lepton pairs can originate from the same meson decay. On the other hand, the like-sign same-event methodology is more sensitive to charge-dependent differences in detector acceptance. In typical tracking setups that involve a magnet field, same-signed pairs with small opening angles are more likely to be affected by the presence of non-instrumented regions between detector segments than unlike-sign pairs.  For both techniques low pair statistics, especially at low-$\sqrt{s_{NN}}$ experiments may necessitate alternative approaches in which the charge asymmetry is calculated \cite{Adamczewski-Musch:2019byl}. It should be noted that the mixed-event technique is also known to not completely remove physical correlations in those cases where the detector's phase space acceptance is small \cite{Zajc:1984vb}.

\paragraph{Hadronic Cocktail}
With the combinatorial background and contributions from detector materials accounted for by a mix of same-event and mixed-event techniques, the remaining backgrounds are of a physical nature and could be either the signal of interest or the physical background to it.  Typically, the physical sources originate in a range of leptonic decay modes of light mesons at low invariant masses ($\pi^0\rightarrow\gamma e^+e^-,\eta\rightarrow\gamma e^+e^-, \eta'\rightarrow\gamma e^+e^-,  \omega\rightarrow\pi^0 e^+e^-, \omega\rightarrow\eta e^+e^-$, and $\phi\rightarrow e^+e^-, \phi\rightarrow\eta e^+e^-$), charmonium ($J/\psi\rightarrow e^+e^-$) at high invariant mass. For collision systems and energies where these sources have well-established, measured yields and momentum spectra these can then be used in  Monte-Carlo simulations. Those calculations will use published dilepton branching ratios \cite{ParticleDataGroup:2020ssz} to generate a dilepton continuum that will take into account the detector's unique features such as acceptance and momentum resolution. For collision energies that do not have such reference data, yield interpolations combined with Tsallis Blast-Wave fits based on other particles can be used \cite{Tang:2008ud}. Furthermore, meson masses spectra for Dalitz decays are handled differently from those for direct decays which follow narrow Breit-Wigner mass distributions. For Dalitz decays the meson mass distribution follows a Kroll-Wada distribution as is discussed in
\cite{Kroll:1955zu, NA60:2009una, Achasov:2000ne, Landsberg:1985gaz, STAR:2012dzw, PHENIX:2009gyd}. 

Dileptons from prompt Drell-Yan processes ($q\bar{q}\rightarrow\gamma^*\rightarrow l^+l^-$) and correlated pairs from (open) heavy-quark decays ($c\bar{c}\rightarrow D\bar{D} \rightarrow X l^+l^-$) will contribute throughout the dilepton continuum. Their respective contributions are typically modeled with PYTHIA-based calculations \cite{Sjostrand:2006za,Sjostrand:2007gs} that are scaled with the number of binary nucleon-nucleon collisions. However, in a heavy-ion environment, this scaling may ignore nuclear modifications to the charm production. Unless such charm contributions can be directly, experimentally identified and removed \cite{Damjanovic:2008ta}, uncertainties in production cross sections and the degree of de-correlation will feed into the systematic uncertainties of the charm component of the hadron cocktail.

As mentioned before, the hadronic cocktail comprises of all known physical sources that contribute to the dilepton continuum. While for $pp$ collisions experiments will often include in this cocktail the $\rho$ meson, this is not necessarily the case for hadron cocktails that are used in $A+A$ data where the spectral function of the $\rho$ meson can be substantially modified by the medium, see Sec.~\ref{sec:vector_mesons_in_medium}. In such cases, the broadened $\rho$ meson yields are part of the excess which may include other experimentally unknown thermal radiation contributions from the hadron gas and QGP phase. Examples of both scenarios are shown in Fig.~\ref{fig:cocktail} where in the left panel for $pp$ collisions the cocktail includes the known vacuum $\rho$ mass spectra, while in the right panel the vacuum $\rho$ meson has been removed and instead theoretical model calculations of its mass spectrum are tested.

\clearpage 
\section{Theoretical aspects of electromagnetic probes}
\label{sec:theoretical_aspects}

Photons and dileptons represent unique probes in heavy-ion collisions that can be used in various ways to learn about the properties of the produced hot and dense medium. In particular, dilepton invariant mass spectra are the only observable which give direct access to the electromagnetic spectral function and its energy dependence. In the low mass regime, the EM spectral function is well described in terms of spectral functions of light vector mesons. This phenomenological finding is referred to as Vector Meson Dominance (VMD) and entails that the electromagnetic-hadronic interaction within a medium can be described by the exchange of vector mesons. The theoretical description of vector mesons and their spectral functions, therefore, plays a central role in the interpretation and computation of dilepton spectra. An important ingredient in the description of such spectral functions at finite temperature and density is chiral symmetry, one of the central properties of QCD. The way in which such spectral functions change with, e.g., temperature, is directly connected to changes in the extent of chiral symmetry breaking and also to changes in the spectral functions of their chiral partners.

In this chapter, we will discuss important theoretical aspects of electromagnetic probes and summarize basic principles of QCD such as chiral symmetry. In particular, we will introduce the electromagnetic spectral function and the concept of VMD. Expressions for photon and dilepton production rates will be presented within the thermal field theory framework, i.e.~in terms of the in-medium EM spectral function, as well as within the microscopic framework of relativistic kinetic theory.

\subsection{QCD phase diagram and symmetries}
\label{sec:QCD_and_symmetries}
\vspace{2mm}

Quantum Chromodynamics is the commonly accepted theory of the strong interaction. It represents a non-Abelian gauge theory with the fundamental degrees of freedom being quarks and gluons. The dynamics of the quarks and gluons are controlled by the gauge-invariant QCD Lagrangian given by
\begin{align}
    \mathcal{L}_{\rm{QCD}}=\sum_q \bar{q}(i\gamma^\mu D_\mu-m_q)q-\frac{1}{4}G_{\mu\nu}^a G^{\mu\nu}_a,
\end{align}
where the sum is over the different quark flavors, $q$ is the quark field, $m_q$ is the quark mass, $\gamma^\mu$ are the Dirac matrices, $D_\mu=\partial_\mu+i g \lambda_a  A_\mu^a/2$ is the covariant derivative with the gluon field $A_\mu^a$ and color index $a$, the strong coupling constant $g$, and the Gell-Mann matrices $\lambda_a$, and $G_{\mu\nu}^a$ is the gluon field strength tensor,
\begin{align}
    G_{\mu\nu}^a = \partial_\mu A_\nu^a-\partial_\nu A_\mu^a+g f^{abc} A_\mu^b A_\nu^c,
\end{align}
with the SU(3) structure constants $f^{abc}$. For more details, see for example the review in \cite{ParticleDataGroup:2020ssz}.

So far, no color-charged particles have been observed in an isolated state. This phenomenon, that all color-charged particles seem to be confined within color-neutral composite states such as hadrons, is referred to as color confinement or simply confinement. Although this phenomenon is not yet understood in terms of an analytic proof, it can be demonstrated by, e.g., lattice QCD calculations where one can show that the gluon field forms a flux tube or `string' between a static quark-antiquark pair which holds them together. The energy stored in such a flux tube is proportional to the separation between the two particles and eventually becomes large enough to create new particles which again form color-neutral objects, see for example \cite{Bali:1994de,Greensite:2011zz}. Color confinement is therefore closely connected to the local non-Abelian SU(3) color gauge symmetry of QCD and does not occur, for example, in the case of the Abelian U(1) gauge symmetry of Quantum Electrodynamics (QED).

In addition to the local SU(3) gauge symmetry, the QCD Lagrangian possesses several global symmetries. The most relevant one in the present context is chiral symmetry which is an exact symmetry in the limit of vanishing current quark masses and refers to transformations associated to left- and right-handed fields
\begin{align}
    q_{L,R}= \frac{1}{2}(1\mp\gamma_5)q,
\end{align}
see for example \cite{Koch:1997ei,Giusti:2015kwf,Sazdjian:2016hrz} for reviews. The symmetry properties of the QCD Lagrangian become more apparent when rewriting it in terms of left- and right-handed fields, which gives
\begin{align}
    \mathcal{L}_{\rm{QCD}}=\sum_q \bar{q}_L i\gamma^\mu D_\mu q_L+\bar{q}_R i\gamma^\mu D_\mu q_R - \bar{q}_L m_q q_R+\bar{q}_R m_q q_L -\frac{1}{4}G_{\mu\nu}^a G^{\mu\nu}_a.
\end{align}

For massless quarks, the QCD Lagrangian is invariant under chiral transformations of the form
\begin{align}
    q_{L,R}\rightarrow e^{-i\vec{\alpha}_{L,R}\cdot \vec{\tau}/2}q_{L,R},
\end{align}
where $\vec{\alpha}$ is a vector of small real angles and $\vec{\tau}$ are the Pauli matrices in isospin space. The corresponding chiral symmetry group is denoted as
\begin{align}
    \text{U}_L(N_f)\otimes \text{U}_R(N_f)=\text{U}_V(1) \otimes \text{SU}_V(N_f) \otimes \text{SU}_A(N_f) \otimes \text{U}_A(1),
\end{align}
where the indices $V$ and $A$ refer to `vector' and `axial-vector' transformations. For small quark masses, as in the case of up and down quarks, the QCD Lagrangian is still approximately invariant under chiral rotations. The $\text{U}_V(1)$ symmetry corresponds to the conservation of baryon number while the $\text{U}_A(1)$ symmetry is broken upon quantization and is connected to the large mass of the $\eta'$ meson, see for example \cite{Weinberg:1996kr}. The remaining subgroup $\text{SU}_V(N_f) \otimes \text{SU}_A(N_f)$, often referred to as chiral symmetry, is broken spontaneously down to the vector subgroup by the dynamical formation of a quark condensate $\langle \bar{q} q \rangle=\langle \bar{q}_L q_R+\bar{q}_R q_L\rangle $. Lattice simulations with dynamical up, down and strange quarks \cite{Fukaya:2010na} show this quark or chiral condensate (of the two lightest quarks) to be 
\begin{align}
    \frac{1}{2}\langle 0 | \bar{u}u+\bar{d}d|0 \rangle \simeq -(234(04)(17) \text{MeV})^3.
\end{align}
The breaking of chiral symmetry has profound consequences on the properties of hadrons. The hadron spectrum, for example, would contain parity partners with degenerate masses in the case of an unbroken chiral symmetry. Due to the explicit and spontaneous breaking of chiral symmetry, however, parity partners such as the sigma meson and the pion, the $\rho(770)$ and the $a_1(1260)$, or the nucleon $N(940)$ and the $N(1535)$ show a significant mass splitting. The pions of course enjoy a special status among the hadronic states since they are the pseudo-Goldstone bosons connected to the spontaneously broken chiral symmetry and thus have a comparably small mass.

Another important effect that determines the non-perturbative structure of the QCD vacuum is given by the interaction of gluons often expressed in terms of the concept of a gluon condensate $G^2$. The gluon condensate is related to the trace of the energy-momentum tensor by
\begin{align}
    \langle T_\mu^\mu \rangle = \epsilon-3P = -\langle G^2 \rangle +m_q \langle \bar{q} q \rangle,
\end{align}
with $G^2=-(\beta(g)/2g)G_a^{\mu\nu}G^a_{\mu\nu}$ where $\beta(g)$ is the renormalization-group beta function which is connected to the breaking of scale invariance, see also \cite{Friman:2011zz,Rapp:2009yu}. This non-perturbative structure of the QCD vacuum gives rise to most of the visible mass in the universe, i.e.~of hadrons like the proton and the neutron, while only a small fraction is generated by the electro-weak sector due to the coupling to the Higgs meson.

\begin{figure}[t]
	\includegraphics[width=0.44\textwidth,height=0.33\textwidth]{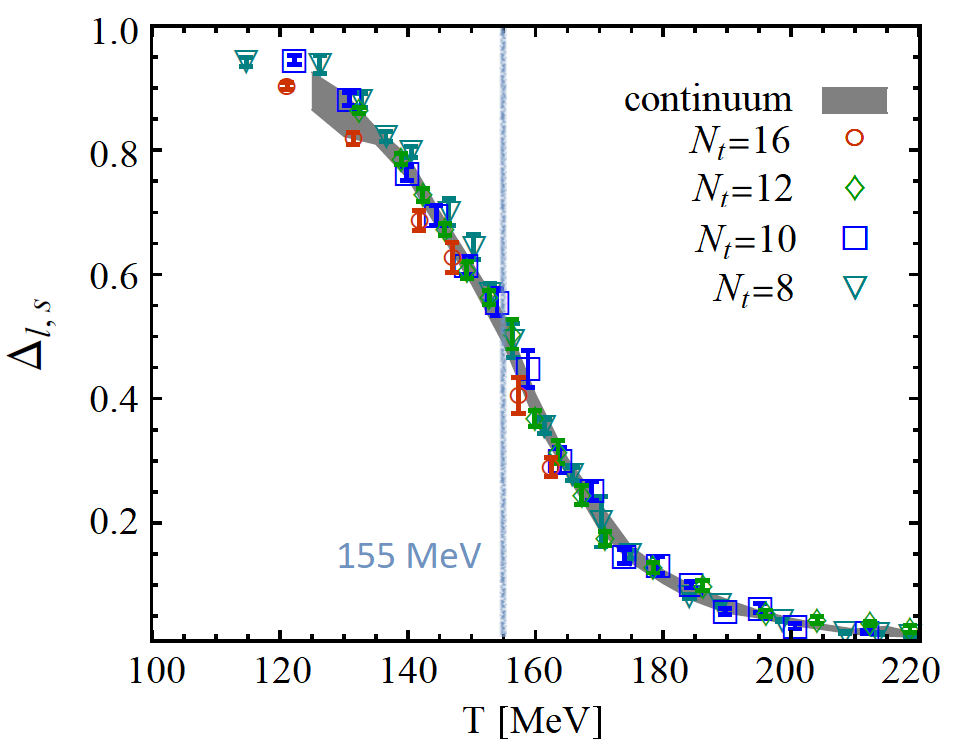}
	\hspace{0.07\textwidth}
	\includegraphics[width=0.44\textwidth, height=0.33\textwidth]{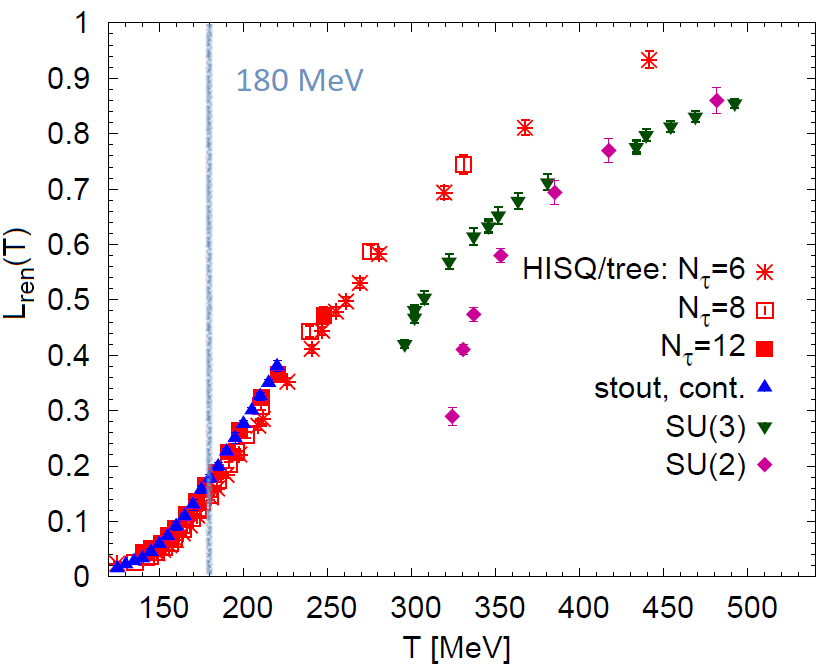}
	\caption{Left: The subtracted chiral condensate $\Delta_{l,s}$ is shown as a function of temperature. The grey band represents the continuum result, obtained with the stout action by the Wuppertal-Budapest collaboration \cite{Borsanyi:2010bp,Borsanyi:2011bn, HotQCD:2018pds}. Right: Temperature dependence of the Polyakov loop in SU(2) and SU(3) pure gauge theory and QCD \cite{Bazavov:2013txa}. Figures adapted from \cite{Borsanyi:2011bn} (left) and \cite{Bazavov:2013txa} (right).}
	\label{fig:order_parameters}
\end{figure}

Numerical simulations of lattice QCD show that the quark condensate decreases with increasing temperature in a smooth fashion which gives rise to a chiral crossover transition. Interestingly, the chiral transition is accompanied by the dissolution of hadrons into quarks, i.e.~the deconfinement transition. While the quark condensate acts as an approximate order parameter for chiral symmetry breaking, the Polyakov loop variable
\begin{align}
    L(\vec{x})=\text{Tr}\,\mathcal{P} \exp \left[ ig\int_0^{1/T} dt A_0 (t,\vec{x})\right],
\end{align}
where $\mathcal{P}$ denotes path ordering of the exponential, represents an order parameter for the confinement-deconfinement transition in the limit of static quarks, and is connected to the center symmetry $Z(3)$ of the QCD action, see also  \cite{Friman:2011zz}. The deconfined and chirally restored phase of strongly interacting matter at high temperatures is commonly referred to as the quark-gluon plasma (QGP). The temperature dependence of the (subtracted) light condensate $\Delta_{l,s}$ as well as of the Polyakov loop is shown in Fig.~\ref{fig:order_parameters} as obtained from lattice QCD \cite{Borsanyi:2011bn,Bazavov:2013txa}, where $\Delta_{l,s}$ is defined as
\begin{align}
    \Delta_{l,s}=\frac{\langle \bar{\psi} \psi \rangle_{l,T}-m_l/m_s \langle \bar{\psi} \psi \rangle_{s,T}}{\langle \bar{\psi} \psi \rangle_{l,0}-m_l/m_s \langle \bar{\psi} \psi \rangle_{s,0}},
\end{align}
with $l=u,d$ and the chiral condensate
\begin{align}
    \langle \bar{\psi} \psi \rangle_{q}=\frac{T}{V}\frac{\partial \ln Z}{\partial m_q.}
\end{align}

For the chiral crossover temperature a value of $T\sim 155$~MeV was found \cite{Borsanyi:2010bp,Borsanyi:2011bn, HotQCD:2018pds} while the deconfinement temperature is more ambiguous and quoted as $T\sim 180$~MeV \cite{Bazavov:2013txa}.

\begin{figure}[t]
	\centering\includegraphics[width=0.55\textwidth]{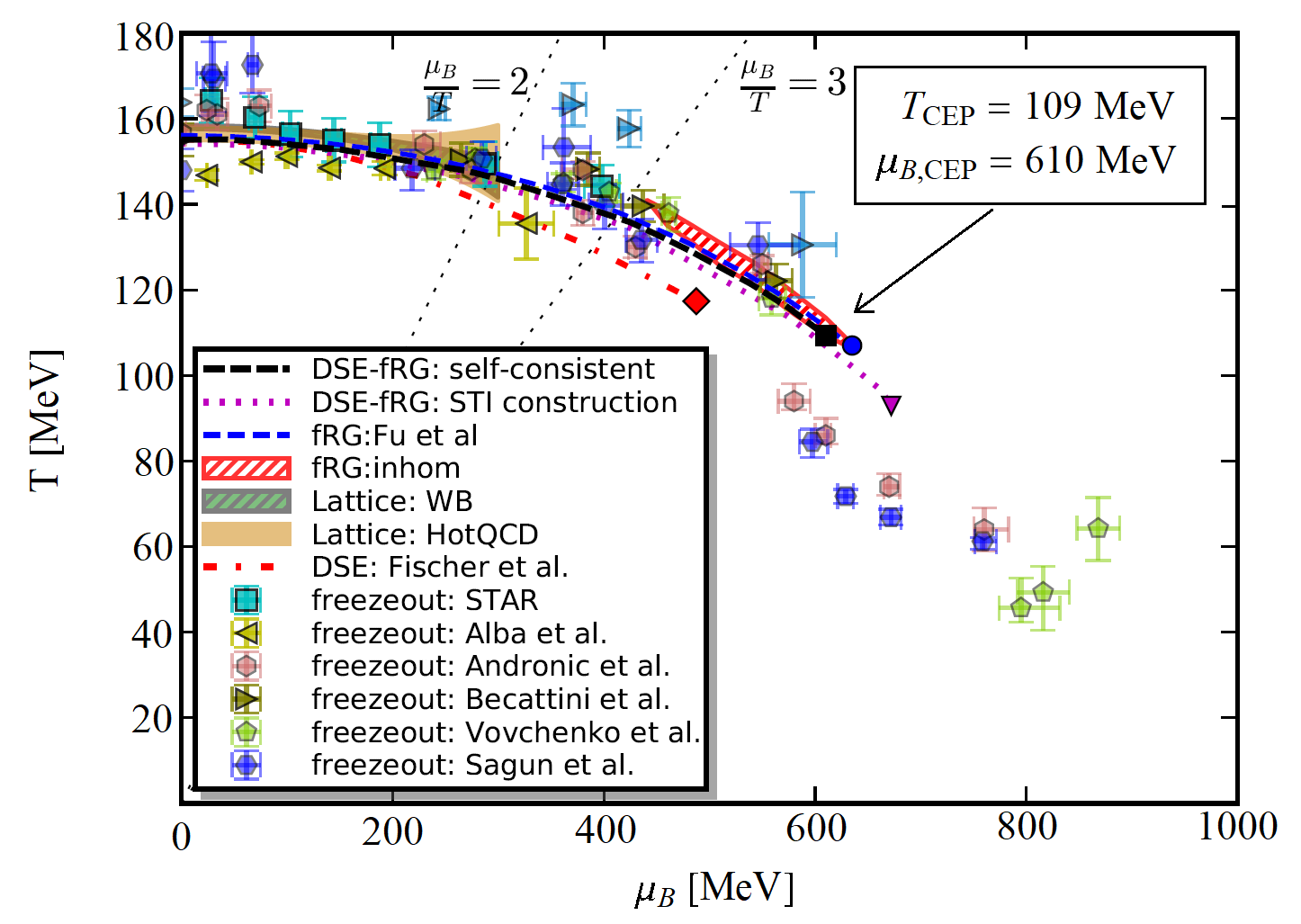}
	\caption{Phase diagram of 2+1 flavor QCD as obtained with functional methods in \cite{Gao:2020fbl} in comparison to other theoretical results and phenomenological freeze-out data. The blue dashed line represents the result obtained in \cite{Gao:2020fbl}. For small chemical potential, the results agree well with the FRG and DSE \cite{Fu:2019hdw,Fischer:2018sdj} as well as with lattice results from the Wuppertal-Budapest collaboration \cite{Borsanyi:2020fev} and the HotQCD collaboration \cite{HotQCD:2018pds}. In the red hatched regime the FRG results from \cite{Fu:2019hdw} show a minimum of the pion dispersion at non-vanishing spatial momentum together with a sizeable chiral condensate which may indicate an inhomogeneous regime. Freeze-out data: \cite{STAR:2017sal} (STAR), \cite{Alba:2014eba} (Alba et al.), \cite{Andronic:2017pug} (Andronic et al.), \cite{Becattini:2016xct} (Becattini et al., without afterburner (light blue) and with afterburner (dark green)), \cite{Vovchenko:2015idt} (Vovchenko et al.), and \cite{Sagun:2017eye} (Sagun et al.). Figure adapted from \cite{Gao:2020fbl}.}
	\label{fig:phase_diagram_2}
\end{figure}

At finite chemical potential, lattice QCD computations are hampered by the fermion sign problem which makes methods like Taylor expansion, re-weighting schemes or extrapolation from imaginary chemical potential necessary in order to obtain results for finite chemical potential, see for example \cite{Karsch:2003jg,deForcrand:2006pv,Kaczmarek:2011zz,Endrodi:2011gv}.
Other options for computations at finite chemical potential are for example given by (resummed) perturbation theory, chiral perturbation theory, BCS theory, or nuclear many-body theory. In recent years, functional methods like the Functional Renormalization Group or Dyson-Schwinger equations have proven to be particularly versatile and useful to study QCD at finite temperature and density. In Fig.~\ref{fig:phase_diagram_2} recent results on the chiral phase diagram of QCD as obtained from functional methods as well as from lattice QCD are shown together with experimental freezeout data \cite{Gao:2020fbl}. Based on the functional approach used in \cite{Gao:2020fbl} the chiral crossover temperature is found to be $T_c=155.1$~MeV while the chiral critical endpoint is located at a baryon chemical potential of $\mu_{B,\text{CEP}}=610$~MeV and a temperature of $T_\text{CEP}=109$~MeV.

\subsection{Photon and dilepton production rates}
\label{sec:photon_and_dilepton_rates}
\vspace{2mm}

In order to investigate the properties of strong-interaction matter in different regions of the QCD phase diagram experimentally, heavy-ion collisions are the only viable possibility. While quantities like the chiral condensate cannot be measured directly, invariant-mass spectra of short-lived resonance decays, $h\rightarrow h_1+h_2$ with a lifetime, $\tau_h$, smaller than the lifetime of the fireball, $\tau_{\text{FB}}\simeq 10$~fm/c, can in principle be used to extract information on the invariant-mass distribution, i.e., the spectral function, at the point of decay. Since hadronic final states are, however, likely to suffer from rescatterings which destroy the invariant-mass information, electromagnetic probes provide an ideal alternative. Photons and dileptons don't interact via the strong force and thus have a long mean-free path as compared to the size of the fireball, $R_\text{FB}\simeq10$~fm. They can therefore carry information on the hot and dense medium directly to the detector.

\begin{figure}[h!]
	\centering\includegraphics[width=0.30\textwidth]{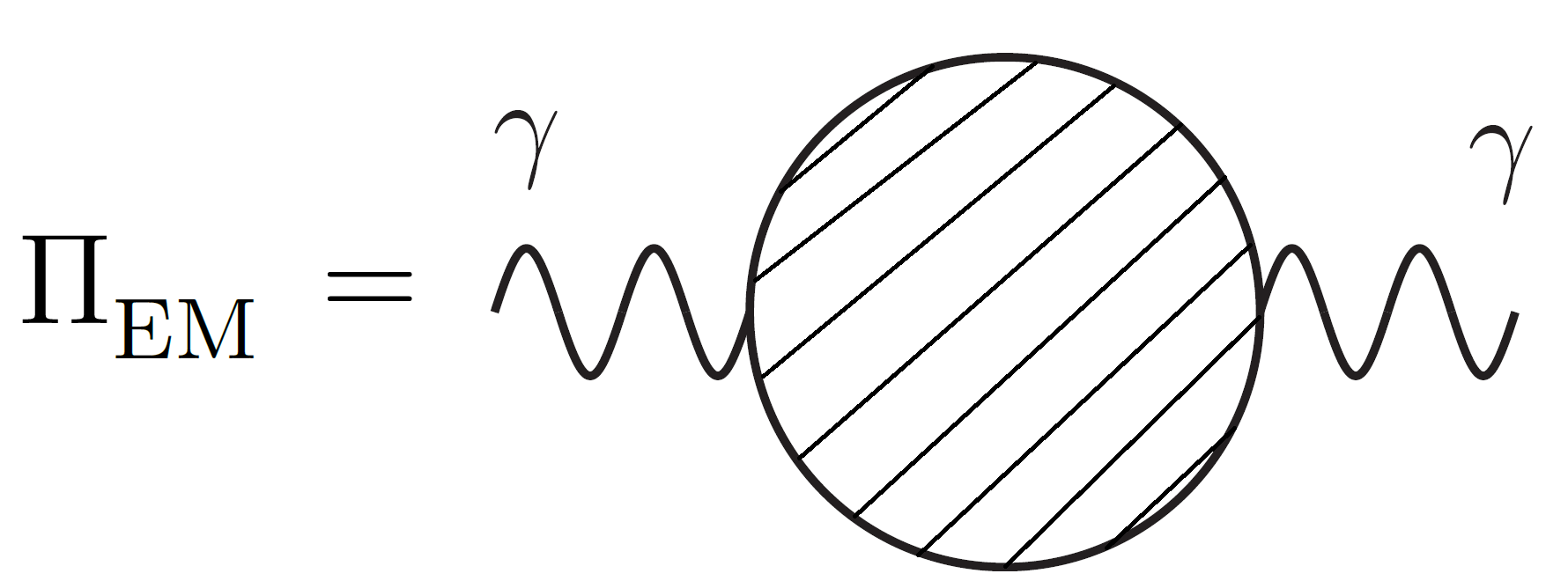}
	\caption{Diagrammatic representation of the EM current-current correlation function $\Pi_{\rm EM}$, cf.~Eq.~(\ref{eq:EM_correlator}).}
	\label{fig:EM_correlator}
\end{figure}

The emission rates of photons and dileptons can be obtained either by using thermal field theory or relativistic kinetic theory. In thermal field theory, the central role is played by the (retarded) electromagnetic (EM) current-current correlation function, defined as
\begin{align}
\Pi_{\rm EM}^{\mu \nu}(M,p;\mu_B,T)  = -\ii \int d^4x \ e^{ip\cdot x} \
\Theta(x_0) \ \tave{[j_{\text EM}^\mu(x), j_{\text EM}^\nu(0)]},
\label{eq:EM_correlator}
\end{align}
where $\langle \langle \cdot \rangle \rangle$ denotes the expectation value at finite temperature, see Fig.~\ref{fig:EM_correlator} for a graphical representation and for example \cite{Friman:2011zz} for a review.

In a partonic basis, considering only the three lightest quarks, the EM current takes the form
\begin{align}
j^\mu_{\text EM}=\frac{2}{3}\bar{u}\gamma^\mu u-\frac{1}{3}\bar{d}\gamma^\mu d - \frac{1}{3}\bar{s}\gamma^\mu s.
\end{align}
This can be rearranged into good isospin states which naturally leads to the hadronic basis according to
\begin{align}
j^\mu_{\text EM}&=
\frac{1}{2}(\bar{u}\gamma^\mu u - \bar{d}\gamma^\mu d) + 
\frac{1}{6}(\bar{u}\gamma^\mu u + \bar{d}\gamma^\mu d) -
\frac{1}{3} \bar{s}\gamma^\mu s \nonumber \\
&=\frac{1}{\sqrt{2}}j_\rho^\mu +\frac{1}{3\sqrt{2}}j_\omega^\mu-\frac{1}{3}j_\phi^\mu,
\end{align}
with the properly normalized hadronic currents $j_v^\mu$ ($v=\rho,\omega,\phi$) with isospin $I=1$ ($\rho$) and $I=0$ ($\omega$ and $\phi$). Converting the isospin coefficients into numerical weights in the EM spectral function gives
\begin{align}
\im \Pi_{\rm EM}\sim \im D_\rho + \frac{1}{9} D_\omega + \frac{2}{9} \im D_\phi,
\end{align}
where $D_v$ are the vector meson propagators. This identifies the isovector $(\rho)$ channel as the dominant source, see also the discussion on the vector dominance model in the next section. Experimentally, the relative contribution is even larger, as given by the electromagnetic decay widths, $\Gamma_{\rho\rightarrow ee}/\Gamma_{\omega \rightarrow ee} \simeq 11$, see also \cite{Rapp:2009yu,Rapp:2011is}.

\begin{figure}[h!]
	\centering\includegraphics[width=0.6\textwidth]{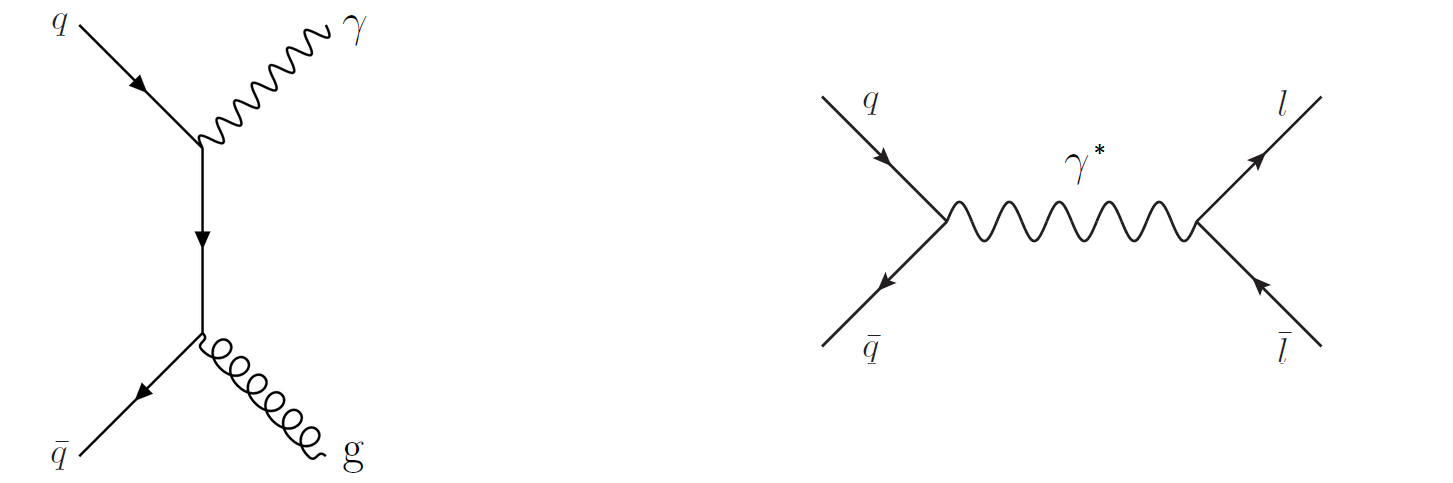}
	\caption{ Feynman diagrams for photon (left) and dilepton (right) production from quark-antiquark annihilation. While the left diagram contains one EM vertex and one strong-interaction vertex, the right diagram contains only EM interactions. The squares of these diagrams (as needed for computing the rate) are therefore proportional to $\alpha_{\text{EM}}\alpha_s$ and $\alpha_{\text{EM}}^2$, respectively. See also Eqs.~(\ref{eq:photon1})-(\ref{eq:dilepton1}).}
	\label{fig:photons_dileptons_alpha}
\end{figure}

The thermal emission rates of photons and dileptons are then given by \cite{McLerran:1984ay, GaleKapusta1991}
\begin{align} 
p_0 \frac{dN_\gamma}{d^4x \ d^3p}& = 
-\frac{\alpha_{\rm EM}}{\pi^2} \
f^B(p_0;T) \  \frac{1}{2}\ g_{\mu\nu} 
\ \im \Pi_{\rm EM}^{\mu\nu} (M=0,p;\mu_B,T),
\label{eq:photon1} \\
\frac{dN_{ll}}{d^4x \ d^4p}& =
-\frac{\alpha_{\rm EM}^2}{\pi^3 M^2}\ L(M)\
f^B(p_0;T) \  \frac{1}{3} \ g_{\mu\nu} 
\ \im \Pi_{\rm EM}^{\mu\nu} (M,p;\mu_B,T),
\label{eq:dilepton1}
\end{align}
where $f^B(p_0;T)=1/(e^{p_0/T}-1)$ denotes the thermal Bose distribution function, $\alpha_{\rm EM}\simeq 1/137$ the EM coupling constant, $p^0$ and $p$ are the energy and momentum of the photon or dilepton in the local rest frame of the medium, and $L(M)$ is the lepton phase-space factor
\begin{align}
L(M)=\sqrt{1-\frac{4m^2}{M^2}}\left( 1+\frac{2m^2}{M^2}\right),
\end{align}
which rapidly reaches 1 above the threshold given by twice the lepton mass $m$. We note that photon and dilepton rates are governed by the same underlying object, i.e., the (retarded) EM spectral function, $\im \Pi_{\rm EM} (M,p;\mu_B,T)$, albeit in different kinematic regimes.  We also note that in terms of the strong coupling $\alpha_s$ the leading order in the photon rate is $\mathcal{O}(\alpha_s)$ while for the dilepton rate we have $\mathcal{O}(\alpha_s^0)$. This can for example be seen by comparing the Feynman diagrams for photon and dilepton production from quark-antiquark annihilation, see Fig.~\ref{fig:photons_dileptons_alpha}.

Alternatively, the emission rate of (virtual) photons can be expressed within the microscopic framework of relativistic kinetic theory as
\begin{align}
p_0 \frac{dN_\gamma}{d^4x \ d^3p}=
\int \frac{d^3q_1}{2(2\pi)^3E_1}
\frac{d^3q_2}{2(2\pi)^3E_2}\frac{d^3q_3}{2(2\pi)^3E_3}
(2\pi)^4
\delta^{(4)}(q_1+q_2\rightarrow q_3+p)
\left|{\cal M}\right|^2\frac{f(E_1)f(E_2)[1\pm f(E_3)]}{2(2\pi)^3},
\label{eq:rel_kin}
\end{align}
where $f(E_i)$ are the distribution functions of the associated particles and ${\cal M}$ is the invariant scattering matrix element, see e.g.~\cite{Kajantie:1986dh}. This microscopic formulation is well suited for non-equilibrium calculations and processes at high momenta, e.g., perturbation theory, while medium effects are more readily implemented within the thermal field theory approach. In the following, we will focus on the latter approach and discuss the EM spectral function in more detail.

\subsection{Electromagnetic spectral function and Vector Meson Dominance}
\label{sec:EM_Spectral_function}
\vspace{2mm}

\begin{figure}[b!]
	\centering\includegraphics[width=0.75\textwidth]{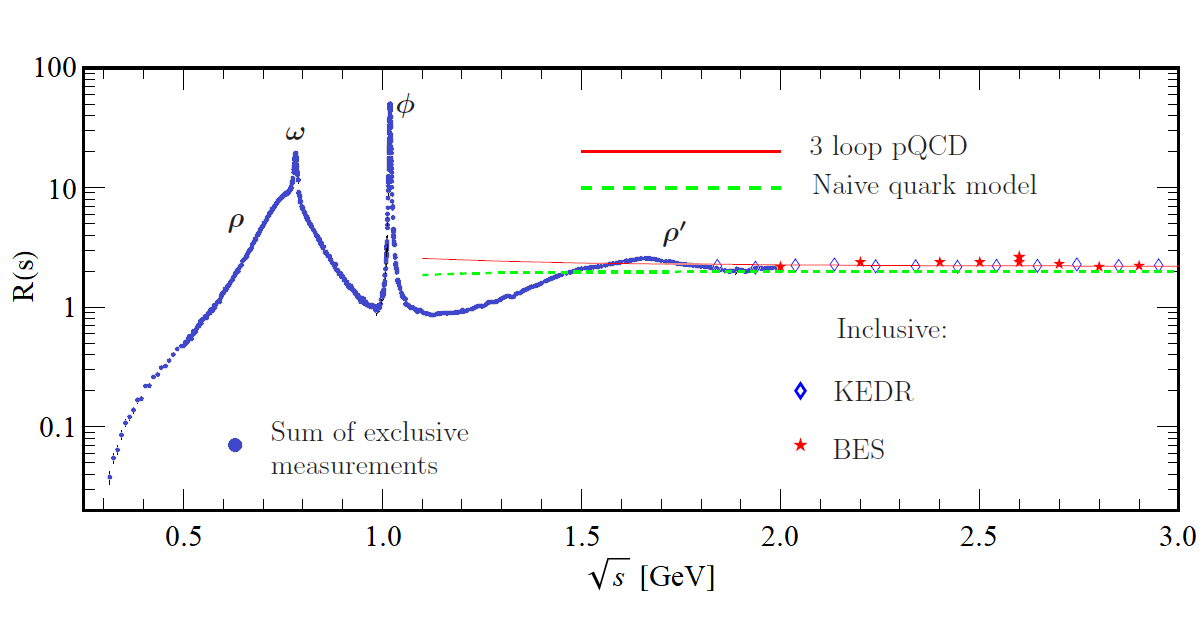}
	\caption{World data on the ratio $R(s)=\sigma(e^+e^-\rightarrow \text{hadrons})/\sigma(e^+e^-\rightarrow \mu^+\mu^-)$ in the light-flavor region where $\sigma(e^+e^-\rightarrow \text{hadrons})$ is the experimental cross section corrected for initial state radiation and with $\sigma(e^+e^-\rightarrow \mu^+\mu^-)=4\pi \alpha_{\text{EM}}^2(s)/3s$. The curves are an educative guide: the dashed green line is a naive quark model prediction and the solid red line is a 3-loop pQCD prediction, see \cite{ParticleDataGroup:2020ssz, Chetyrkin:2000zk} for details. Figure adapted from \cite{ParticleDataGroup:2020ssz}.}
	\label{fig:EM_spectral_function}
\end{figure}

The EM spectral function is well known in the vacuum where it can be obtained experimentally from $e^+ e^-$ annihilation. In fact, it is directly proportional to the corresponding cross section into hadronic final states,
\begin{align}
R(s)
=\frac{\sigma(e^+e^-\rightarrow \text{hadrons})}{\sigma(e^+e^-\rightarrow \mu^+\mu^-)}
=\frac{(-12\pi)}{s}\text{Im}\Pi_{\text{EM}}^{\text{vac}}(s),
\end{align}
where $\sigma(e^+e^-\rightarrow \mu^+\mu^-)=4\pi \alpha_{\text{EM}}^2/3s$. Experimental results on this ratio are shown in Fig.~\ref{fig:EM_spectral_function}. 

\begin{figure}[t]
	\centering\includegraphics[width=0.38\textwidth]{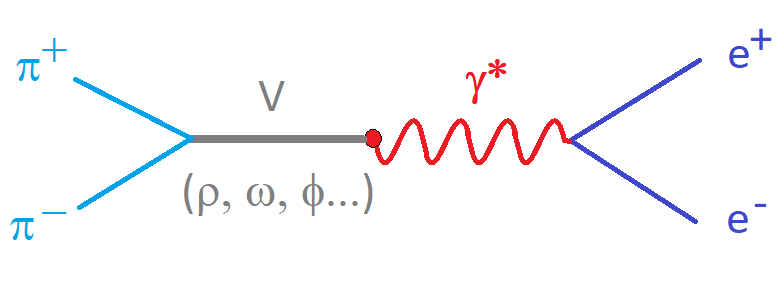}
	\caption{Graphical representation of Vector Meson Dominance: Interactions between photons and hadronic matter occur by the exchange of a vector meson between the dressed photon and the hadronic target.}
	\label{fig:VDM}
\end{figure}

The data exhibit a nonperturbative resonance regime up to $\sqrt{s}\simeq 1.1$~GeV which is dominated by the light vector mesons $\rho(770)$, $\omega(782)$ and $\phi(1020)$, and an almost structureless perturbative regime for $\sqrt{s}\gtrsim 1.5$~GeV. The former is in agreement with Vector Meson Dominance \cite{Sakurai} which asserts that the coupling of a (real or virtual) photon to $\it{any}$ EM hadronic current exclusively proceeds via an intermediate vector meson, see Fig.~\ref{fig:VDM} for a graphical illustration.

Using Vector Meson Dominance, the EM current can be expressed in terms of the light vector mesons at low energies, giving rise to the following current-field identity,
\begin{align}
j_{\text{EM}}^\mu(M\leq 1 \,\text{GeV})=
\frac{m_\rho^2}{g_\rho}\rho^\mu + 
\frac{m_\omega^2}{g_\omega}\omega^\mu + 
\frac{m_\phi^2}{g_\phi}\phi^\mu,
\end{align}
with the vector-meson (quantum-)fields $\rho^\mu(x)$, $\omega^\mu(x)$ and $\phi^\mu(x)$ as the relevant degrees of freedom in this regime. At higher energies, one uses a partonic description,
\begin{align}
j_{\text{EM}}^\mu(M> 1.5 \,\text{GeV})=
\sum_q e_q \bar{q}\gamma^\mu q,
\end{align}
with the quark and anti-quark fields $q(x)$ and $\bar{q}(x)$ and the fractional quark charge $e_q$. We note that resonance formation becomes important again in the vicinity of the heavy-quark, i.e., charm and bottom, thresholds.
The EM spectral function, cf.~Eq.~(\ref{eq:EM_correlator}), is then given by 
\begin{align}
\text{Im}\Pi_{\text{EM}}^{\text{vac}}(M\leq 1 \,\text{GeV})=\sum_{V=\rho,\omega,\phi}
\frac{m_V^4}{g_V^2}\im D_V^{\text{vac}}(M)
\label{eq:VMD_1}
\end{align}
at low energies where $D_V^{\text{vac}}(M)$ are the vector meson propagators and by
\begin{align}
\text{Im}\Pi_{\text{EM}}^{\text{vac}}(M> 1.5 \,\text{GeV})=\sum_q 
-e_q^2 \frac{M^2}{12\pi}(1+\frac{\alpha_s(M)}{\pi}+\dots)N_c
\end{align}
at higher energies where $N_c$ is the number of colors and $\alpha_s=g_s^2/4\pi$.

\begin{figure}[t!]
	\centering\includegraphics[width=0.40\textwidth,height=0.33\textwidth]{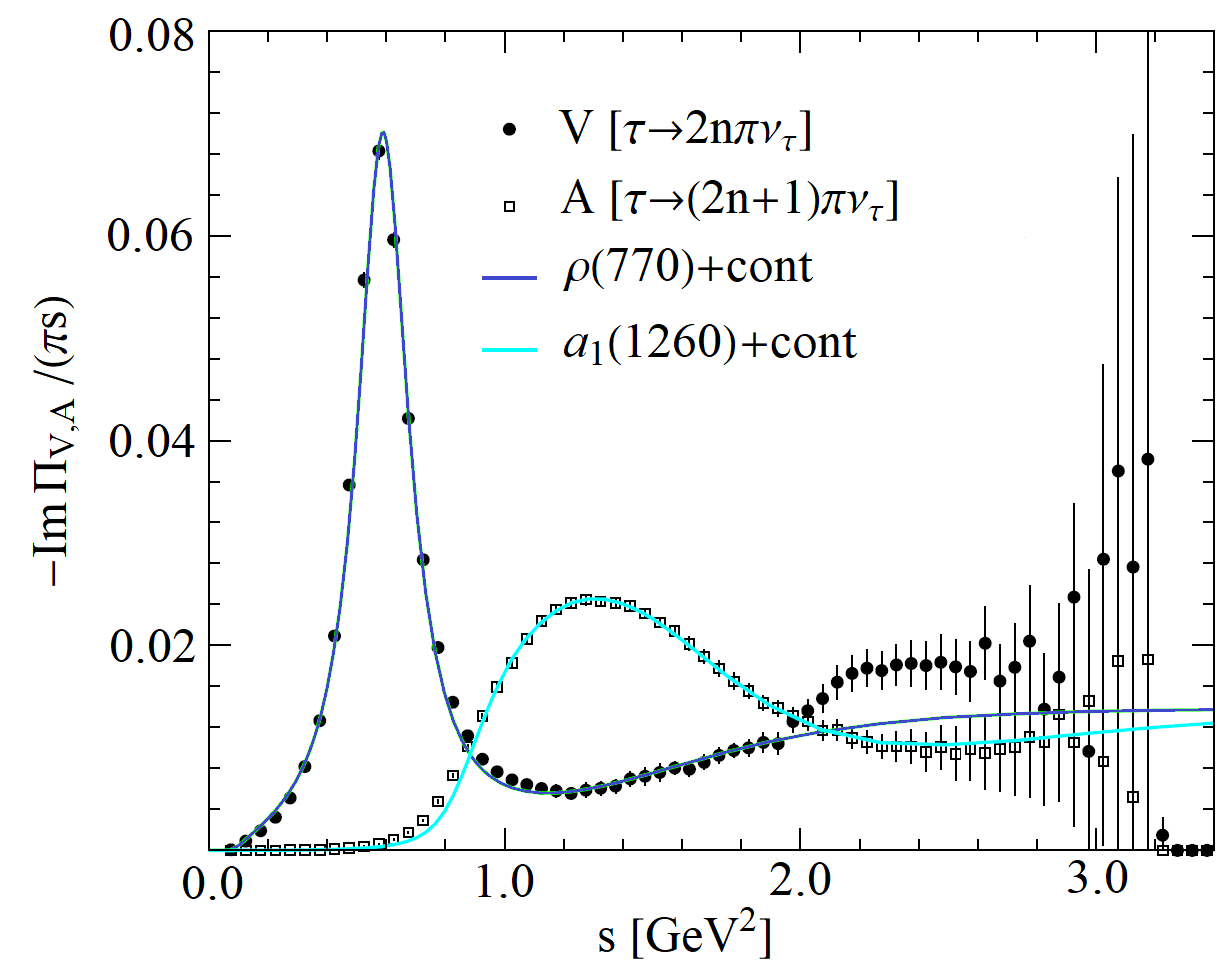}
	\hspace{0.1\textwidth}
	\includegraphics[width=0.30\textwidth,height=0.33\textwidth]{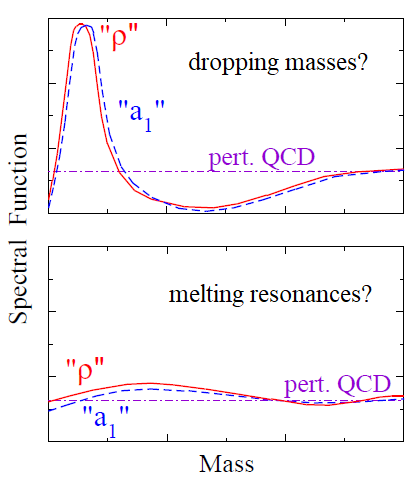}
	\caption{Left panel: Vector and axial-vector spectral functions as measured in hadronic $\tau$ decays \cite{ALEPH:1998rgl} with model fits using vacuum $\rho$ and $a_1$ strength functions supplemented by perturbative continua \cite{Rapp:2002tw}. Right panel: Scenarios for the effects of chiral symmetry restoration on the in-medium vector and axial-vector spectral functions. Figure adapted from \cite{Rapp:2009yu}.}
	\label{fig:rho_a1_exp}
\end{figure}

The low-mass strength of the EM spectral function is dominated by the iso-vector $\rho^0$ meson. In the vacuum, the EM spectral function in the isovector-vector channel ($IJ^P=11^+$) and the isovector-axial-vector channel ($IJ^P=11^-$) have been measured with excellent precision at the Large Electron Positron (LEP) collider in hadronic $\tau$ decays by the ALEPH \cite{ALEPH:1998rgl} and OPAL \cite{OPAL:1998rrm} collaborations, see Fig.~\ref{fig:rho_a1_exp}, giving access to the $\rho$ and the $a_1$ spectral functions. The difference in mass and width of these resonances is in fact one of the best pieces of empirical evidences for dynamical chiral symmetry breaking. This connection can be quantified in terms of sum rules, cf.~Sec.~\ref{sec:sum_rules}.

At finite temperatures and densities, medium effects are expected to change the shape of the spectral functions. Eventually, the $\rho$ and $a_1$ spectral functions will become degenerate due to the restoration of chiral symmetry. Historically, several scenarios were envisioned as of how this generation happens. Two possible scenarios are sketched on the right-hand side of Fig.~\ref{fig:rho_a1_exp}, i.e., the `dropping mass' scenario and the `melting resonances' scenario. As discussed in the following, the exact way in which this degeneration happens is still subject to ongoing research, but the `melting resonances' scenario is likely the correct description. Due to the importance of the $\rho$ spectral function for the EM spectral function and thus photon and dilepton rates, most of the efforts in investigating medium effects in dilepton rates have therefore focused on developing a realistic description of the $\rho$ meson at finite temperature and density, as discussed in more detail in the next chapter.

\clearpage
\section{Vector mesons in medium}
\label{sec:vector_mesons_in_medium}

In this section, we will give an overview of different approaches to describe vector- and axial-vector meson spectral functions in a thermal medium. In particular, we will discuss low-density expansions and chiral mixing, results from lattice QCD, chiral and QCD sum rules, hadronic many-body theory and the massive Yang-Mills approach, as well as the analytically-continued FRG (aFRG) method.

\subsection{Low-density expansions and chiral mixing}
\label{sec:chiral_mixing}
\vspace{2mm}

At low temperatures and densities, i.e.~for a dilute pion gas, one can apply chiral reduction and current algebra to find the following `mixing theorem' for the vector and axial-vector correlation functions \cite{Dey:1990ba},
\begin{align}
\Pi_{V}(q)&=(1-\varepsilon)\, \Pi_{V}^0(q)+\varepsilon \,\Pi_{A}^0(q),\\
\Pi_{A}(q)&=(1-\varepsilon)\, \Pi_{A}^0(q)+\varepsilon \,\Pi_{V}^0(q),
\end{align}
where $\Pi_{V}^0(q)$ and $\Pi_{A}^0(q)$ denote the $T=0$ (vacuum) values of the respective correlation functions and the mixing parameter is given by $\varepsilon=T^2/6 f_\pi^2$. This mixing theorem holds in the chiral limit of vanishing pion mass and when neglecting any momentum transfer from thermal pions in the heat bath. With increasing temperature, we, therefore, observe an increased mixing of the correlators of the chiral partners which is mediated by pion-exchange processes of the type $\pi+V\leftrightarrow A$ and $\pi+A\leftrightarrow V$.

\begin{figure}[b!]
	\centering\includegraphics[width=0.42\textwidth]{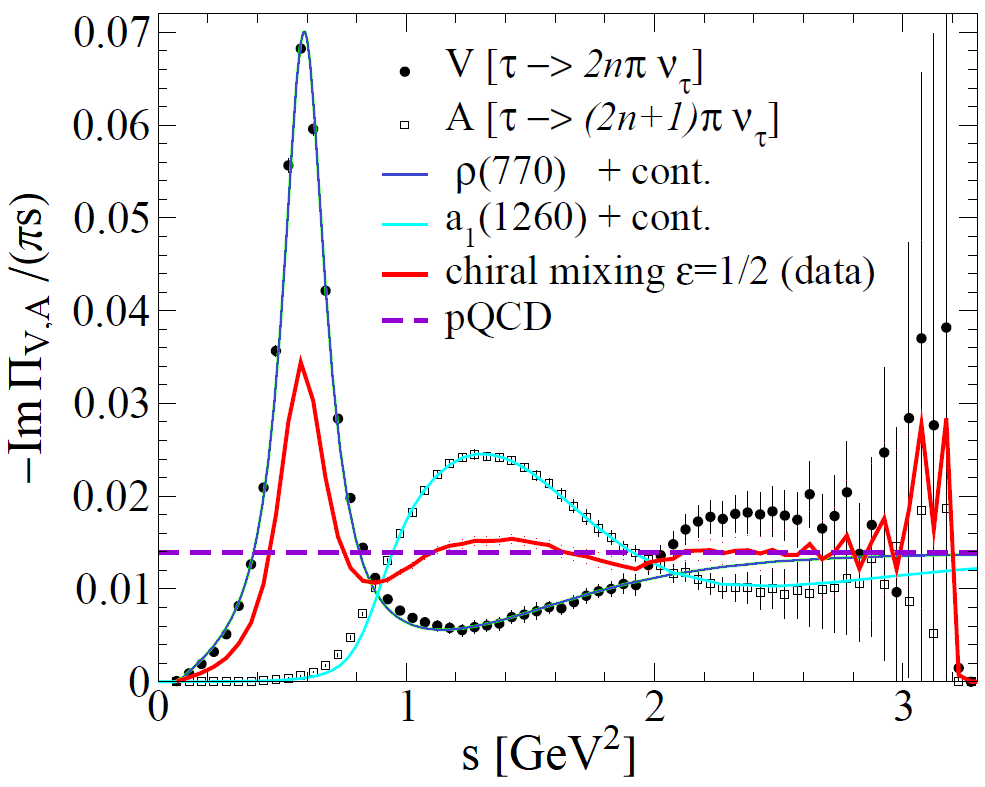}
	\caption{Effect of chiral mixing \cite{Dey:1990ba} on vector ($V$) and axial-vector ($A$) spectral functions. When extrapolated to the chiral restoration point ($\epsilon=1/2$), the degenerate $V$ and $A$ spectral functions also degenerate with the perturbative continuum (dashed line) down to $s\simeq 1~\text{GeV}^2$, thus filling the ``dip-region'' in the vacuum $V$ spectral function. Figure adapted from \cite{Rapp:2011is}.}
	\label{fig:chiral_mixing}
\end{figure}

In terms of the $\rho$ and $a_1$ spectral functions, chiral mixing, therefore, leads to a reduced strength of the vector spectral function near the $\rho$ pole and an increased strength near the $a_1$ resonance, see Fig.~\ref{fig:chiral_mixing}. For a more detailed discussion of chiral mixing and results at different temperatures, we refer to \cite{Holt:2012wr}.

Chiral mixing of course also has direct consequences on the thermal dilepton rate which was found in \cite{Huang:1995dd} to be given by
\begin{align}
\frac{dN_{ll}}{d^4x d^4q}=\frac{4\alpha_{\text{EM}}^2f^B}{(2\pi)^2}
\left\{\rho_{\text{EM}}-(\varepsilon-\frac{\varepsilon^2}{2})(\rho_V-\rho_A))\right\},
\end{align}
where $\rho_{\text{EM}}=-\im \Pi_{\text{EM}}/(\pi s)$ is the inclusive EM spectral function, see also \cite{Friman:2011zz}. Therefore, also the dilepton rate is expected to be suppressed near the $\rho$ pole and enhanced in the energy range of the $a_1$ resonance. The effect of chiral mixing was for example used in \cite{vanHees:2006ng,vanHees:2006iv} to supplement the computation of dilepton spectra which were then compared to NA60 data \cite{NA60:2006ymb} with good agreement. For further details on chiral mixing in cold nuclear matter as well as at finite temperature we refer to \cite{Krippa:1997ss,Chanfray:1999me} and  \cite{Urban:2001uv,Harada:2008hj}, respectively.

\subsection{Lattice QCD}
\label{sec:lattice}
\vspace{2mm}

Thermal photon rates can for example be computed from first principles by using lattice QCD, see \cite{Ding:2010ga} for one of the first works in this direction. Lattice QCD is a well-established non-perturbative approach that is formulated on a grid, or lattice, of points in space and time. When the size of the lattice is taken infinitely large and its sites infinitesimally close to each other, continuum QCD is recovered. The numerical evaluation of the QCD path integral is facilitated by transforming the action to imaginary (Euclidean) time, which converts the oscillatory behavior of the integrand in the partition function into an exponential damping. This makes the use of Monte Carlo importance-sampling techniques for the selection of gauge configurations possible.

The main challenge is then to extract the EM vector spectral function from numerical data on the Euclidean vector-channel propagator,
\begin{align}
	G^E_V(\tau,\vec{p})=\int_0^{\infty}\frac{dp_0}{2\pi}\rho_V(p_0,\vec{p})
	\frac{\cosh[p_0(\tau-1/2T)]}{\sinh(p_0/2T)}.
\end{align}
Inverting this integral relation, i.e.~reconstructing the spectral function,  based on a finite set of data points on the correlation function in imaginary time $\tau$ is known to be an ill-conditioned problem. There are several numerical continuation methods available in the literature that aim at obtaining the best possible reconstruction of spectral functions. For example, the Maximum Entropy Method (MEM) \cite{Jaynes:1957zza,Jaynes:1957zz,PressTeukolskyVetterlingEtAl2002}, the Backus-Gilbert (BG) method \cite{BackusGilbert1968,G.Backus1970,PressTeukolskyVetterlingEtAl2002}, the Schlessinger Point Method (SPM) \cite{Schlessinger1966, Tripolt:2016cya}, or a Tikhonov regularization \cite{Dudal:2013yva} which allows to probe unphysical (non positive-definite) spectral densities, see also \cite{Hobson:1998bz}, have been proposed. They all have different strengths and different regimes of applicability. The question as to which of the methods will give the best reconstruction, therefore, depends on the particular problem to which they are applied. For a direct comparison of such methods, both in QCD and in condensed matter systems, we refer to \cite{Tripolt:2018xeo}.

Once the spectral function $\rho_V(p_0,\vec{p})$ has been extracted from the data on the Euclidean correlation function the photon rate can be obtained as \cite{McLerran:1984ay}
\begin{align}
	p_0\frac{dR_\gamma}{d^3p}=\sum_f Q_f^2 \frac{\alpha_{\text{EM}}}{4\pi^2}\frac{\rho_V(p_0=|\vec{p}|)}{e^{p_0/T}-1},
\end{align}
where the sum is over the number of flavors and $Q_f$ is the fractional charge of the corresponding quarks. In the so-called hydrodynamical regime, which is parametrically given by $p_0, p \lesssim \alpha_s^2 T$, the general theory of statistical fluctuations applies and the vector spectral function can be related to the effective diffusion coefficient $D_{\text{eff}}$ through a Kubo formula as
\begin{align}
D_{\text{eff}}(p_0) = \frac{\rho_V(p_0=|\vec{p}|)}{4 p_0 \chi_s},
\end{align}
where the susceptibility $\chi_s=\int d^4x \langle V_0^{\text{EM}}(x)V_0^{\text{EM}}(0)\rangle$ determines the value of the conserved charge correlator at zero momentum and is also known in the continuum. For $N_f=3$, the thermal photon rate is then given by
\begin{align}
	\frac{dR_\gamma}{d^3p}=\frac{2\alpha_{\text{EM}}\chi_s}{3\pi^2}n_B(p_0)D_{\text{eff}}(p_0),
\end{align}
where $n_B(p_0)=1/(e^{p_0/T}-1)$ is the thermal Bose-Einstein distribution or occupation number. We also note that there is a direct connection to the electrical conductivity which is given by
\begin{align}
\sigma_{\text{el}}=e^2 \sum_f Q_f^2 \chi_s D_{\text{eff}}.
\end{align}

\begin{figure}[t]
	\centering\includegraphics[width=0.44\textwidth,height=0.35\textwidth]{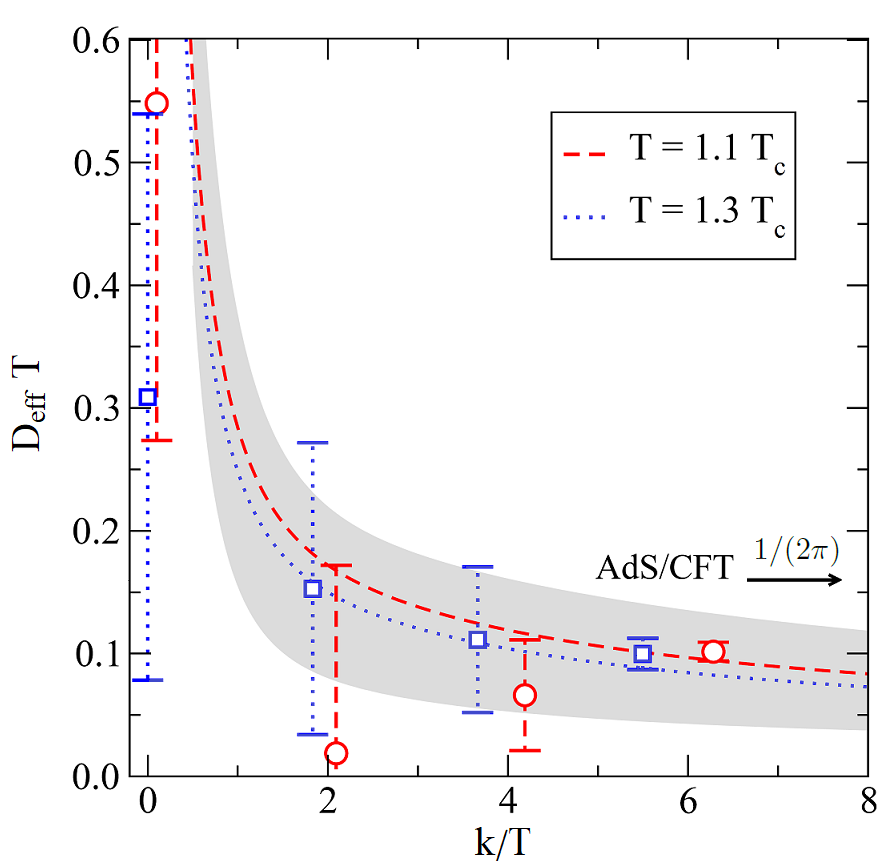}
	\hspace{0.03\textwidth}
	\centering\includegraphics[width=0.51\textwidth,height=0.35\textwidth]{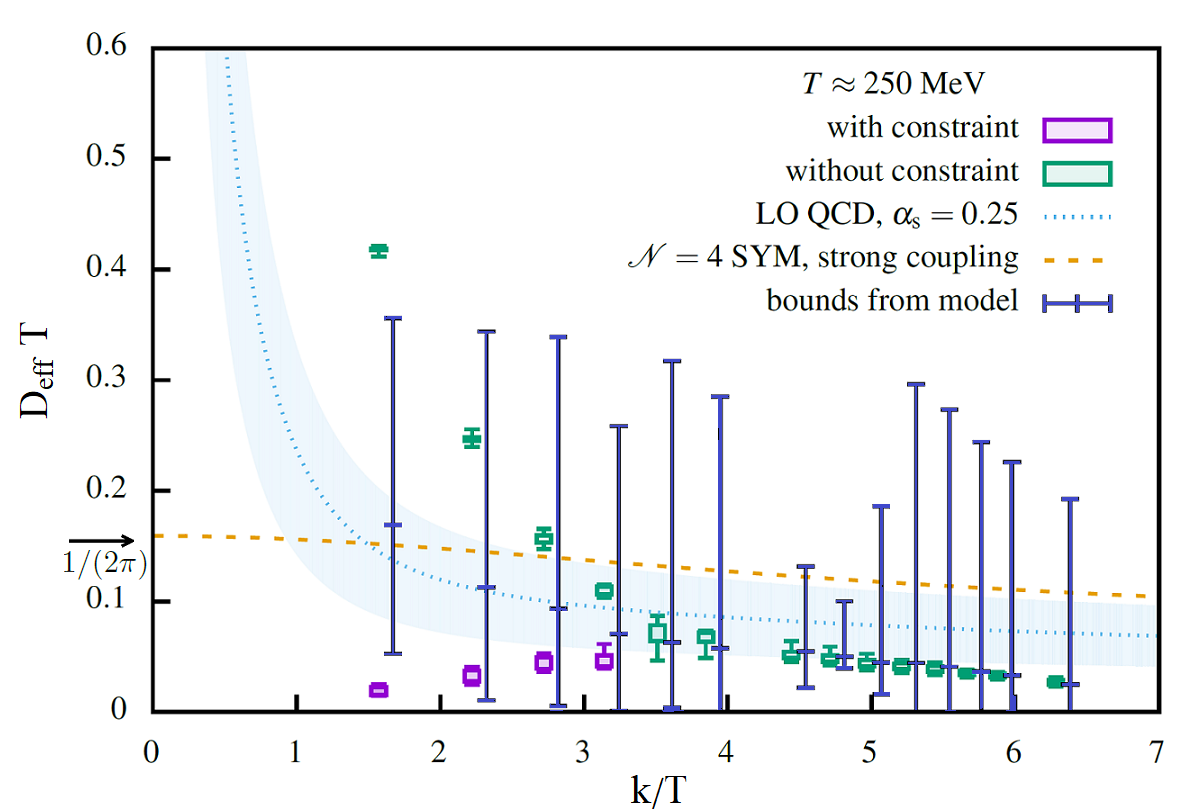}
	\caption{Left: The diffusion coefficient $D_{\text{eff}}$, which is proportional to the thermal photon production rate, divided by temperature is shown vs.~photon momentum $k$ for two different temperatures near the crossover transition as obtained from quenched lattice QCD ($N_f=0$) \cite{Ghiglieri:2016tvj}. The NLO perturbative prediction from \cite{Ghiglieri:2013gia} is shown as continuous curves. The AdS/CFT value is $D_{\text{eff}}\,T=1/(2\pi)$ \cite{Policastro:2002se}. Right: Same as left but for dynamical QCD with two flavors of Wilson clover fermions at $T\approx 250$~MeV \cite{Brandt:2017vgl}. Results from the Backus-Gilbert method are plotted as purple and green dots, corresponding to whether or not a constraint on the resolution function within the BG method was used. Bounds obtained from a model on the spectral function are shown as black bars. In addition, the strong-coupling result from $\mathcal{N}=4$ SYM and a weak-coupling result from leading-order perturbative QCD with $\alpha_s=0.25$ are shown. Figures adapted from \cite{Ghiglieri:2016tvj} (left) and \cite{Brandt:2017vgl} (right).}
	\label{fig:lattice}
\end{figure}

Recent lattice QCD results for the thermal QGP photon rate were presented in \cite{Ghiglieri:2016tvj} and \cite{Brandt:2017vgl}. In \cite{Ghiglieri:2016tvj}, lattice results for the vector-current correlator for quenched QCD were analyzed with the help of a polynomial interpolation for the spectral function, which vanishes at zero frequency and matches to high-precision perturbative results at large invariant masses. The corresponding results on the diffusion coefficient are shown in Fig.~\ref{fig:lattice} for two different temperatures near the crossover transition. The lattice results agree well with the pQCD result, in particular at larger photon momenta with $k/T>3$. This also supports the program of implementing pQCD rates into hydrodynamical codes. The theoretical uncertainties could be as low as $\sim 20\%$, save for soft photon momenta with $k<2T$ where the pQCD results represent an overestimate. Given that the diffusion coefficient is a decreasing function of $k$, the soft photon production rate increases at small $k$ even faster than the naive estimate $dR/d^3k\sim \alpha_{\text{EM}}\, T\, n_B(k)$.

In \cite{Brandt:2017vgl}, the Euclidean vector-current correlation function was computed for dynamical QCD with two flavors of Wilson clover fermions and analyzed using the Backus-Gilbert method as well as a model ansatz for the spectral function. For the BG method, two different results have been obtained, depending on whether or not a constraint on the resolution function that ensures that the result does not contain contributions from the spectral function at $p_0=0$ has been used. The corresponding results on the diffusion coefficient are shown in Fig.~\ref{fig:lattice} for a temperature of $T=250$~MeV. The results are in agreement with expectations from perturbation theory, however, the uncertainties remain large. Also, the two BG estimators resulting from implementing or not implementing this constraint do not agree with each other. Future lQCD calculations are therefore necessary, for example using larger lattices or higher statistics, in order to obtain photon rates with small uncertainties that can be used in comparisons to experimental data.

\subsection{Chiral and QCD sum rules}
\label{sec:sum_rules}
\vspace{2mm}

Sum rules allow to connect the nonperturbative physics encoded in spectral functions to the condensates of QCD. In particular, the chiral, or Weinberg, sum rules connect moments of the difference between vector and axial-vector spectral functions with chiral order parameters like the pion `pole strength', or pion decay constant, $f_\pi$. As has been shown in \cite{Kapusta:1993hq}, the Weinberg sum rules remain valid at finite temperatures except for two important modifications induced by the breaking of Lorentz invariance caused by the heat bath which defines a preferred rest frame. The sum rules then apply for a fixed spatial momentum and separately for the longitudinal and transverse parts of the vector and axial-vector spectral functions,
\begin{align}
\Pi_{V}^{\mu\nu}=\Pi_{V,A}^T P_T^{\mu\nu} + \Pi_{V,A}^L P_L^{\mu\nu},
\end{align}
where $P_L$ and $P_T$ are the usual projection operators. The chiral sum rules are then given by
\begin{align}
-\int_0^\infty\frac{dq_0^2}{\pi (q_0^2-\vec{q}^{\,2})}[\im \Pi^L_V(q_0,\vec{q})-\im \Pi^L_A(q_0,\vec{q})]&=0,\\
-\int_0^\infty\frac{dq_0^2}{\pi}[\im \Pi^{L,T}_V(q_0,\vec{q})-\im \Pi^{L,T}_A(q_0,\vec{q})]&=0,\\
-\int_0^\infty q_0^2\frac{dq_0^2}{\pi}[\im \Pi^{L,T}_V(q_0,\vec{q})-\im \Pi^{L,T}_A(q_0,\vec{q})]&=-2\pi \alpha_s \langle \langle \mathcal{O}_4\rangle \rangle,
\end{align}
see also \cite{Rapp:2009yu}. The in-medium chiral sum rules give rise to constraints on both the energy and momentum dependence of in-medium spectral functions. They also demonstrate that chiral restoration requires degeneracy of the entire spectral functions.

QCD sum rules, on the other hand, have been devised by Shifman et al.~\cite{Shifman:1978bx} as a nonperturbative method to evaluate empirical properties of hadronic current correlation functions in QCD. For a given hadronic channel $\alpha$ and for space-like momenta $q^2=-Q^2<0$ we have
\begin{align}
\Pi_\alpha(Q^2)=\Pi_\alpha(0)+\Pi'_\alpha(0)Q^2+Q^4\int \frac{ds}{\pi s^2}\frac{\im \Pi_\alpha(s)}{s+Q^2},
\label{eq:QCD_sum_rule}
\end{align}
where $\Pi(0)$ and $\Pi'(0)$ are subtraction constants. The basic idea is now to evaluate both sides of this equation using different techniques thereby establishing a link between spectral functions, which are usually related to observables or evaluated in model calculations, and ground state properties, i.e.~condensates. For sufficiently large momenta, the left-hand side of Eq.~(\ref{eq:QCD_sum_rule}) can be expanded in inverse powers of $Q^2$ according to Wilson's operator product expansion (OPE),
\begin{align}
\frac{\Pi(Q^2)}{Q^2}=-c_0 \log \frac{Q^2}{\mu^2}+\sum_{j=1}^\infty \frac{c_j}{Q^{2j}},
\label{eq:OPE}
\end{align}
where the coefficients $c_j$ are composed of perturbatively calculable (Wilson) coefficients and $Q^2$-independent expectation values of matrix elements of quark and gluon field operators, i.e.~the condensates. 

\begin{figure}[t]
	\centering\includegraphics[width=0.88\textwidth]{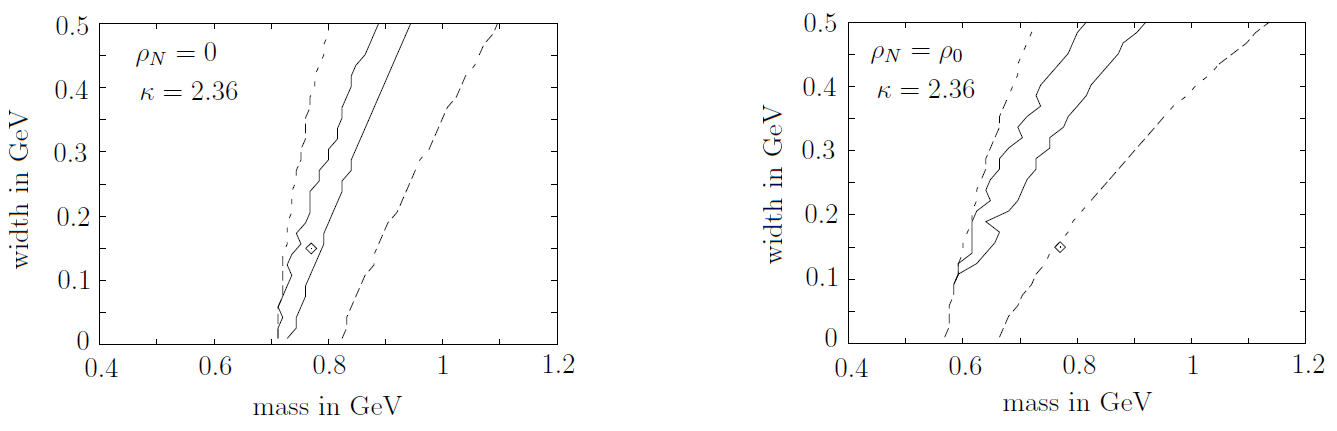}
	\caption{QCD sum rule constraints on the $\rho$ meson mass and width as inferred from Breit-Wigner parametrizations of the $\rho$ spectral function \cite{Leupold:1997dg} in the vacuum (left) and for cold nuclear matter at saturation density (right). Dashed and solid lines indicate allowed regions of mass and width as defined by maximum deviations of $0.2\%$ and $1\%$ of the l.h.s.~and the r.h.s.~of the QCD sum rule. The diamond marks the vacuum parameters. Figure adapted from \cite{Rapp:2009yu}.}
	\label{fig:QCD_sum_rules}
\end{figure}

For explicit OPE expansions of the vector and axial-vector correlation functions in terms of quark and gluon condensates we refer to \cite{Rapp:2009yu,Leupold:2009kz}. Results from quantitative studies are based on Breit-Wigner model spectral functions and which also include effects of non-scalar condensates induced by the hadron structure of the heat-bath particles are shown in Fig.~\ref{fig:QCD_sum_rules}. Therein, the allowed regions of the mass and the width of the $\rho$ meson are shown in the vacuum and at finite density. These results, however, don't allow for a clear prediction of the behavior of the mass and/or width of the rho meson in a medium. For further details and an overview of different in medium QCD sum rule applications we refer to \cite{Friman:2011zz} and references therein.

\subsection{Massive Yang Mills and hadronic many-body theory}
\label{sec:MYM_and_HMBT}
\vspace{2mm}

Model-independent approaches as discussed in the preceding sections provide valuable constraints on the vector and axial-vector correlation functions. However, quantitative calculations suitable for comparison with experiments, in particular at finite density, require the construction of effective models. Hadronic chiral Lagrangians are therefore a suitable starting point where vector mesons can for example be introduced via a local gauging procedure. The most common approaches are based on non-linear realizations of chiral symmetry, i.e.~without explicit sigma meson, within the Hidden Local Symmetry (HLS) \cite{Bando:1984ej} or Massive Yang Mills (MYM) \cite{Gomm:1984at} schemes. In this section we will focus on the latter and discuss recent developments of hadronic many-body theory within MYM \cite{Rapp:1999ej, Rapp:2009yu, Rapp:2011is}.

The basic building block of the MYM Lagrangian is the chiral pion Lagrangian based on the unitary pion field
\begin{align}
U=\exp(i\sqrt{2}\phi/f_\pi), \qquad \phi\equiv\phi_a \frac{\tau_a}{\sqrt{2}}.
\end{align}
Hadronic gauge fields, $A_{L,R}^\mu$, are introduced via the covariant derivative,
\begin{align}
D^\mu U=\partial^\mu - ig(A_L^\mu U-U A_R^\mu)
\end{align}
which leads to a MYM Lagrangian of the form
\begin{align}
\mathcal{L}_{\rm MYM} &= \frac{1}{4} f_\pi^2 \
\tr \left[ D_\mu U D^\mu U^\dagger\right]
-\frac{1}{2} \tr \left[ (F_L^{\mu\nu})^2+(F_R^{\mu\nu})^2\right]
+m_0^2 \ \tr \left[(A_L^\mu)^2+(A_R^\mu)^2\right]
\nonumber\\
 & \quad -i\xi \ \tr \left[D_\mu U D^\mu U^\dagger F_L^{\mu\nu} +
D_\mu U D^\mu U^\dagger F_R^{\mu\nu} \right] +
\sigma \, \tr \left[F_L^{\mu\nu} U F_{R \mu\nu} U^\dagger \right], 
\end{align}
with the bare mass $m_0$ and where the last two terms are necessary to achieve a satisfactory phenomenology in the vacuum. The $\rho$ and $a_1$ fields can be expressed in terms of the gauge fields as $\rho^\mu=A^\mu_R+A_L^\mu$ and $a_1^\mu\sim A^\mu_R-A_L^\mu$ where the latter includes a field redefinition to remove the $\partial^\mu\vec{\pi}A^\mu$ term. The MYM Lagrangian is then given by
\begin{align}
\mathcal{L}_{\rm MYM}=& \frac{1}{2} m_\rho^2 \vec{\rho}_\mu^2
+\frac{1}{2}\left[m_\rho^2+g^2 f_\pi^2\right] \vec{a}^2_{1,\mu} +
g^2 f_\pi \vec{\pi} \times \vec{\rho}^\mu \cdot \vec{a}_{1,\mu} +
\nonumber\\
  & g_{\rho\pi\pi}^2 \left[ \vec{\rho}_\mu^2 \vec{\pi}^2 
-\vec{\rho}^\mu \cdot \vec{\pi} \ \vec{\rho}_\mu \cdot \vec{\pi} \right] 
+ g_{\rho\pi\pi} \vec\rho_\mu \cdot (\vec\pi\times\partial^\mu\vec\pi) 
+ \dots 
\end{align}
where the couplings are related by $g_{\rho\pi\pi}=g^2/2$ and the masses of the $\rho$ and $a_1$ meson are given by
\begin{align}
m_\rho^2=m_0^2, \qquad m_{a_1}^2=m_0^2+g^2 f_\pi^2,
\end{align}
where $m_0$ is an external parameter. We note that the mass {\it splitting} is generated by dynamical chiral symmetry breaking, i.e.~via $f_\pi$. The photon field $B_\mu$ can be introduced into the MYM Lagrangian in terms of the vector dominance coupling \cite{Gomm:1984at},
\begin{align}
\mathcal{L}_{\rho\gamma}=\frac{em_\rho^2}{g_{\rho\pi\pi}}B_\mu \rho_3^\mu,
\end{align}
which describes the mixing of the photon with the neutral $\rho$ meson.

\begin{figure}[t]
	\centering\includegraphics[width=0.70\textwidth]{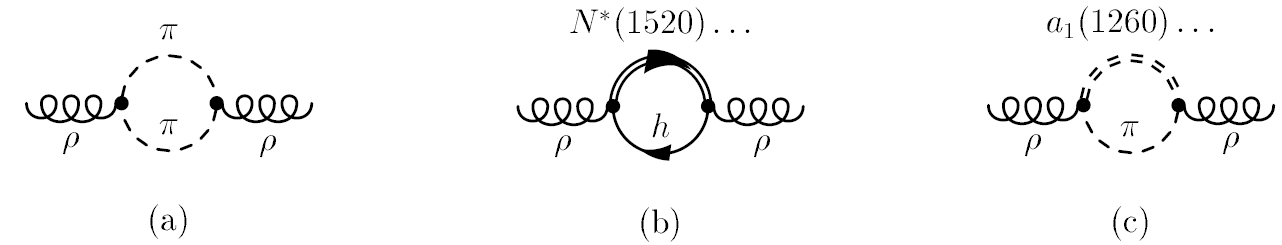}
	\caption{Diagrammatic representation of the self energies contributing to the interactions of the $\rho$ meson in hot and dense hadronic matter: (a) renormalization of its pion cloud due to modified pion propagators, (b) and (c) direct interactions of the $\rho$ meson with baryons and mesons, respectively, typically approximated by baryon- and mesons-resonance excitations \cite{Urban:1999im,Rapp:1999us}. Figure adapted from \cite{Rapp:2009yu}.}
	\label{fig:HMBT_self_energies}
\end{figure}

Medium effects can then be described by computing the self energies of the $\rho$ and $a_1$ meson at finite temperature and density. The $\rho$ meson propagator can for example be written as
\begin{align}
D_\rho^{L,T}(q_0,q)=\frac{1}{M^2-m_V^2-\Sigma_{\rho\pi\pi}^{L,T}-\Sigma_{\rho M}^{L,T}-\Sigma_{\rho B}^{L,T}},
\end{align}
with the various transverse and longitudinal contributions to the self energy, see also Fig.~\ref{fig:HMBT_self_energies}.

In particular, $\Sigma_{\rho\pi\pi}$ accounts for the pion cloud of the $\rho$ meson which for example accounts for its finite width in the vacuum due to the process $\rho\rightarrow\pi\pi$. In the medium, the pion is itself modified by interactions with hadrons from the heat bath such as $\pi N \rightarrow \Delta$. The contribution $\Sigma_{\rho M}$ describes direct mesonic interactions ($M=\pi,K,\rho,\dots$) and $\Sigma_{\rho B}$ describes interactions with baryons ($B=N,\Lambda, \Delta, \dots$) from the heat bath. These contributions vanish in the vacuum.

\begin{figure}[t]
	\centering\includegraphics[width=0.40\textwidth]{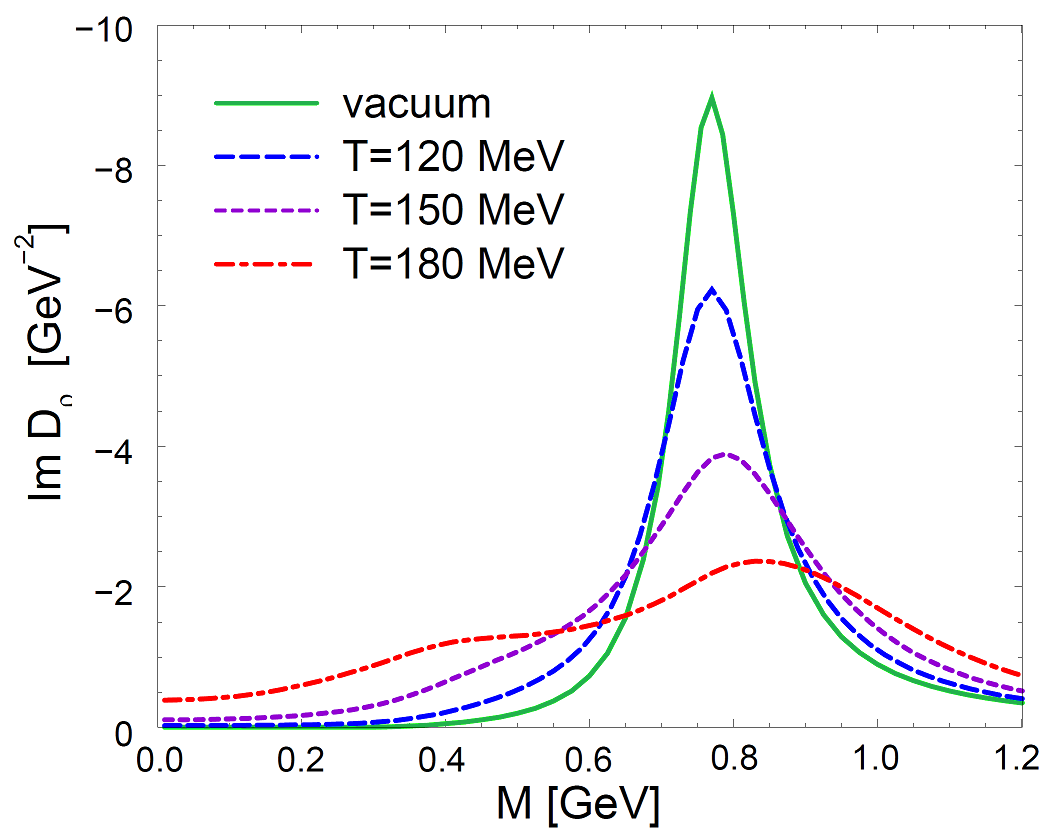}\hspace{0.1\textwidth}
	\centering\includegraphics[width=0.40\textwidth]{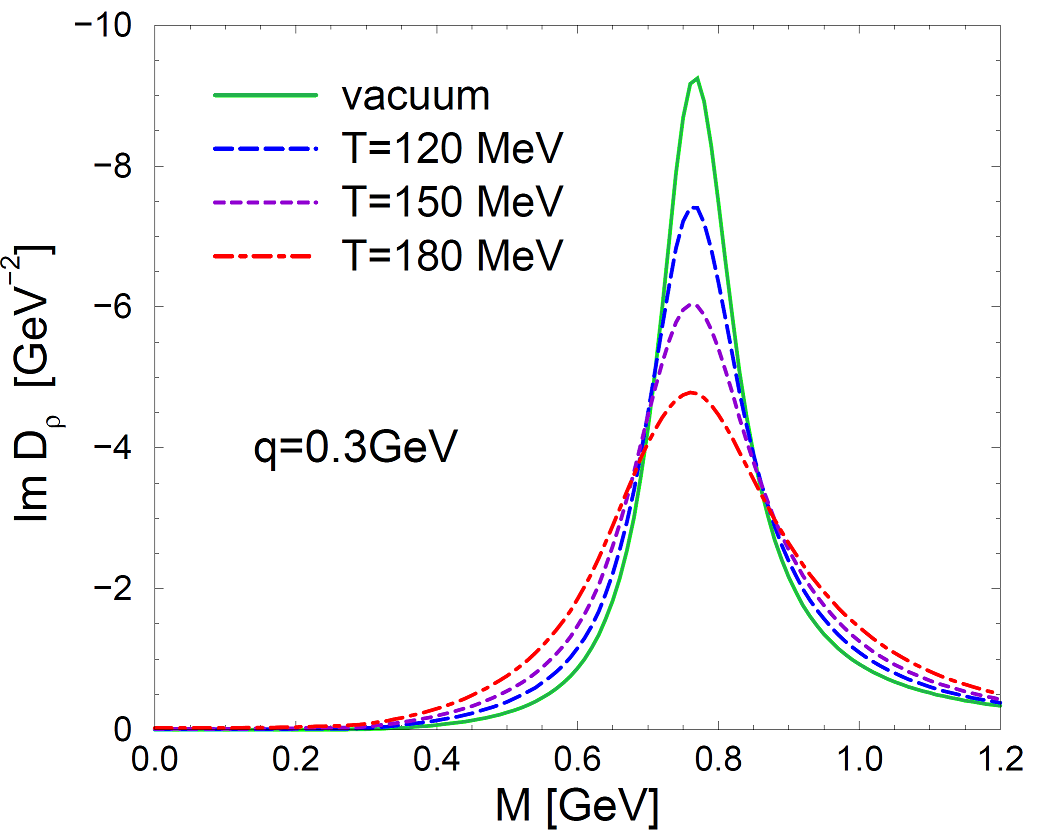}
	\caption{ Left: The $\rho$ meson spectral function as obtained from hadronic many-body theory within MYM \cite{ Rapp:2011is} is shown for different finite temperatures at a baryon chemical potential of $\mu_B=330$~MeV. The different temperatures then correspond to densities of $0.1\rho_0$, $0.7\rho_0$ and $2.6\rho_0$, respectively, with the nuclear saturation density $\rho_0=0.16\,\text{fm}^{-3}$. Right: Same as left but with all baryon-induced effects switched off. Figure adapted from \cite{Rapp:2011is}.}
	\label{fig:HMBT_rho}
\end{figure}

The involved parameters are either constrained by chiral (or gauge) symmetry or fixed by phenomenological information on e.g.~hadronic decay widths of resonances, radiative decays, and form factors. For the case of cold nuclear matter, different calculations of the in-medium $\rho$ spectral function have reached an agreement at a semi-quantitative level, see for example \cite{Friman:2011zz} and references therein. An example for the $\rho$ spectral function in hot and dense matter as relevant for heavy-ion collisions is shown in Fig.~\ref{fig:HMBT_rho}. It turns out that the $\rho$ resonance peak undergoes a strong broadening with increasing temperature, indicative for its ultimate {\it melting} near the phase transition. Fig.~\ref{fig:HMBT_rho} also shows that the medium modifications, in particular the enhancement in the low-mass regime, are largely due to baryonic processes. Baryons also play an essential role in conditions as created in heavy-ion collisions at higher collision energies as achieved at RHIC or LHC energies since the relevant quantity is the {\it sum} of baryon and antibaryon densities. 

\begin{figure}[b!]
	\centering\includegraphics[width=0.99\textwidth]{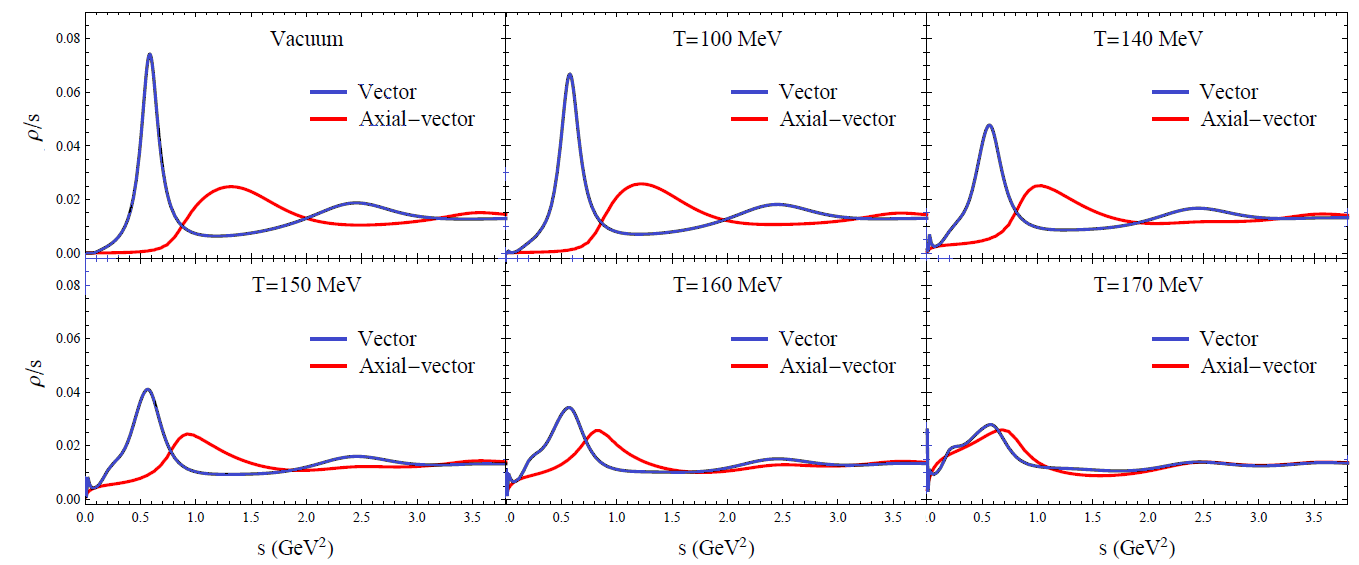}
	\caption{ The $\rho$ and $a_1$ spectral functions as obtained from hadronic many-body theory (for the $\rho$) and from a combined study based on chiral and QCD sum rules (for the $a_1$) are shown at different temperatures and vanishing baryo-chemical potential \cite{Hohler:2013eba}. The spectral functions `melt' and the $a_1$ mass decreases with increasing temperature until the spectral functions become nearly degenerate near the chiral crossover transition. Figure adapted from \cite{Rapp:2016xzw}.}
	\label{fig:MYM_sum_rules_spectral}
\end{figure}

In \cite{Hohler:2013eba}, a combined analysis of finite-temperature QCD \cite{Hatsuda:1992bv} and Weinberg \cite{Kapusta:1993hq} sum rules has been carried out to test the $\rho$ spectral function that describes dilepton spectra \cite{Rapp:1999ej} with respect to chiral restoration at different temperatures and vanishing baryo-chemical potential. Therein, the $\rho$ spectral function as obtained from hadronic many-body theory was used as input for the sum rules, together with input from lattice QCD on quantities like the pion decay constant and the quark condensate, to identify viable in-medium $a_1$ spectral functions that satisfy both QCD and sum rules within in accuracy of $\sim 0.5\%$. The resulting $\rho$ and $a_1$ spectral functions are shown in Fig.~\ref{fig:MYM_sum_rules_spectral}. The spectral functions gradually degenerate with increasing temperature as chiral symmetry gets restored. This analysis also suggests a mechanism of chiral restoration by which the broadening of both $\rho$ and $a_1$ is accompanied by a reduction of the $a_1$ mass moving toward the $\rho$ mass, while the latter approximately stays constant.

These findings are in good agreement with a recently conducted microscopic study of the in-medium $\pi-\rho-a_1$ system within the MYM framework \cite{Hohler:2015iba}. Therein, the notorious difficulties of the MYM approach to describe the vacuum axial-vector spectral function could be overcome by introducing a broad $\rho$ propagator into the $a_1$ self energy, accompanied by vertex corrections to maintain PCAC \cite{Hohler:2013ena}. A one-loop calculation at finite temperature showed a broadening of both the $\rho$ and $a_1$ peaks accompanied by a downward mass shift of the $a_1$ and therefore corroborates the `burning' of the chiral mass splitting as the mechanism of chiral restoration. The same behavior was also found in recent lattice QCD calculations \cite{Aarts:2015mma} which investigated the nucleon correlation function and its chiral partner, the $N(1535)$ resonance, at finite temperature. Also here it was found that the mass of the ground state, i.e.~of the nucleon, remains essentially constant while the mass of the excited state approaches the former and degenerates with it near the chiral crossover temperature.

\clearpage
\section{Vector mesons with the analytically-continued FRG (aFRG) method}
\label{sec:aFRG}
\vspace{2mm}

In recent years, considerable progress has been made in developing a formalism for the description of vector mesons and their spectral functions in hot and dense strong-interaction matter. This formalism is referred to as analytically-continued Functional Renormalization Group (aFRG) method and enjoys several distinctive advantages. First, as it is based on the non-perturbative FRG approach, it is capable of including effects from quantum and thermal fluctuations and thus goes beyond mean-field approaches. Second, it can be applied at finite temperature and chemical potential without complications such as the fermion sign problem encountered in lattice QCD. In addition, a particular benefit of the aFRG method is that it allows for a consistent description of both thermodynamical quantities as well as real-time quantities like spectral functions and transport coefficients. The analytic continuation from imaginary to real energies is therein performed on the level of the flow equations, thus avoiding the need for any numerical reconstruction techniques. Within the last years, the aFRG method has been applied to effective descriptions of QCD involving either quark-meson or baryon-meson systems. In these approaches, chiral symmetry was used as the underlying construction principle which in particular allowed to study the resulting chiral phase structure, corresponding critical effects in spectral functions as well as different mechanisms of chiral symmetry breaking and restoration. In the following, we will give a brief overview of the FRG framework, the analytic continuation procedure of the aFRG method, and of recent results obtained on vector and axial-vector meson spectral function in nuclear matter within this framework.

\subsection{Flow equations and analytic continuation}
\label{sec:flow_equations}
\vspace{2mm}

\begin{figure}[h!]
	\centering\includegraphics[width=0.2\textwidth]{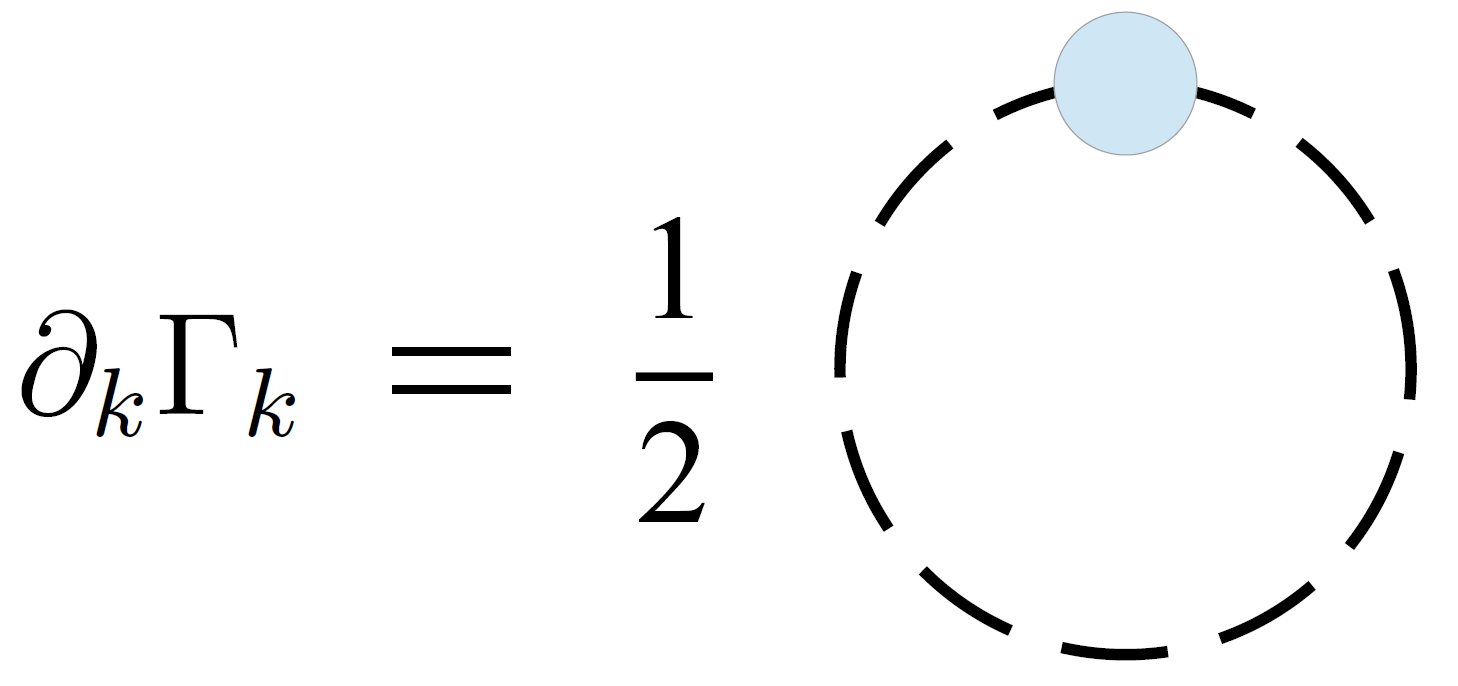}
	\caption{Diagrammatic representation of the flow equation for the effective average action $\Gamma_k$, cf.~Eq.~(\ref{eq:Wetterich_equation}). The dashed line represents the full inverse scale-dependent propagator, $(\Gamma_k^{(2)}[\phi]+R_k)^{-1}$, and the circle symbolizes the regulator insertion $\partial_k R_k$.}
	\label{fig:flow_Gamma}
\end{figure}

The FRG is a powerful and versatile non-perturbative framework with applications ranging from statistical physics over condensed matter theory to quantum field theory, for reviews see for example \cite{Berges:2000ew,Polonyi:2001se,Pawlowski:2005xe,Schaefer:2006sr,Kopietz2010,Braun:2011pp, Gies:2006wv,Dupuis:2020fhh}. It is based on Wilson's coarse-graining idea of successively integrating out quantum fluctuations in momentum space. In the formulation pioneered by C.~Wetterich \cite{Wetterich:1992yh} 
(within Euclidean space-time)
, the RG-scale dependence of the effective average action $\Gamma_k$, which interpolates between the bare action $S\simeq \Gamma_\Lambda$ at an ultraviolet (UV) scale $\Lambda$ and the full quantum effective action $\Gamma\equiv\Gamma_{k=0}$ in the infrared (IR), is given by the following flow equation, see also Fig.~\ref{fig:flow_Gamma},
\begin{align}
\label{eq:Wetterich_equation}
\partial_k\Gamma_k[\phi]=\frac{1}{2}\Tr{\frac{\partial_k R_k}{\Gamma_k^{(2)}[\phi]+R_k}}.
\end{align}

Therein, the regulator function $R_k$ acts as a mass term and suppresses fluctuations of low-momentum modes, $p \lesssim k$, while the high-momentum modes, $p \gtrsim k$, are already integrated out and included in $\Gamma_k$. For a discussion of how to devise optimized regulators in a particular truncation where this can be quite non-trivial, see \cite{Pawlowski:2015mlf}. Apart from the regulator function, the Wetterich equation only depends on the second functional field derivative of the effective action, which is denoted as $\Gamma_k^{(2)}[\phi]$. The trace in Eq.~(\ref{eq:Wetterich_equation}) represents a summation over internal indices as well as an integration over momentum space, which gives rise to a simple one-loop structure of the Wetterich equation since $(\Gamma_k^{(2)}[\phi]+R_k)^{-1}$ represents the full scale-dependent propagator $D_k$. 

\begin{figure}[t]
	\centering\includegraphics[width=0.2\textwidth]{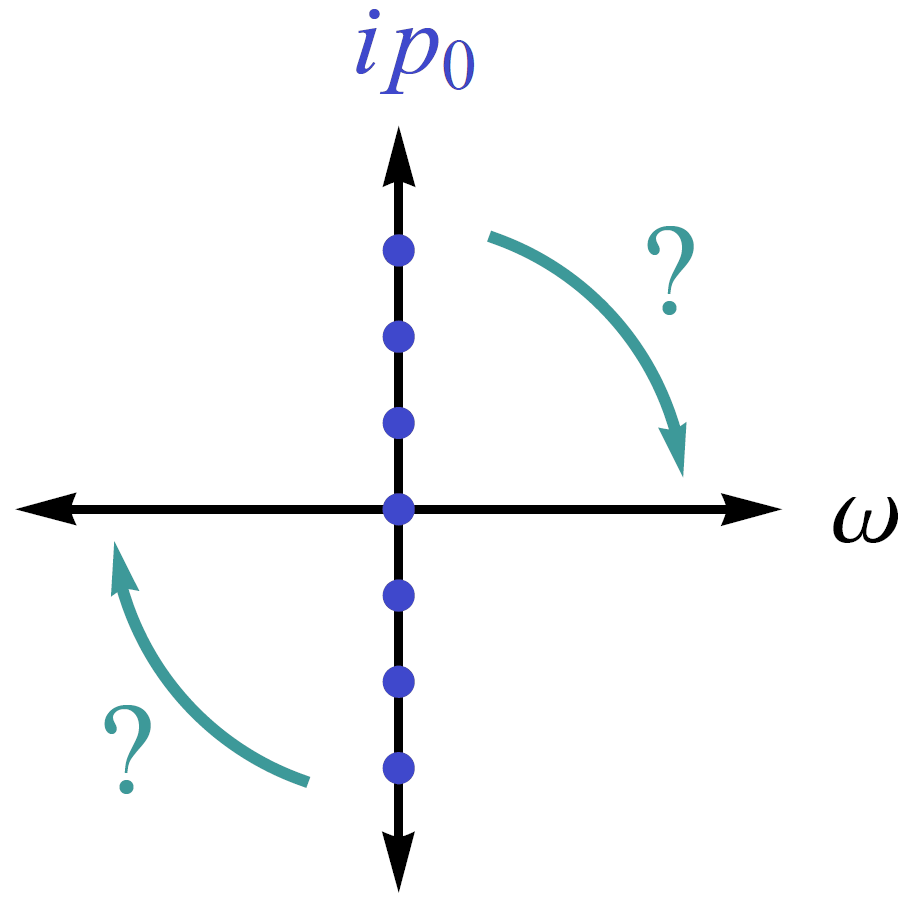}
	\caption{Graphical representation of the analytic-continuation problem: Information on real-time quantities (as a function of a real frequency $\omega$) needs to be inferred from Euclidean data (given at discrete imaginary energies $ip_0$), see text for details.}
	\label{fig:analytic_continuation}
\end{figure}

As already discussed in Sec.~\ref{sec:lattice}, one of the main challenges that need to be overcome in the calculation of real-time quantities like spectral functions within Euclidean approaches to thermal field theory is the analytic-continuation problem. 
As the FRG is usually formulated in Euclidean space-time, this problem also occurs here and needs to be addressed.
It describes the difficulty to reconstruct the real-time part of a function that is only given at imaginary frequencies, see Fig.~\ref{fig:analytic_continuation} for a graphical illustration. Knowing the correct analytic continuation of a correlation function is equivalent to knowing its spectral function. However, the reconstruction of correlators from Euclidean data is an exponentially hard inverse problem. In fact, it represents an ill-posed problem where no unique solution exists without making further assumptions. Even when dealing with infinitely many data points of infinite precision, additional constraints are necessary. These are given by the Baym-Mermin boundary conditions which require the correlator to be analytic outside the real axis and to be bounded in the limit of infinitely large energies \cite{Baym1961}.

Within the FRG, several techniques have been developed recently to circumvent the numerical analytic-continuation problem at finite temperatures. In \cite{Floerchinger:2011sc}, an approach where the analytic continuation from imaginary Matsubara frequencies to real frequencies is done on the level of the flow equations was proposed. By choosing an ansatz for the analytic structure of the renormalized propagator it was possible to solve flow equations for its parameters, e.g. masses and decay widths. Direct calculations of spectral functions from real-time correlations with the FRG on the Schwinger-Keldysh closed-time contour have been performed for self-interacting scalar fields \cite{Mesterhazy:2015uja,Roth:2021nrd}. Moreover, in \cite{Pawlowski:2015mia} an approach based on four-dimensional regulator functions, where one needs to take into account poles in the complex energy plane numerically, was proposed along with a real-time computation on the Keldysh contour with general spatial momentum regulators. 

As a particularly innovative and promising alternative, the so-called aFRG method, i.e., analytically continued FRG, was developed in \cite{Kamikado:2013sia, Tripolt:2013jra}. 
The aFRG method allows to calculate real-time quantities like spectral functions and transport coefficients within Euclidean FRG approaches by applying an analytic-continuation procedure on the level of the flow equations in two steps.
First, at finite temperature, the periodicity of bosonic and fermionic occupations numbers with respect to the internal Euclidean energy, $p_0$, is exploited,
\begin{align}
n_{B,F}(E+i p_0)\rightarrow n_{B,F}(E).
\end{align}
In a second step, the Euclidean energy $p_0$ is replaced by a continuous real frequency $\omega$,
\begin{align}
\Gamma^{(2),R}(\omega,\vec p)=-\lim_{\epsilon\to 0} \Gamma^{(2),E}(p_0=-i(\omega+i\epsilon), \vec p).
\end{align}
This leads to flow equations for the real and imaginary parts of equilibrium correlation functions in the real-frequency domain. When performed in the given order, this procedure also fulfills the physical Baym-Mermin boundary conditions \cite{Baym1961}. Another distinct advantage of this approach is that it is thermodynamically consistent since the thermodynamic potential, as obtained from the flow equation for the effective average action, is used as input for the calculation of the two-point functions and the spectral functions. 

The aFRG method has been successfully applied in different situations, for example to calculate in-medium spectral functions of pions and the scalar $\sigma$ meson \cite{Tripolt:2013jra, Tripolt:2014wra, Tripolt:2016cey}, the quark spectral function \cite{Tripolt:2018qvi,Tripolt:2020irx} as well as vector- and axial-vector meson spectral functions at finite temperature and density in extended linear-sigma models with quarks \cite{Jung:2016yxl, Jung:2019nnr}, together with the corresponding electromagnetic spectral function and thermal dilepton rates \cite{Tripolt:2018jre} inside quark matter. More recently, it was also used to compute vector and axial-vector spectral functions in nuclear matter based on the parity-doublet model, \cite{Tripolt:2021jtp}, as discussed in more detail in the following.

\subsection{Mass generation of nucleons and the parity-doublet model}
\label{sec:PDM_theory}

The natural mass scale of nuclear physics is given by the proton mass of $m_{p}\approx 1$~GeV. While the mass of the electron, for example, is attributed to the Higgs boson in the Standard Model, the mass of the proton is mainly due to non-perturbative effects of QCD. When treated as a classical model, QCD is a non-Abelian local gauge theory that does not possess a mass scale in the absence of Lagrangian masses for the matter fields, i.e.~in the chiral limit. In this limit, the theory is scale-invariant which entails that the energy-momentum tensor is traceless. Upon quantization, however, the regularization and renormalization of ultraviolet divergences introduce a mass scale. This phenomenon is known as `dimensional transmutation' and leads to the trace anomaly of the energy-momentum tensor, already in the chiral limit of vanishing current quark masses \cite{Collins:1976yq, Nielsen:1977sy}. In full QCD, the trace of the energy-momentum tensor is given by
\begin{align}
\label{eq:T}
    T_\mu^\mu=\frac{\beta(g)}{2g}G^{\mu\nu a} G_{\mu\nu a}+\sum_{f}m_f(1+\gamma_{m_f})\bar{q}_f q_f
\end{align}
where $\beta(g)$ is the $\beta$ function of QCD, with the QCD coupling $g$, and $\gamma_{m_f}$ is the anomalous mass dimension of a quark with flavor $f$. We note that in view of the Einstein equation \cite{Einstein1915}
\begin{align}
    R_{\mu\nu}-\frac{1}{2}g_{\mu\nu}R=8\pi G T_{\mu\nu},
\end{align}
where $g_{\mu\nu}$ is the metric tensor, $R_{\mu\nu}$ is the Ricci curvature tensor, $R$ is the scalar curvature (Ricci scalar) and $G$ is the Newton constant, the trace of the energy-momentum tensor is also referred to as `scalar gravitational form factor'. In fact, the scalar gravitational form factor $G(q^2)$ is directly proportional to the matrix element of the energy-momentum tensor,
\begin{align}
    \langle {\bf p}_1 |T^\mu_\mu| {\bf p}_2\rangle=\left( \frac{M^2}{p_{01} p_{02}}\right)^{1/2} 
    \bar{u}(p_1,s_1)u(p_2,s_2)G(q^2),
\end{align}
with the momentum transfer $q^2$, see \cite{Kharzeev:2021qkd} for details. The mass of the proton can be obtained from this form factor as 
\begin{align}
    G(0)=m_p.
\end{align}
As evident from Eq.~(\ref{eq:T}), the proton mass receives contributions from two terms, i.e.~the gluonic term and the quark term, where the latter is directly connected to spontaneous chiral symmetry breaking in terms of the quark condensate. In the chiral limit of massless quarks, the forward matrix element of Eq.~(\ref{eq:T}) contains only the gluon term and the mass of the proton is entirely due to gluons. The contribution of the second term, also known as `$\sigma$ term', for non-vanishing quark masses, can be extracted from experimental data on pion and kaon scattering amplitudes, see for example \cite{RuizdeElvira:2017stg}, or computed in lattice QCD \cite{Yang:2015uis}. It is found that the $\sigma$ term contributes about $80$~MeV, or about $8\%$, to the total proton mass. Therefore, the bulk of the proton mass is expected to be untouched by chiral symmetry restoration at high temperatures or densities. Effective theories involving nucleons and chiral symmetry clearly need to account for this behavior in order to allow for a realistic description of nuclear matter. 

A promising candidate for a consistent non-perturbative description of nuclear matter is given by the parity-doublet model which describes nucleons along with their parity partners \cite{Detar:1988kn}. The particular strength of the parity-doublet model is that it can account for a finite nucleon mass in a chirally-invariant fashion in contrast to, e.g., the chiral Walecka model where the nucleon mass is predominantly generated by dynamical chiral symmetry breaking and hence gives rise to massless Lee-Wick matter in the chirally restored phase. The parity-doublet model therefore also provides a natural description for the parity-doubling structure of the low-lying baryons observed in recent lattice-QCD calculations \cite{Glozman:2012fj,Aarts:2017rrl}. 

On the mean-field level, the parity-doublet model is known to provide a phenomenologically successful description of nuclear matter \cite{Gallas:2009qp} that exhibits two sequential phase transitions at low temperatures and is able to provide realistic estimates for the equation of state of dense matter under neutron-star conditions \cite{Marczenko:2018jui}. Within the FRG, the parity-doublet model for nucleons interacting with pions, sigma, and omega mesons was studied in \cite{Weyrich:2015hha} with a focus on describing the liquid-gas transition of nuclear matter together with chiral symmetry restoration in the high-density phase. It was found that within the FRG it is more difficult to reproduce known values for observables like the binding energy per nucleon, the nuclear saturation density, and the nucleon sigma term all at the same time \cite{Weyrich:2015hha, Tripolt:2021jtp}.

\subsection{Results for the parity-doublet model}
\label{sec:PDM_results}
\vspace{2mm}

In \cite{Tripolt:2021jtp} the $\rho$ and $a_1$ spectral functions were studied in nuclear matter at finite temperature and baryon-chemical potential. As a low-energy effective theory, we use a chiral baryon-meson model, namely a parity-doublet model, which contains pions, sigma mesons, $\rho$ and $a_1$ mesons as well as nucleons and their parity partners, chosen to be the $N^*(1535)$. The vector and axial-vector mesons were introduced using a novel FRG formulation for massive vector fields based on (anti-)self-dual field strengths \cite{Jung:2019nnr}. Explicitly, the ansatz made in \cite{Tripolt:2021jtp} for the effective average action reads 
\begin{align}
\label{eq:Gamma_PDM}
\Gamma_{k}=
\int d^{4}x \:\Big\{&\bar{N_1}\left(\slashed\partial-\mu_B \gamma_0+h_{s,1}(\sigma+i\vec{\tau}\cdot\vec{\pi}\gamma^{5})
+h_{v,1}(\gamma_\mu \vec{\tau}\cdot\vec{\rho}_\mu+\gamma_\mu\gamma^5\vec{\tau}\cdot{\vec a}_{1,\mu})\right)N_1
\nonumber\\
&+\bar{N_2}\left(\slashed\partial-\mu_B \gamma_0+h_{s,2}(\sigma-i\vec{\tau}\cdot\vec{\pi}\gamma^{5})
+h_{v,2}(\gamma_\mu \vec{\tau}\cdot\vec{\rho}_\mu-\gamma_\mu\gamma^5\vec{\tau}\cdot{\vec a_{1,\mu}}\right)N_2
\\
&+m_{0,N}\left(\bar{N_1}\gamma^{5}N_2-\bar{N_2}\gamma^{5}N_1\right)
+U_{k}(\phi^2)-c\sigma + \frac{1}{2} (D_\mu \phi)^\dagger D_\mu \phi 
-\frac{1}{4} \,\tr\, \partial_\mu \rho_{\mu\nu} \partial_\sigma \rho_{\sigma\nu} +\frac{m_v^2}{8} \, \tr\, \rho_{\mu\nu}\rho_{\mu\nu}
\Big\},\nonumber
\end{align}
where the nucleon fields $N_1$ and $N_2$ are defined to have opposite parity and respectively represent the iso-doublet of nucleons, $(p,n)$, and their parity partners, i.e.~the $N^*(1535)$. The chirally-invariant bare nucleon mass is given by $m_{0,N}$ and the iso-triplet vector and axial-vector fields, $\vec \rho_\mu$ and ${\vec a}_{1\mu}$, are obtained from the anti-symmetric rank-2 tensor fields as
\begin{align}
   	\vec{\rho}_{\mu} = \frac{1}{2m_v}\text{tr}(\partial_\sigma \rho_{\sigma\mu}\vec{T}_V),
	\qquad \qquad
	\vec{a}_{1\mu} = \frac{1}{2m_v}\text{tr}(\partial_\sigma \rho_{\sigma\mu}\vec{T}_A).
\end{align}
This formulation avoids the known problems of the Proca formalism, where the propagator is only transversal on-shell, and of the Stueckelberg formalism, where massless single-particle contributions appear when restoring transversality. For more details on the ansatz given by Eq.~(\ref{eq:Gamma_PDM}) we refer to \cite{Tripolt:2021jtp}.

\begin{figure}[t]
	\centering\includegraphics[width=0.4\textwidth]{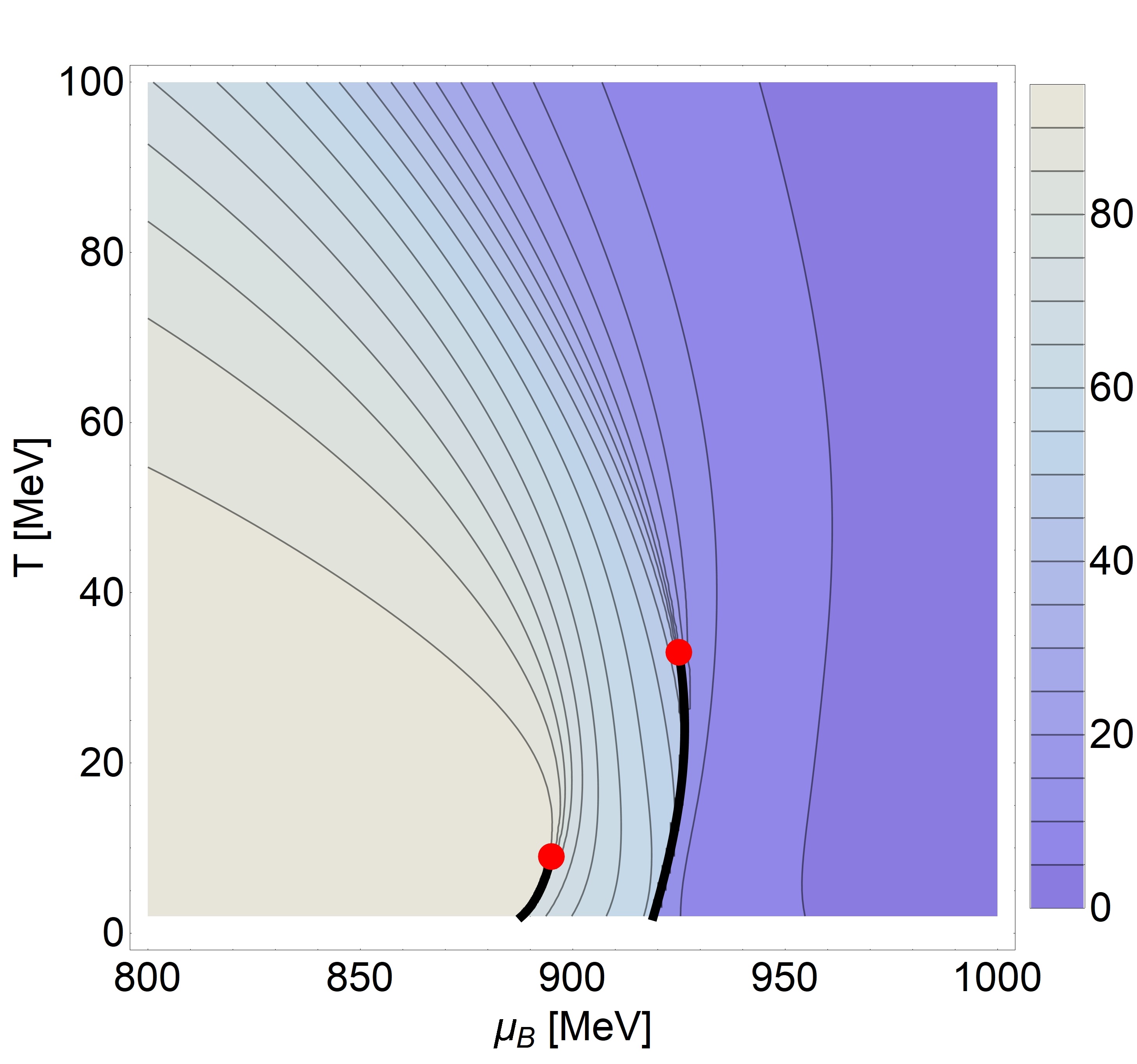}
	\caption{The phase diagram of the parity-doublet model as obtained in \cite{Tripolt:2021jtp} represented as a contour plot of $\sigma_0(\mu_B,T)$ with darker colors indicating smaller values, as shown in the legend bar. Two distinct first-order phase transitions are observed at low temperatures which end in a critical point at ($\mu_B\approx 896$~MeV, $T\approx 10$~MeV) and at ($\mu_B\approx 925$~MeV, $T\approx 33$~MeV), respectively. Figure adapted from \cite{Tripolt:2021jtp}.}
	\label{fig:PDM_phase_diagram}
\end{figure}

This extended parity-doublet model captures the essential features of mass generation in QCD, in that hadron masses only partially result from the spontaneous breaking of chiral symmetry. On the other hand, the degeneracy in the spectral functions of parity partners in the restored phase is entirely driven by the evolution of the chiral condensate. In the following, we summarize the most important results obtained in \cite{Tripolt:2021jtp} on the phase diagram of the parity-doublet model as well as on the $\rho$ and the $a_1$ spectral function. The phase diagram is shown in Fig.~\ref{fig:PDM_phase_diagram} as obtained from solving the flow equation for the effective potential for different combinations of temperature $T$ and baryon-chemical potential $\mu_B$ and identifying its global minimum $\sigma_0(\mu_B,T)$. 
Two critical endpoints as well as two first-order lines are observed in the low-temperature regime at high chemical potentials. These transitions are identified as the nuclear liquid-gas transition (at lower chemical potentials) and a chiral phase transition (at higher chemical potentials). We note that for a phenomenologically acceptable description of the thermodynamics of nuclear matter as well as of the binding energy per nucleon, the nuclear saturation density and the equation of state (EoS) the inclusion of a short-range vector-like repulsive effect, often modeled in terms of the $\omega$ vector meson, in the calculation of the thermodynamic grand potential will be necessary. In addition, we note that the employed parity-doublet model does not contain quarks or gluons and is therefore not suited to describe matter at high temperatures. It is therefore also not surprising that the location of the chiral critical endpoint in this computation is different from the one shown in Fig.~\ref{fig:phase_diagram_2} for QCD.

\begin{figure}[t]
	\centering\includegraphics[width=0.9\textwidth]{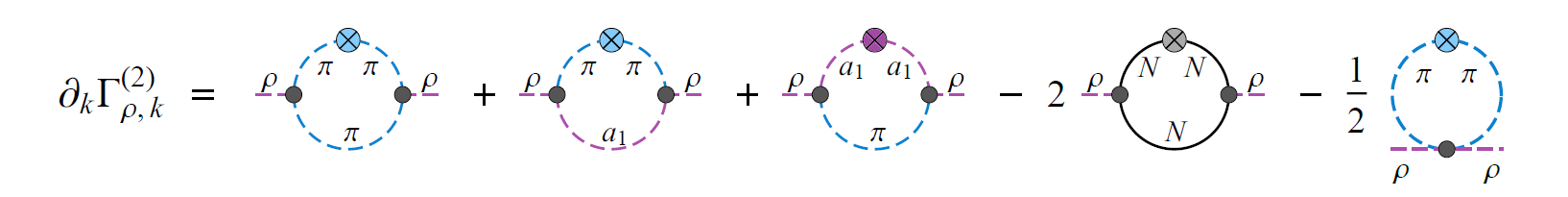}\\[-1mm]
	\centering\includegraphics[width=0.9\textwidth]{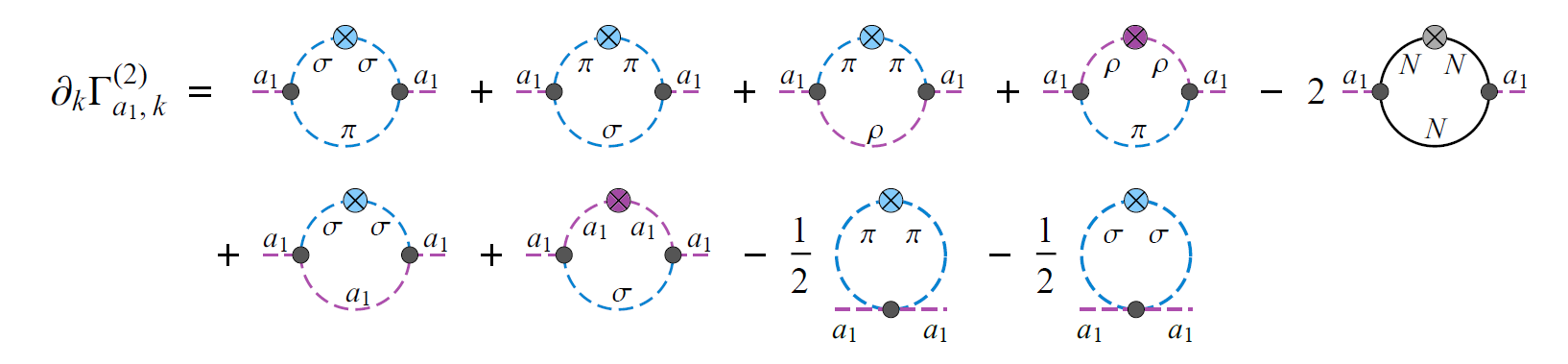}
	\caption{Flow equations of the $\rho$ and the $a_1$ two-point function in diagrammatic form. Dashed (solid) lines represent bosonic (fermionic) propagators while crossed circles indicate regulator insertions. Figure adapted from \cite{Tripolt:2021jtp}.}
	\label{fig:2PF}
\end{figure}

The spectral functions are obtained from the real-time two-point functions $\Gamma_k^{(2)}$ which are computed by solving their respective flow equations. These flow equations can be derived from the Wetterich equation, Eq.~(\ref{eq:Wetterich_equation}), by taking two functional field derivatives which results in the following general structure,
\begin{align}
\label{eq:2PF}
\partial_k\Gamma_{k}^{(2)}(p)=\STr \Big\{ (\partial_k R_k) D_k(q) \Gamma^{(3)}_k D_k(q+p)\Gamma^{(3)}_k D_k(q) \Big\}
 -\frac{1}{2}\STr \Big\{(\partial_k R_k) D_k(q)\Gamma^{(4)}_k D_k(q)\Big\},
\end{align}
where $D_k$ is the scale-dependent propagator and $\Gamma^{(3)}_k$ and $\Gamma^{(4)}_k$ are the three- and four-point vertex functions which are in turn derived from the ansatz for the effective action, Eq.~(\ref{eq:Gamma_PDM}). A diagrammatic representation of the resulting flow equations for the $\rho$ and the $a_1$ two-point function is given in Fig.~\ref{fig:2PF}. These flow equations are analytically continued using the aFRG method and then solved numerically for real energies $\omega$. The zero-crossings of the real parts of the two-point functions are used as an approximation for the pole masses of the respective resonances, which are fixed to be $m_\rho^p\approx 775$~MeV and $m_{a_1}^p\approx 1230$~MeV.

The spectral functions are then obtained from the retarded IR propagators as
\begin{equation}
\rho(\omega,\vec p)=-\frac{1}{\pi}\text{Im}\, D^R(\omega,\vec p),
\end{equation}
which can be expressed in terms of the retarded two-point function as
\begin{equation}
\rho(\omega,\vec p)=\frac{1}{\pi}\frac{\text{Im}\,\Gamma^{(2),R}(\omega,\vec p)}{\left(\text{Re}\,
	\Gamma^{(2),R}(\omega,\vec p)\right)^2+\left(\text{Im}\,\Gamma^{(2),R}(\omega,\vec p)\right)^2}.
\end{equation}
For the results shown in this section, the external spatial momentum $\vec{p}$ was set to zero which makes an additional splitting of the spectral functions into a part transverse and longitudinal to the medium unnecessary.

\begin{figure}[b!]
	\centering\includegraphics[width=0.4\textwidth]{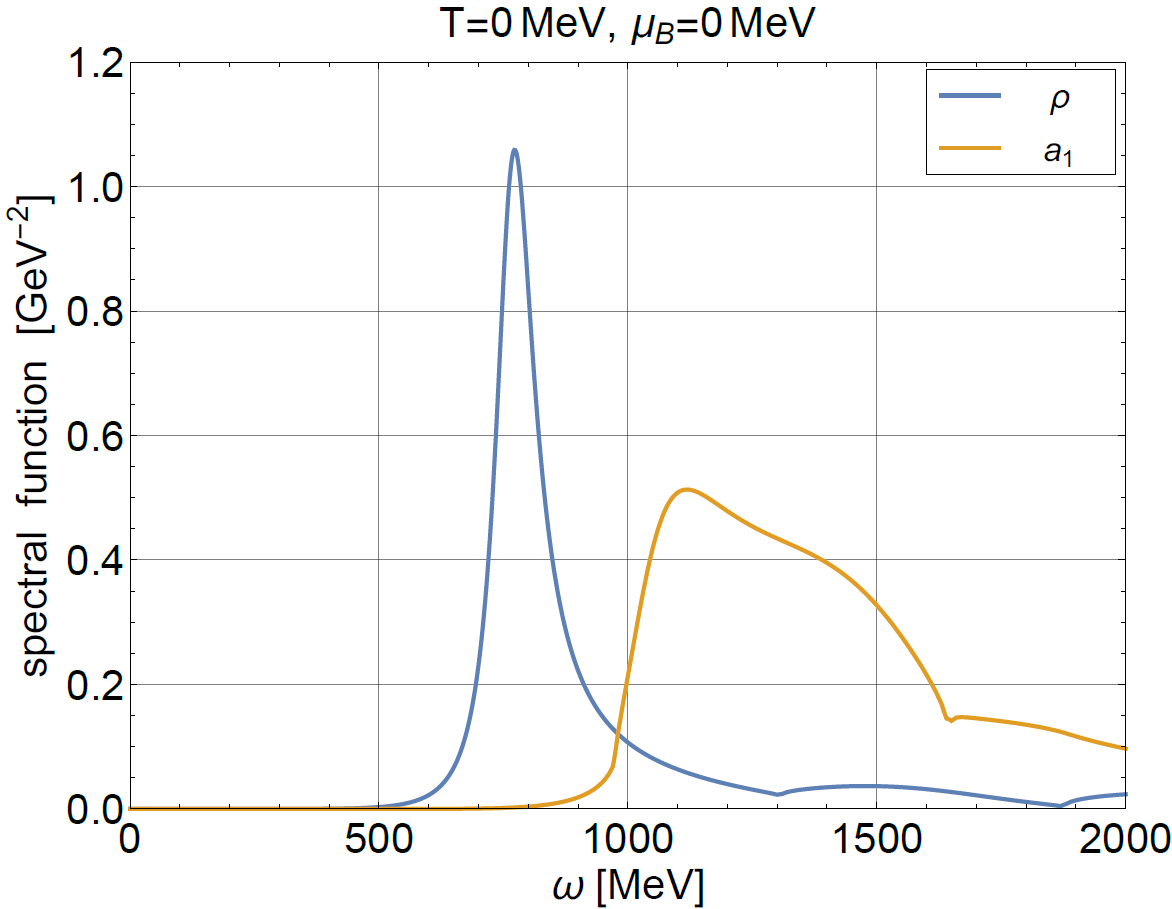}\hspace{0.1\textwidth}
	\includegraphics[width=0.4\textwidth]{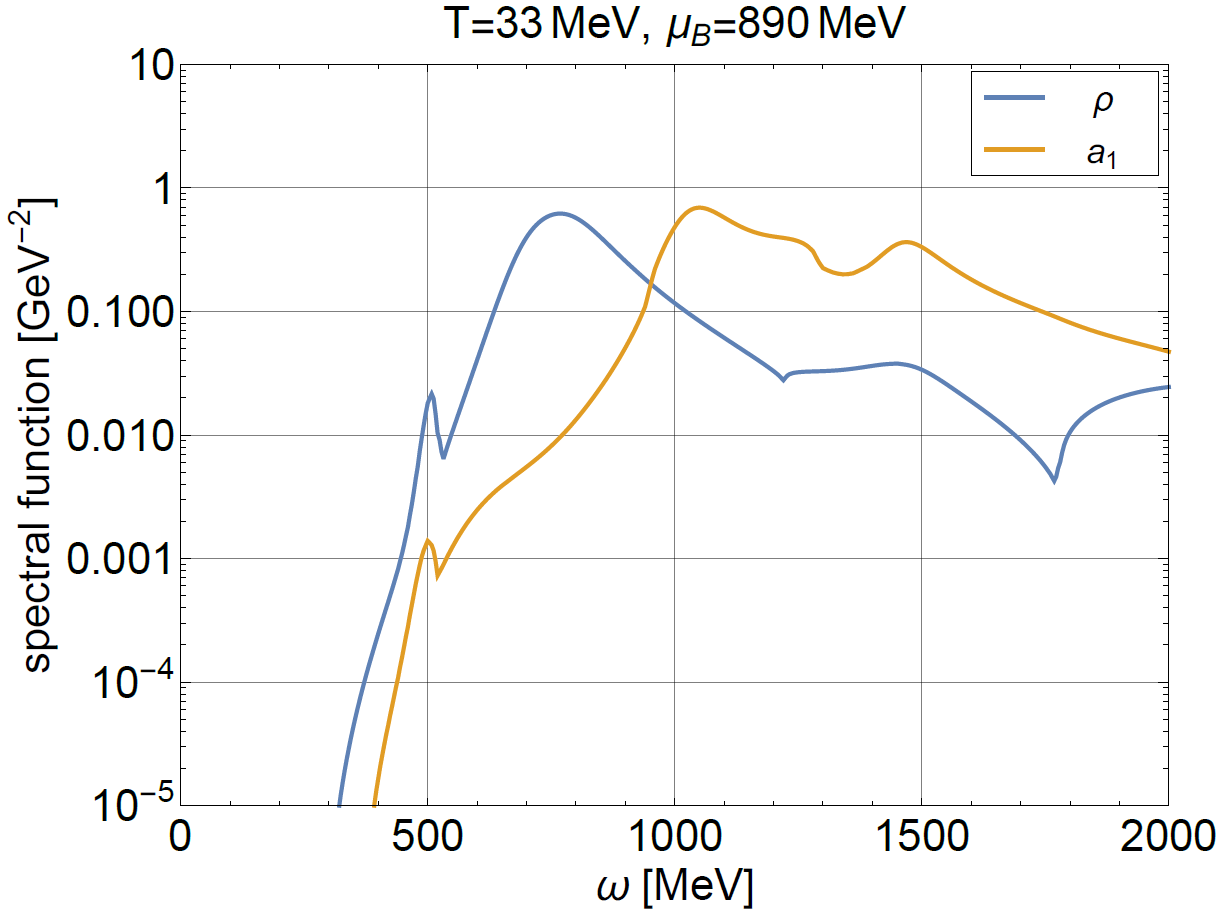}
	\caption{Spectral functions of the $\rho$ and the $a_1$ meson in the vacuum (left) as well as at $\mu_B=890$~MeV and $T=33$~MeV (right). In the vacuum, the $\rho$ spectral function shows a prominent peak at its pole mass of $m_\rho^p\approx 775$~MeV while the $a_1$ spectral function exhibits a broader maximum. At finite temperature and density, the spectral functions show complicated in-medium modifications due to the various decay and capture processes, e.g.~from the $a_1^*+N_1\rightarrow N_2$ process at $\omega\approx 500$~MeV here. Figure adapted from \cite{Tripolt:2021jtp}.}
	\label{fig:spectral_FRG_1}
\end{figure}

Fig.~\ref{fig:spectral_FRG_1} shows the $\rho$ and the $a_1$ spectral function as obtained in \cite{Tripolt:2021jtp} in the vacuum as well as at $\mu_B=890$~MeV and $T=33$~MeV. In the vacuum, the $\rho$ spectral function shows a prominent peak at its pole mass of $m_\rho^p\approx 775$~MeV. The only process contributing in this energy regime is the decay into two pions, $\rho^*\rightarrow \pi+\pi$, while at higher energies the decay channels $\rho^*\rightarrow a_1+\pi$ and $\rho^*\rightarrow N_1+ \bar N_1$ give rise to additional thresholds at around $1300 $~MeV and $1880$~MeV. The $a_1$ spectral function shows a broad maximum near $\omega\approx 1100$~MeV where the width is due to the processes $a_1^*\rightarrow \sigma+\pi$ and $a_1^*\rightarrow \rho+\pi$. At higher energies one observes the $a_1^*\rightarrow a_1+\sigma$ threshold while the $a_1^*\rightarrow N_1+\bar N_1$ contribution is very small below $\omega\approx 2$~GeV. We note in particular that this is the first time that the $\rho$ and $a_1$ spectral functions have been obtained within an aFRG setting without suffering from unphysical decay thresholds into quark-antiquark pairs since here a hadronic effective theory was used which contains nucleons and their parity partners in the place of the quarks in chiral quark models such as the Nambu-Jona-Lasinio or quark-meson model, no matter whether these are enhanced by Polyakov-loop variables to model confinement or not.

\begin{figure}[t!]
	\centering\includegraphics[width=0.4\textwidth]{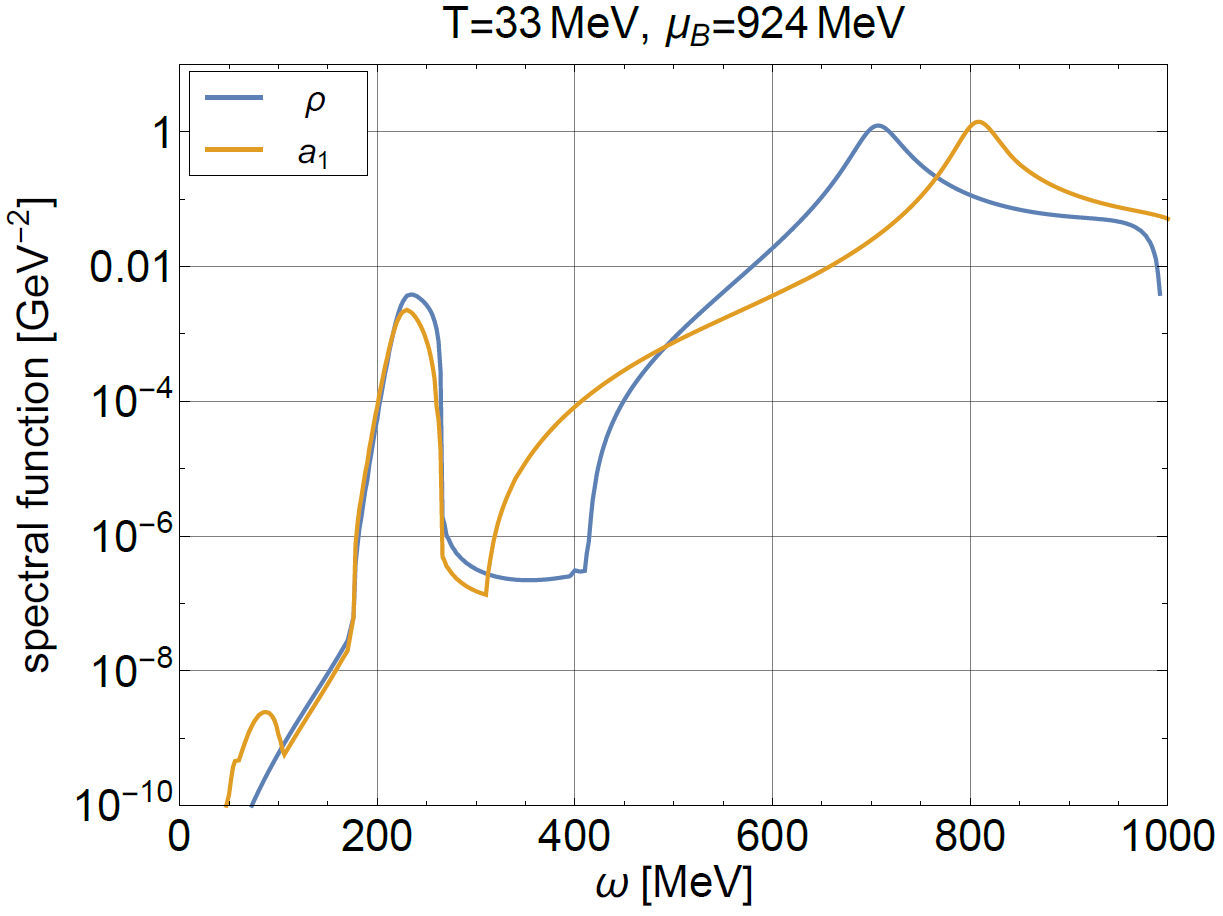}
	\caption{Critical spectral functions of the $\rho$ and the $a_1$ meson at $T=33$~MeV and $\mu_B=924$~MeV, close to the chiral CEP. The most prominent low-energy contributions to both spectral functions arise from baryon-resonance formation $\rho/a_1 + N_1 \to N_2$ which gives rise to prominent peaks around $\omega \approx 250$~MeV where the critical spectral functions have basically no support otherwise. Figure adapted from \cite{Tripolt:2021jtp}.}
	\label{fig:spectral_CEP}
\end{figure}

\begin{figure}[t!]
	\centering\includegraphics[width=0.4\columnwidth]{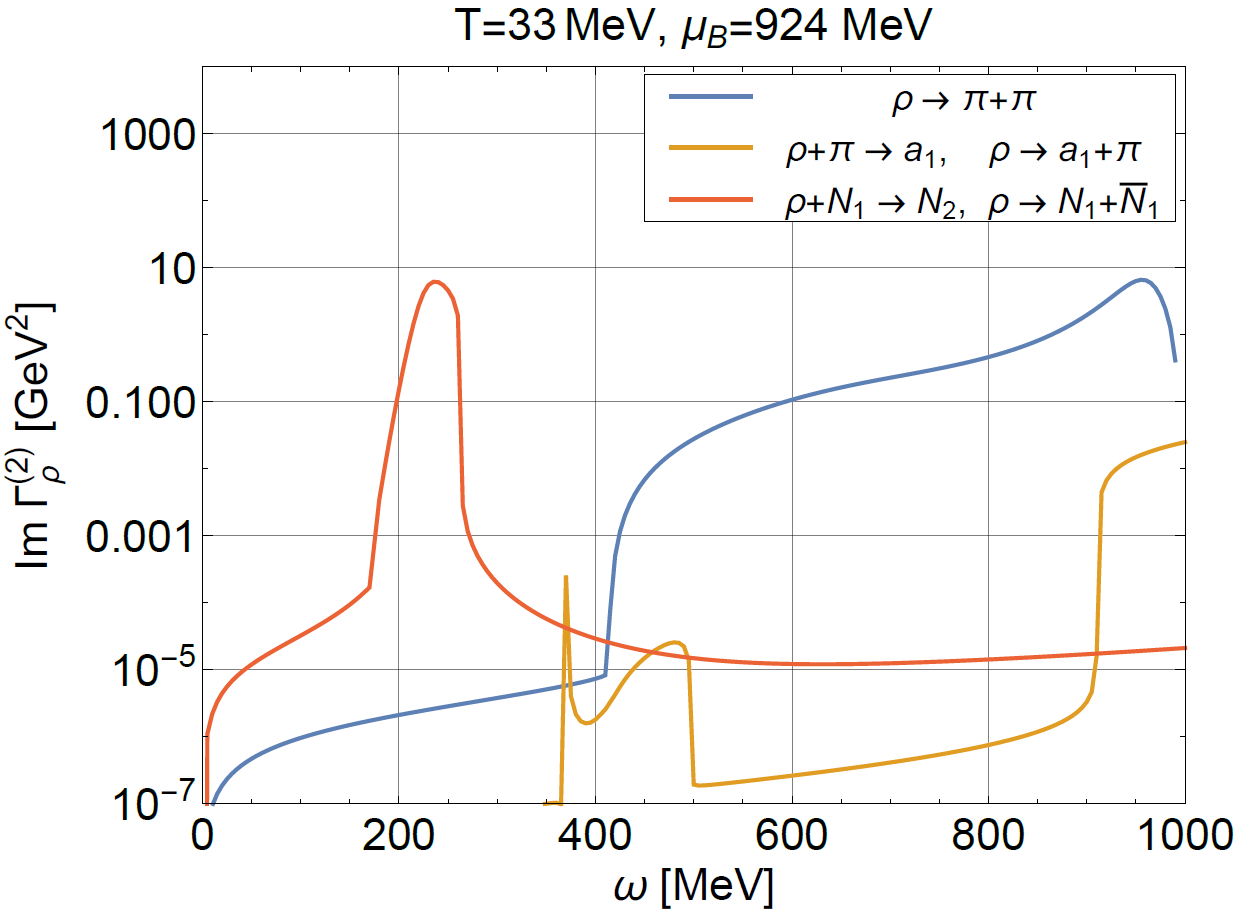}
	\hspace{0.1\textwidth}
	\includegraphics[width=0.4\columnwidth]{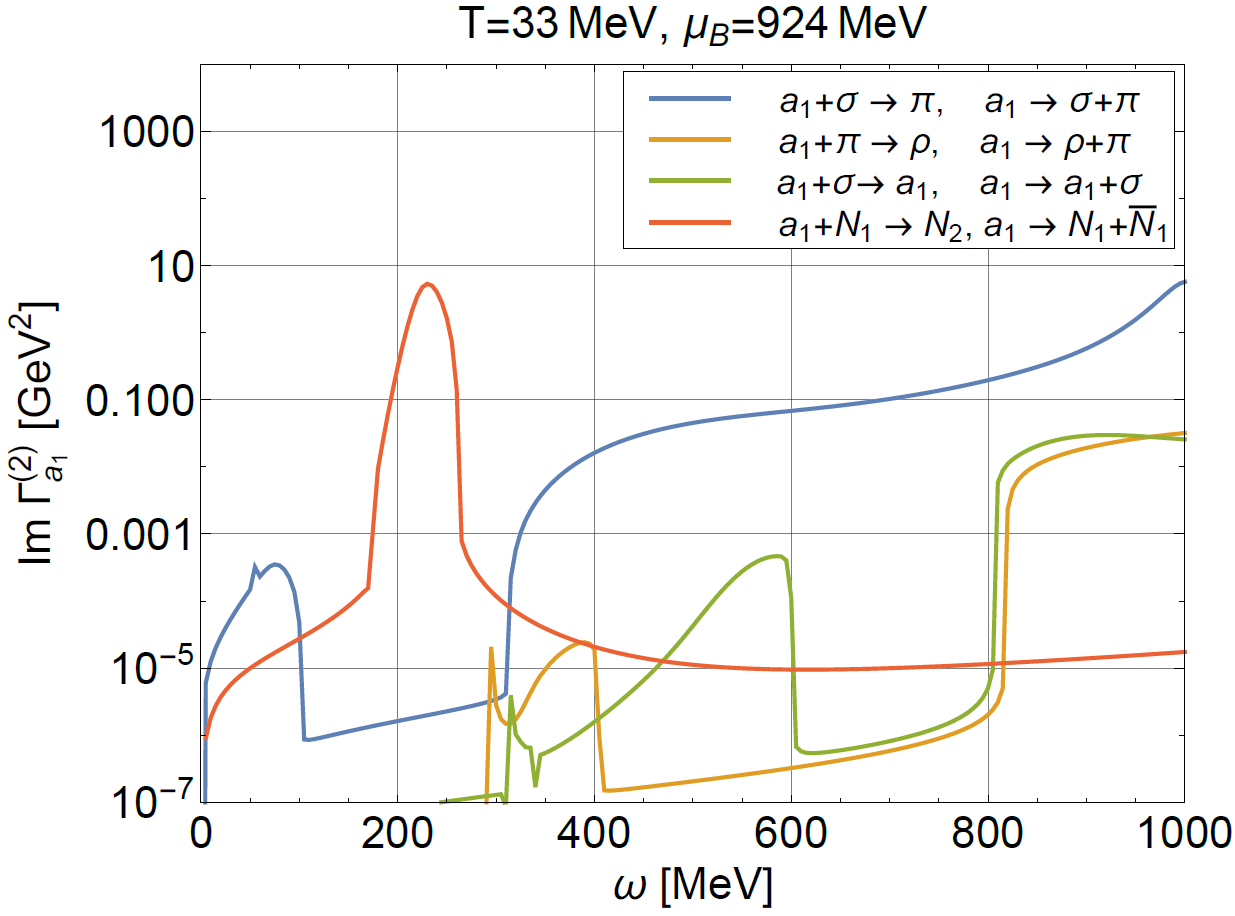}
	\caption{Imaginary part of the $\rho$ (left) and the $a_1$ (right) two-point functions at $T=33$~MeV and $\mu_B=924$~MeV, close to the chiral CEP. Figure adapted from \cite{Tripolt:2021jtp}.}
	\label{fig:ImGamma2_CEP}
\end{figure}

In Fig.~\ref{fig:spectral_CEP} the $\rho$ and $a_1$ spectral functions are shown very close to the chiral CEP, i.e.~at $\mu_B=924$~MeV and $T=33$~MeV. Although the chemical potential changes only by $\sim 30$~MeV as compared to Fig.~\ref{fig:spectral_FRG_1}, the spectral functions show drastic changes, the most prominent being the emergence of a peak structure at low energies of about $\omega\approx 250$~MeV. This is due to the fact that the chiral condensate changes considerably in this regime, as can be seen from the contour lines in Fig.~\ref{fig:PDM_phase_diagram}. This change of the chiral condensate is accompanied by a decrease of the $N^*(1535)$ mass, see also the more detailed discussion in \cite{Tripolt:2021jtp}, which in turn moves the energy threshold for the baryon-resonance formation processes $\rho^* + N_1 \to N_2$ and $a_1^* + N_1 \to N_2$ to lower energies. This can also be seen from Fig.~\ref{fig:ImGamma2_CEP}, where the imaginary parts of the $\rho$ and $a_1$ two-point functions are shown at the same temperature and chemical potential as used in Fig.~\ref{fig:spectral_CEP}. 

Another interesting effect is observed in the $a_1$ spectral function in Fig.~\ref{fig:spectral_CEP}. Here, a small peak is observed at energies below $\omega\approx 100$~MeV due to the critical capture process $a_1^* +\sigma \to \pi$, cf.~also Fig.~\ref{fig:ImGamma2_CEP}. The location of this peak is determined by the mass difference of the pion and the sigma meson, with the pion being the heavier particle here as the sigma meson becomes almost massless near the CEP. However, the observed peak in the $a_1$ spectral function is about six orders of magnitude smaller than the baryon-resonance production peak discussed before and thus turns out to be far too weak to be potentially significant.

We note that the occurrence of the baryon-resonance production peaks is a unique prediction of the baryonic mirror assignment and its observation through enhanced dilepton pair production in the vicinity the chiral CEP would be an important confirmation of this picture of mass generation in QCD. However, in order to make a clear theoretical prediction, the current framework needs to be improved in several ways. In particular, more baryonic states (to which the $\rho$ meson couples) as well as four-point interactions such as 4$\pi$ scattering processes need to be included. Such improvements will likely result in a smearing of the observed peak structures. In addition, when interested in the resulting dilepton spectra in heavy-ion collisions, the dilepton rates obtained from these spectral functions will have to be convoluted with a space-time evolution of the produced fireball. This will give rise to an additional source of smearing and make any peak structures more difficult to observe experimentally.

A first study in this direction was presented recently in \cite{Larionov:2021ycq} where mean-field masses from the parity-doublet model were used for the nucleon and the $N^*(1535)$ in a transport simulation. Therein, an increased dilepton yield was indeed observed at lower energies, see Sec.~\ref{sec:thermal_dilepton_rates} for a more detailed discussion.

\clearpage
\section{Thermal photon and dilepton rates}
\label{sec:thermal_photon_and_dilepton_rates}

In this section, we give an overview of results on thermal photon and dilepton rates which are used in the computation of photon and dilepton spectra in heavy-ion collisions. The underlying theoretical frameworks range from perturbation theory over effective hadronic descriptions like massive Yang-Mills and hadronic many-body theory to the FRG and lattice QCD which are also briefly discussed.

\subsection{Thermal photons from the QGP}
\label{sec:thermal_photons_QGP}
\vspace{2mm}

\begin{figure}[b!]
	\centering\includegraphics[width=0.65\textwidth]{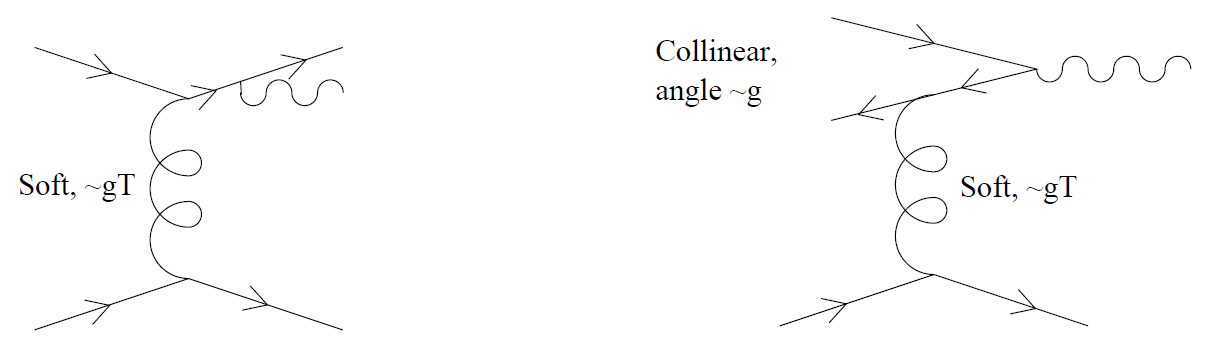}
	\caption{Bremsstrahlung (left) and inelastic pair annihilation (right) processes as relevant for the computation of the thermal photon emission rate within perturbation theory. The momentum exchange is soft, i.e.~of order $gT$, while the outgoing photon is hard, with energy and momentum $\sim T$ which leads to nearly collinear external states. Figure adapted from \cite{Arnold:2001ba}.}
	\label{fig:photons_PT_processes}
\end{figure}

The photon emission rate of an equilibrated, hot QCD plasma at leading order in the logarithm of the strong coupling constant has been available for 20 years \cite{Arnold:2001ba, Arnold:2001ms}. In this seminal work by Arnold, Moore, and Yaffe (AMY), it was shown that the correct leading-order description requires inclusion of near-collinear bremsstrahlung and inelastic pair annihilation contributions as well as of Landau-Pomeranchuk-Migdal suppression effects, see Fig.~\ref{fig:photons_PT_processes}. The photon emission, which is sensitive to the interference of unscattered and scattered waves, thus occurs over a region of spatial extent $1/g^2T$, which is the same as the mean free path for additional scatterings of the quark. Therefore, self-energy resummation is required. We also note that the AMY rates are for an infinite medium, while finite-size effects have been investigated in \cite{Caron-Huot:2010qjx}.

\begin{figure}[t]
	\centering\includegraphics[width=0.4\textwidth]{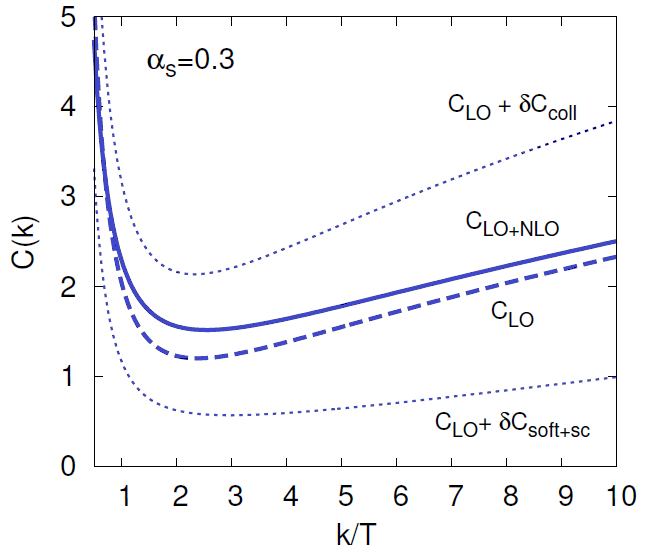}\hspace{0.1\textwidth}
	\centering\includegraphics[width=0.37\textwidth]{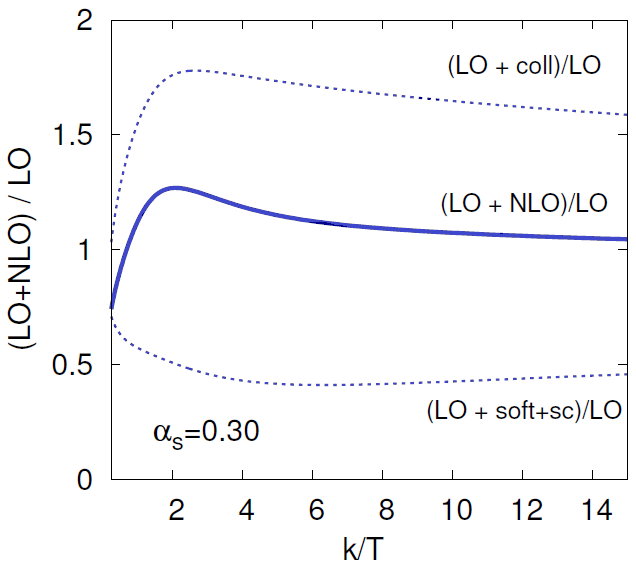}
	\caption{The function $C(k)$ (left) and the normalized thermal photon rate $dR_\gamma/dk$ (right) of a QCD plasma, cf.~Eq.~(\ref{eq:PT_photon_rate}), are shown for $\alpha_s=0.3$ vs.~$k/T$ with photon momentum $k$ and temperature $T$. The NLO result contains contributions from three different regimes: soft, collinear (coll) and semi-collinear (sc). Figure adapted from \cite{Ghiglieri:2013gia}.}
	\label{fig:photons_PT_NLO}
\end{figure}

The next-to-leading order (NLO) $\mathcal{O}(g_s)$ correction to the thermal photon production rate in a QCD plasma was obtained in \cite{Ghiglieri:2013gia}. For the phenomenologically interesting value of $\alpha_s=0.3$ it was found that the NLO correction represents a $20\%$ increase and has a functional form similar to the LO result, see Fig.~\ref{fig:photons_PT_NLO}. Therein, the photon production rate is related to the function $C(k)$ by
\begin{align}
    (2\pi)^3\frac{dR_\gamma}{d^3k}=\mathcal{A}(k) C(k),
    \label{eq:PT_photon_rate}
\end{align}
where $\mathcal{A}(k)$ is the leading-log coefficient. For QCD with up, down, and strange quarks it is given by
\begin{align}
    \mathcal{A}(k)=\frac{4\alpha_{\textbf{EM}}n_F(k)g_s^2T^2}{3k},
\end{align}
see \cite{Ghiglieri:2013gia} for details. The NLO rate arises from distinct kinematic regions and can be separated into contributions from the soft, collinear and semi-collinear regions. It turns out that the NLO contribution from the collinear regime is largely canceled by the contribution from the soft and semi-collinear region, cf.~Fig.~\ref{fig:photons_PT_NLO}. Since the overall correction to the LO result is rather small, the LO rates are still widely used in computations of thermal photon production rates, see also Fig.~\ref{fig:photons_T_paquet}.

\subsection{Thermal photons from hadrons}
\label{sec:thermal_photons_hadrons}
\vspace{2mm}

At high energies, the charged particle multiplicity is dominated by mesons. In order to compute the thermal photon rate of a gas consisting of hadrons, for example of light pseudo-scalar, vector, and axial-vector mesons ($\pi$, $K$, $\rho$, $K^*$, $a_1$), the massive Yang-Mills (MYM) approach can be used \cite{Turbide:2003si}. This approach has the advantage of being able to describe hadronic phenomenology at tree level with a rather limited set of adjustable parameters. Therein, vector and axial-vector fields are introduced into an effective nonlinear $\sigma$-model Lagrangian as massive gauge fields of the chiral $\text{U(3)}_L\times \text{U(3)}_R$ symmetry \cite{Gomm:1984at, Song:1993ae}. The thermal photon production rate can then be obtained within relativistic kinetic theory, cf.~Eq.~(\ref{eq:rel_kin}), by evaluating all possible photon-generating processes given by the MYM Lagrangian.

As pointed out in Sec.~\ref{sec:photon_and_dilepton_rates}, the thermal emission rates of photons and dileptons are closely connected. Both are expressed in terms of the EM spectral function, albeit evaluated in different kinematic regimes, i.e.~the lightlike regime with $M^2=0$ for real photons and the timelike regime with $M^2=q_0^2-q^2>0$ for dileptons. In the case of dilepton rates it has long been known that baryonic effects are of particular importance in order to describe the observed spectra \cite{Rapp:1999ej, Rapp:2002tw}. The same is essentially true for photon production at low energies, see for example \cite{Turbide:2003si} where it was found that the photon emissivity is dominated by baryonic effects in the low-energy regime, $q_0\lesssim 1$~GeV.

\begin{figure}[t]
	\centering\includegraphics[width=0.9\textwidth]{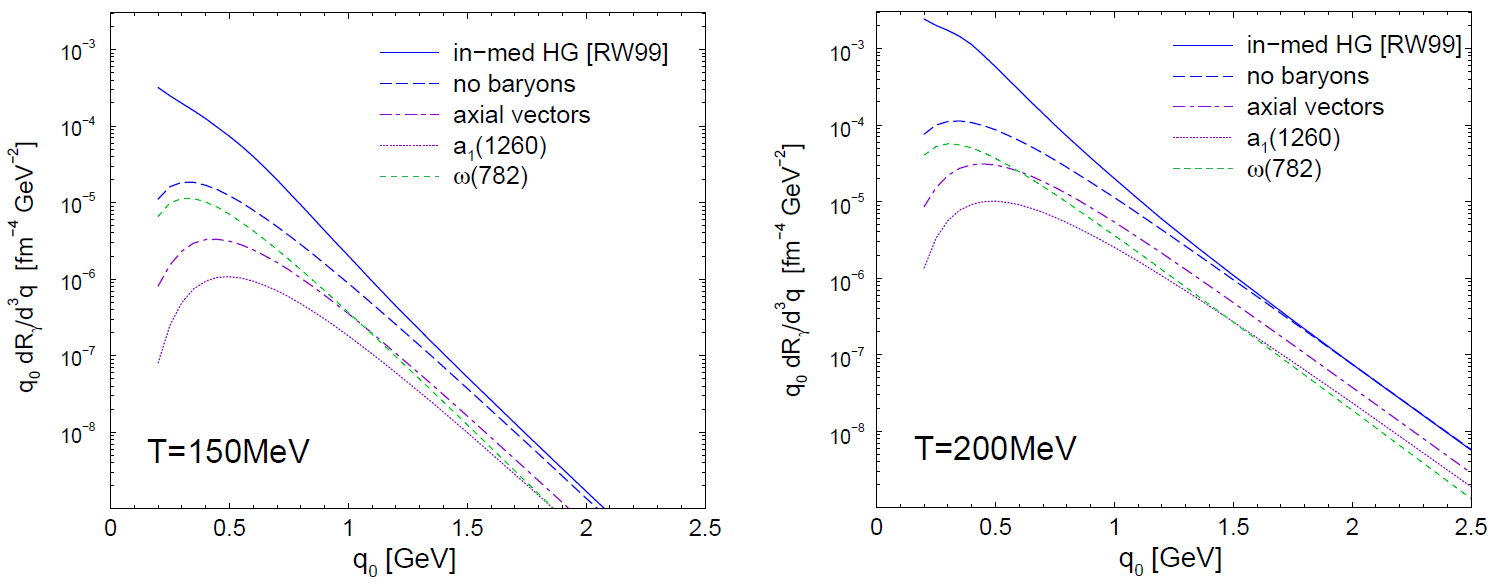}
	\caption{Thermal photon production rate as obtained from the hadronic many-body approach of \cite{Rapp:1999qu, Rapp:1997fs,Rapp:1999us} based on an in-medium $\rho$ spectral function. The left panel corresponds to $\mu_B=340$~MeV and $T=150$~MeV and the right panel to $\mu_B=220$~MeV and $T=200$~MeV, i.e.~conditions resembling CERN-SPS energies. Figure adapted from \cite{Turbide:2003si}.}
	\label{fig:thermal_photons_baryons}
\end{figure}

The results from the hadronic many-body approach are summarized in Fig.~\ref{fig:thermal_photons_baryons} for two temperature-density values characteristic for meson-to-baryon ratios at full CERN-SPS energy (160 GeV). The solid curve shows the net photon spectrum obtained by taking the full $\rho$-meson spectral function to the photon point, whereas the long-dashed curve represents the non-baryonic component. The low-energy regime is clearly dominated by baryonic effects, similar to the case of low-mass dileptons. These effects are mostly due to direct $\rho N$ resonances such as $\Delta(1232)$, $N(1520)$, as well as $\Delta(1232)N^{-1}$ and $NN^{-1}$ excitations in the two-pion cloud of the $\rho$.

Beyond $q_0\simeq 1$~GeV, mesonic (resonance) states become the dominant source of photons in the many-body approach, which includes radiative decays of $\omega(782)$, $h_1(1170)$, $a_1(1260)$, $f_1(1285)$, $\pi(1300)$, $a_2(1320)$, $\omega(1420)$, $\omega(1650)$, $K^*(892)$ and $K_1(1270)$. In particular the $\omega\rightarrow \pi \gamma$ decay is also relevant at lower energies, cf.~Fig.~\ref{fig:thermal_photons_baryons}, while beyond energies of $q_0\simeq 2$~GeV the $\omega$ $t$-channel exchange in $\pi\rho\rightarrow \pi\gamma$ is the most important process, see \cite{Turbide:2003si}.

\begin{figure}[b!]
	\centering\includegraphics[width=0.35\textwidth]{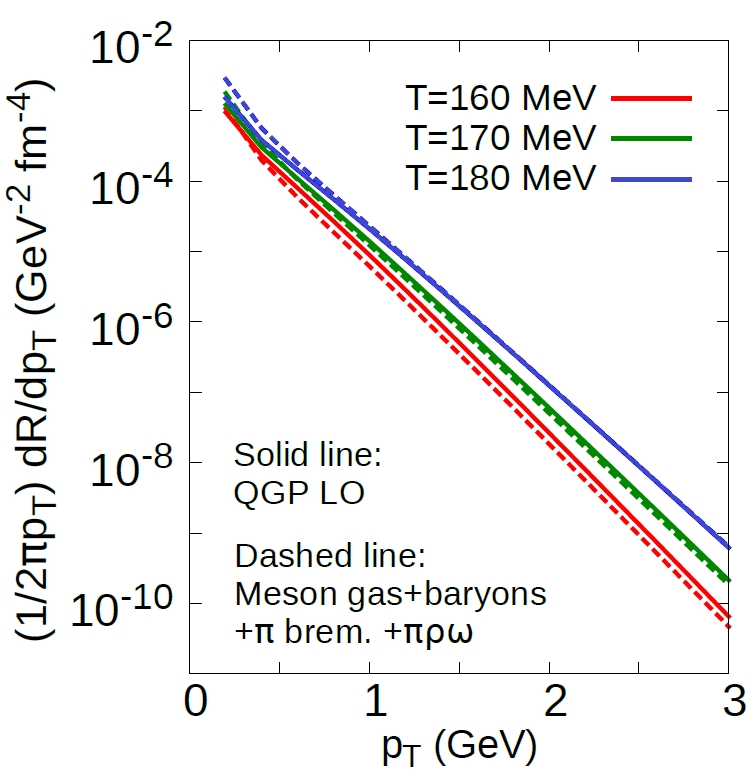}
	\caption{The thermal photon rate vs. the transverse photon momentum $p_T$ as obtained from LO perturbative QCD \cite{Arnold:2001ba, Arnold:2001ms} in comparison to the hadronic photon rate which contains contributions from a hot meson gas and baryons \cite{Turbide:2003si}, from $\pi\pi$ bremsstrahlung \cite{Heffernan:2014mla} and from reactions involving pions, $\rho$ mesons and $\omega$ mesons \cite{Holt:2015cda}, see text for details. Figure adapted from \cite{Paquet:2015lta}.	}
	\label{fig:photons_T_paquet}
\end{figure}

The different contributions from the various hadronic sources then need to be combined in order to obtain the total emission rate, taking care of double-counting and coherence issues. In Fig.~\ref{fig:photons_T_paquet} the hadronic photon rate near the crossover rate is shown together with the ideal QGP rate. In addition to the meson gas contribution obtained from MYM and the baryonic contributions obtained from the $\rho$ spectral function approach, the hadronic rates here also contain estimates of $\pi\pi$ bremsstrahlung contributions \cite{Heffernan:2014mla} and of the reactions $\pi\rho\rightarrow\omega\gamma$, $\pi\omega\rightarrow\rho\gamma$ and $\pi\omega\rightarrow\rho\pi$ \cite{Holt:2015cda}. For a detailed study on baryonic sources of thermal photons we refer to \cite{Holt:2020mwf}. It is instructive to compare the total hadronic rate with the LO QGP rate near the crossover region, as also shown in Fig.~\ref{fig:photons_T_paquet}. One observes that these two rates are nearly identical, thus substantiating the hadron-parton duality idea which entails that the physical system near the crossover can be described in terms of both sets of degrees of freedom, i.e.~hadrons and partons, in this regime. On the one hand, this implies an approach to chiral restoration, since the perturbative rate is chirally restored. On the other hand, it also implies an approach to confinement due to the transition to partonic degrees of freedom.

We note that the photon rates shown in Fig.~\ref{fig:photons_T_paquet} were obtained for an ideal medium without taking effects from a non-zero shear and bulk viscosity into account. The bulk dynamics of strongly interacting matter is, however, sensitive to these transport coefficients. In recent years, progress has been made by including shear and/or bulk corrections to the description of photon rates, see for example \cite{Paquet:2015lta} for a summary of the current status. At present, however, not all photon sources known are amenable to a calculation of viscous corrections.

\subsection{Thermal dilepton rates}
\label{sec:thermal_dilepton_rates}
\vspace{2mm}

The in-medium vector meson spectral functions discussed in the previous sections directly figure into the (low-mass) dilepton rates. The most common assumption therein is Vector Meson Dominance which works well in the vacuum (at least in the purely mesonic sector), as discussed in Sec.~\ref{sec:EM_Spectral_function}. In the baryonic sector, modified versions of VMD \cite{Kroll:1967it} are suitable to describe photo-absorption reactions on the nucleon and on nuclei, i.e., up to at least nuclear saturation density \cite{Rapp:1997ei}. Combining these two results from the mesonic and the baryonic sector, it appears reasonable to assume that VMD holds at least to some extent also in matter in general (composed of mesons and/or baryons at finite temperature and density). However, its ultimate fate in the medium, especially when approaching phase transitions, is not settled.

\begin{figure}[b!]
	\centering\includegraphics[width=0.45\textwidth,height=0.3\textwidth]{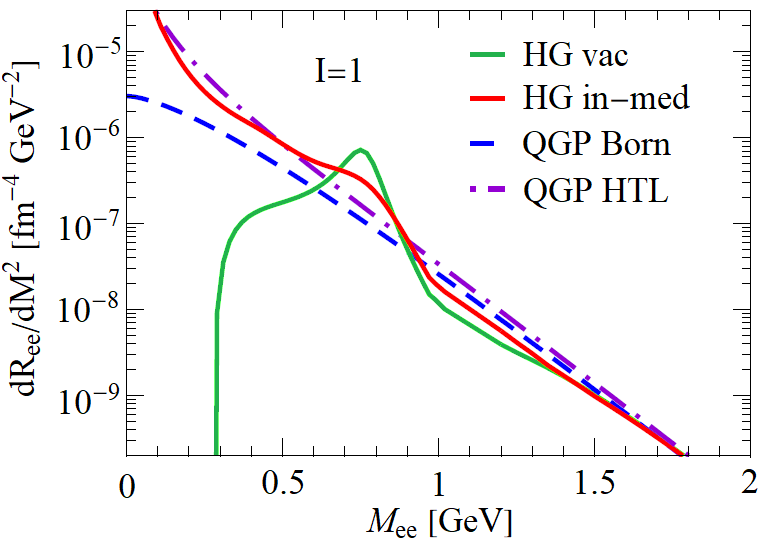}
	\centering\includegraphics[width=0.46\textwidth,height=0.31\textwidth]{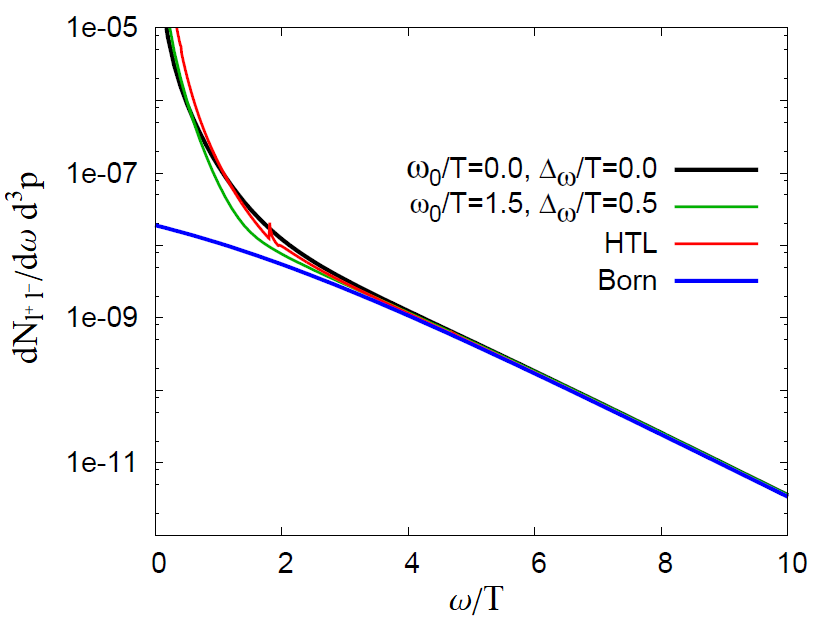}
	\caption{Thermal dilepton rates. Left: as obtained from a hadron gas with vacuum (solid green line) and in-medium \cite{Rapp:1999us} (solid red line) EM spectral function at $T=150$~MeV, compared to $q\bar{q}$ annihilation in leading-order (dashed line) and with hard-thermal loop corrections \cite{Braaten:1990wp} (dashed-dotted line). Right: as obtained from quenched lattice QCD at $1.45 T_c$ \cite{Ding:2010ga} (black and green), also compared to LO and HTL calculations, see text for details. Figure adapted from \cite{Rapp:2011is}.}
	\label{fig:thermal_dileptons}
\end{figure}

In Fig.~\ref{fig:thermal_dileptons} thermal dilepton rates as obtained from hadronic many-body theory, perturbation theory and lattice QCD are shown. Rather than choosing a particular 3-momentum, it is more convenient (and more closely related to mass spectra as observed in experiment) to display the rates in momentum integrated form, i.e.~in terms of
\begin{align}
    \frac{dR_{ll}}{dM^2}=\int \frac{d^3q}{2q_0}\frac{dR_{ll}}{d^4q}=\int d^4x \frac{d^3q}{2q_0} \frac{dN_{ll}}{d^4x \ d^4q}.
\end{align}
The left panel of Fig.~\ref{fig:thermal_dileptons} confirms that the strong broadening of the $\rho$ spectral function, together with the chiral mixing in the dip region, make the hadronic rate approach the partonic-based calculations, in particular the HTL-improved result, where HTL stands for Hard Thermal Loop effective theory \cite{Braaten:1989mz, Andersen:1999fw,Andersen:2002ey,Andersen:2003zk}. At higher temperatures, i.e.~at $T=1.45\, T_c$ as shown in the right panel of Fig.~\ref{fig:thermal_dileptons}, the dilepton rate is rather structureless and one observes a good agreement of the non-perturbative lQCD rate with the HTL rate. These lattice results were obtained for 2-flavor quenched QCD with massless quarks. Therein, the vector spectral functions were modeled by an ansatz consisting of a Breit-Wigner part at low energies and a continuum part at higher energies. The continuum part is smoothly switched off at low energies by a cutoff function that involves the parameters $\omega_0$ and $\Delta_\omega$ which determine the energy at which the continuum part is switched off and the `width' of the cutoff function, respectively. For example, in the limit $\Delta_\omega \rightarrow 0$, the continuum part is simply set to zero below $\omega_0$. 

We note that at energies $\omega \lesssim T$ perturbative calculations, as well as the resummation of certain subsets of diagrams (HTL), become complicated as several scales of order $g^n T$ become important. In fact, the straightforward HTL-resummation \cite{Braaten:1989mz} is known to lead to an infrared divergent spectral function. However, in order to give rise to a non-vanishing, finite transport coefficient, the spectral function needs to be linear in $\omega$, see also Sec.~\ref{sec:dileptons_conductivity}. Fig.~\ref{fig:thermal_dileptons} shows the resulting dilepton rates for the different frameworks. We note that a strong increase at low energies is of course expected due to the Bose factor in the definition of the thermal dilepton rate, cf.~Eq.~(\ref{eq:dilepton1}).

\begin{figure}[t!]
	\centering\includegraphics[width=0.4\textwidth]{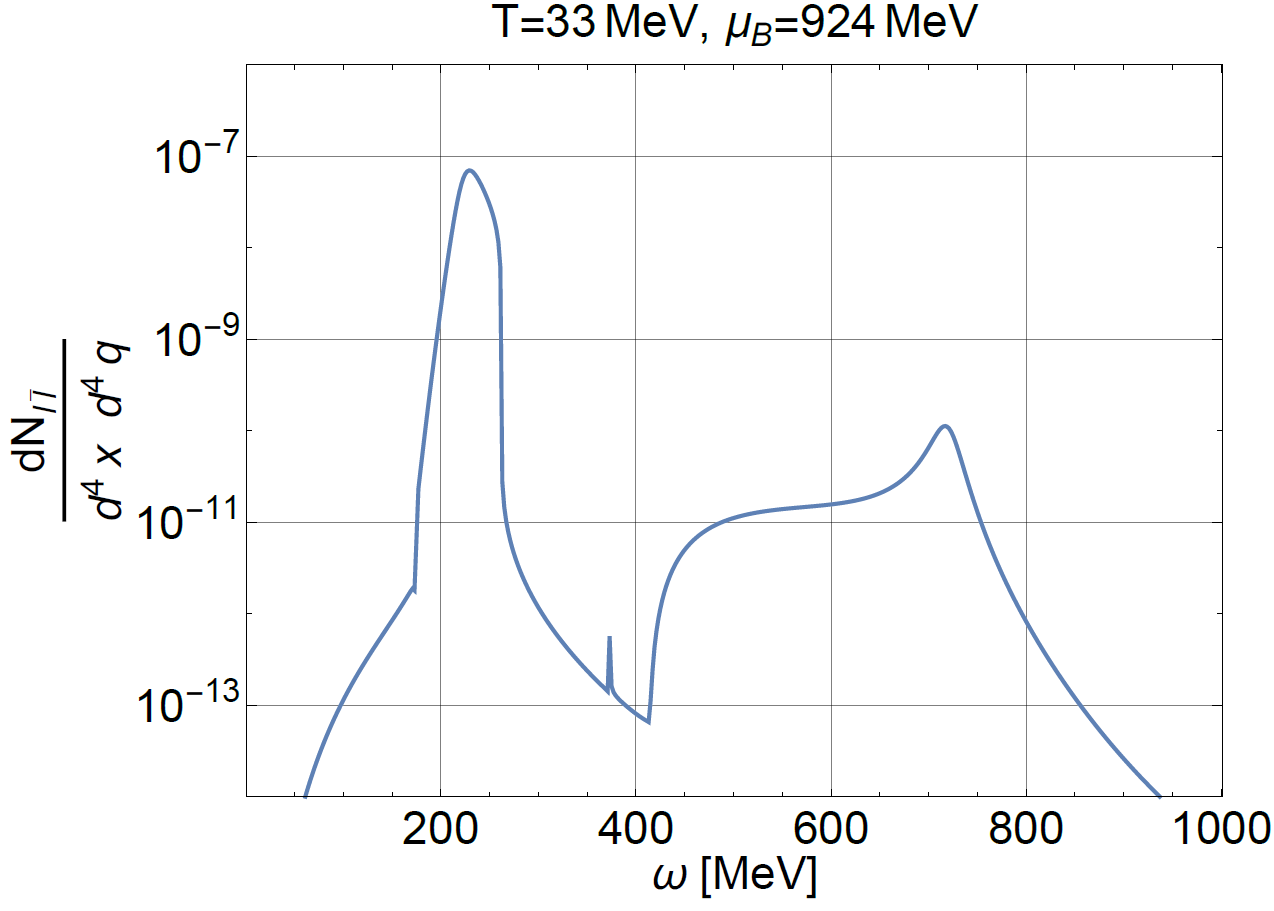}
	\caption{Preliminary result on the thermal dilepton rate as obtained from the Weldon formula, Eq.~(\ref{eq:Weldon}), in combination with the result for the $\rho$ spectral function shown in Fig.~\ref{fig:spectral_CEP}. A strong enhancement near $\omega\approx 220$~MeV is observed which is due to the nucleon resonance peak predicted by the parity-doublet model for partial chiral symmetry restoration, see text for details.}
	\label{fig:PDM_dileptons_CEP}
\end{figure}

Results on the thermal dilepton rate can also be obtained from the vector meson spectral functions presented in Sec.~\ref{sec:aFRG} based on the aFRG method. In Fig.~\ref{fig:PDM_dileptons_CEP} we show a preliminary estimate for the thermal dilepton rate based on the $\rho$ spectral function shown in Fig.~\ref{fig:spectral_CEP} near the chiral CEP. This was obtained by using the Weldon formula \cite{Weldon:1990iw},
\begin{align}
\frac{d^8N_{l\bar{l}}}{d^4x\,d^4q}=\frac{\alpha}{12\pi^3}
\left(1+\frac{2m^2}{q^2}\right)
\left(1-\frac{4m^2}{q^2}\right)^{1/2}
q^2 (2\rho_T+\rho_L)\, n_B(q_0),
\label{eq:Weldon}
\end{align}
where $\alpha\approx 1/137$ is the fine-structure constant, $m$ the lepton mass, and $n_B$ the bosonic occupation number. It expresses the dilepton production rate per space-time volume $d^4x$ and per 4-momentum interval $d^4q$ in terms of the longitudinal and transverse EM spectral function in a thermal medium. 

As a first approximation, the EM spectral function can be obtained using Vector Meson Dominance, cf.~Eq.~(\ref{eq:VMD_1}), with the rho spectral function from Fig.~\ref{fig:spectral_CEP} and phenomenological values for the vector meson mass and coupling as found in \cite{Hohler:2015iba}, i.e.~$m_V=0.86$~GeV and $g_V=6.01$. By further simplifying the Weldon formula by neglecting the dilepton mass $m$ and setting the spatial momentum to zero which entails $\rho_T=\rho_L$, we obtain the result shown in Fig.~\ref{fig:PDM_dileptons_CEP}. We observe an enhancement of the nucleon-resonance peak at lower energies while the high-energy part of the spectral function is suppressed. This is of course a direct consequence of the presence of the bosonic occupation number factor in Eq.~(\ref{eq:Weldon}). At lower energies the dilepton rate strongly decreases, despite the presence of the occupation number factor which becomes very large here. This is due to the employed FRG truncation which does not (yet) include all relevant processes in this energy regime. Similar behavior can also be seen in Fig.~\ref{fig:thermal_dileptons} for the vacuum line.

These results suggest that the nucleon-resonance production peak, which is strongest in regimes of the phase diagram where the chiral condensate is small, might be observed experimentally in the vector channel through an increased dilepton yield at correspondingly low invariant masses measured in heavy-ion collisions at a few GeV/nucleon with high statistics. A more detailed study including effects from additional resonances as well as a convolution with the space-time evolution of the collision process will be necessary in order to see whether the nucleon resonance production peak survives and is still visible in the final dilepton spectrum. In particular dileptons from the Dalitz decay may mask such a signal. High statistics will likely be necessary to observe this enhancement but its detection would yield strong evidence in support of the parity-doubling scenario as providing the mechanism for chiral symmetry restoration inside dense nuclear matter. 

A first study in this direction was recently performed in \cite{Larionov:2021ycq} where the parity-doublet model was used on the mean-field level for the masses of the nucleon and the $N^*(1535)$ within the GiBUU microscopic transport model. Within this setup it was found that the strong dropping of the Dirac mass of the $N^*(1535)$ in the higher-density stage of a collision leads to a considerable enhancement in the production of this resonance as compared to the standard Walecka model. The resulting dilepton yields at low and intermediate invariant masses were found to be slightly enhanced due to these chiral effects and to be in good agreement with HADES data for C+C collisions 1A GeV. We note that it would be interesting to repeat such simulations for heavier nuclei which will lead to the production of a medium with higher densities and thus perhaps to larger effects in the dilepton spectra.

\clearpage 
\section{Photons in heavy-ion collisions}
\label{sec:photons_in_HICs}

\subsection{Classification and production channels}
\label{sec:photons_classification}
\vspace{2mm}

\begin{figure}[b!]
	\centering\includegraphics[width=0.8\textwidth]{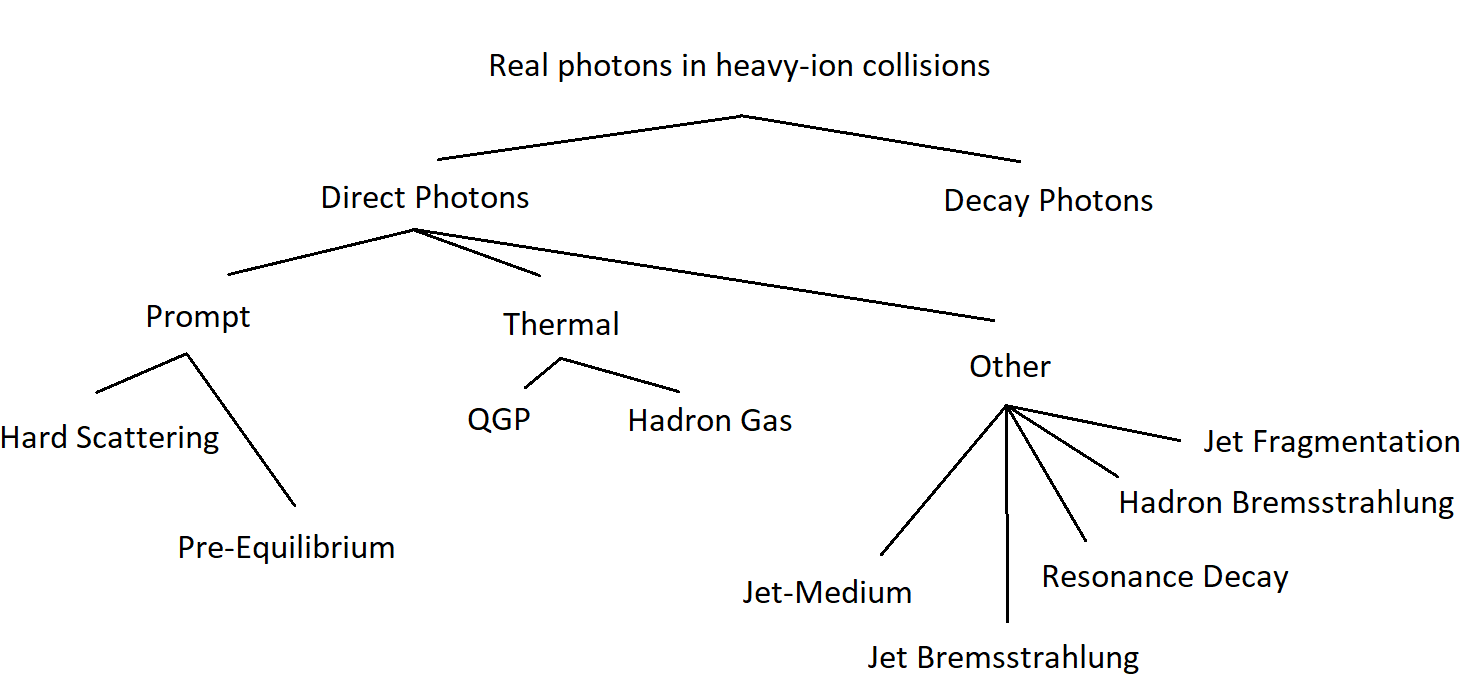}
	\caption{Prevalent terminology for real photons in heavy-ion collisions.}
	\label{fig:photons_terminology}
\end{figure}

Photons are produced at all stages of a heavy-ion collision and can be classified into decay photons and direct photons. Decay photons originate from the decay of long-lived resonances such as pions, eta mesons and omega mesons after freeze-out, while all other photons are called direct photons. Experimentally, the direct-photon contribution is obtained by subtracting the decay-photon contribution from the inclusive (total) spectra:
\begin{align}
    \gamma_{\text{direct}}=\gamma_{\text{inc}}-\gamma_{\text{decay}}=\left( 1-\frac{\gamma_{\text{decay}}}{\gamma_{\text{inc}}}\right)\cdot \gamma_{\text{inc}},
\end{align}
where $\gamma$ denotes some generic quantity proportional to the number of photons, as for example the photon yield in Eq.~(\ref{eq:photon_slope}).
This so-called subtraction method, see e.g.~\cite{WA98:2000ulw}, is based on the measurement of the inclusive photon yield via the reconstruction of their conversion products while the decay photons are obtained by a cocktail calculation. This calculation is based on yield parametrizations of mesons with photon decay branches. The main source of decay photons ($\sim 80\%$) is $\pi^0\rightarrow \gamma\gamma$ and followed by the decay $\eta \rightarrow \gamma\gamma$ ($\sim 18\%$). Direct photons can be classified as follows, see also Fig.~\ref{fig:photons_terminology}:
\begin{itemize}
	\item prompt photons (usually with high transverse momenta $p_T$, i.e.~`hard') from initial hard-scattering processes, see for example \cite{Owens:1986mp}, or the pre-equilibrium phase which includes photons from the `hot glue', created before local thermalization \cite{Shuryak:1992bt} or the Glasma \cite{McLerran:2014hza,Berges:2017eom}, photons from the strong magnetic field \cite{Tuchin:2010gx,Ayala:2017vex}, and from synchrotron radiation \cite{Zakharov:2016kte}
    \item thermal photons from the QGP as well as from the hot and dense hadron-gas phase \cite{Kapusta:1991qp,Baier:1991em} (including short-lived resonances like $\omega$, $a_1$, $\Delta$, $N^*$, ...)
	\item other sources: jet-medium interaction \cite{Fries:2002kt,Renk:2013kya}, hadronic bremsstrahlung \cite{Haglin:2003sh,Liu:2006imd}, jet bremsstrahlung \cite{Owens:1986mp}, jet fragmentation \cite{Owens:1986mp,Klasen:2014xfa}, other resonance decays, $B$-field induced photons, ...
\end{itemize}

We note that it is not always easy to separate photons from the various (sub-)categories. For example, photons from sources like bremsstrahlung are often emitted in the thermal range and are thus indistinguishable from truly `thermal' photons. Even the distinction between direct and decay photons can be ambiguous. Short-lived `resonances', like $\omega$, $\phi$, $a_1$ are sources of decay photons \cite{Xiong:1992ui} but are usually not subtracted by the experiments from the inclusive photon yields - typically, only $\pi^0$ and $\eta$ decays are considered. Thermal photons from the QGP and the hadron gas are of particular interest since they contain information on the hot and dense equilibrium phase of the collision.

\subsection{Interpretation of photon spectra}
\label{sec:photons_interpretation}
\vspace{2mm}

\subsubsection*{Photons as a thermometer and barometer}
\label{sec:photons_thermometer_barometer}
\vspace{2mm}

\begin{figure}[b!]
	\centering\includegraphics[width=0.6\textwidth]{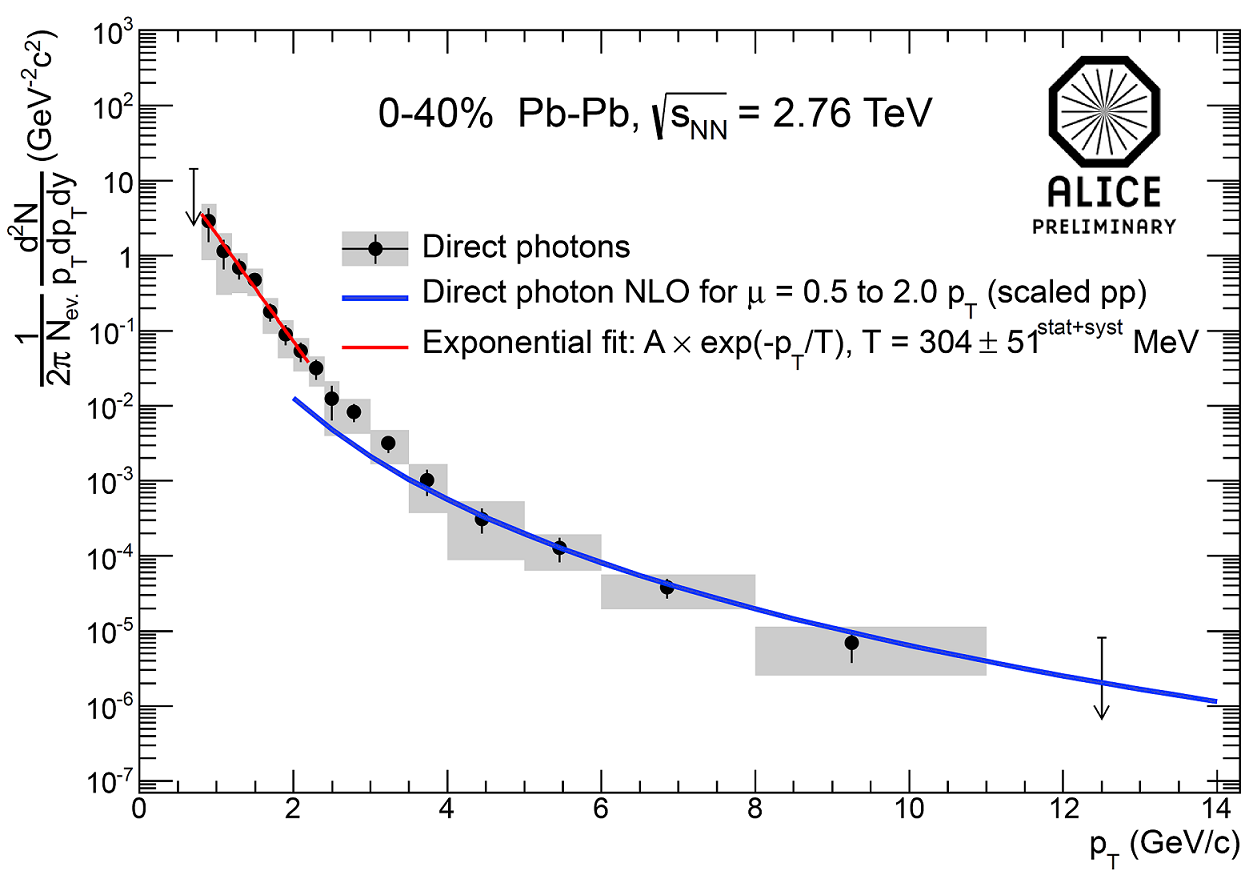}
	\caption{Direct photon spectrum as measured by ALICE in Pb-Pb collisions at $\sqrt{s_{NN}}=2.76$~TeV for $0-40\%$ centrality \cite{Wilde:2012wc, ALICE:2015xmh} with an exponential fit at lower energies and NLO pQCD predictions at higher energies (where $\mu$ is the renormalization scale and the results from p-p reactions are appropriately `scaled' to describe Pb-Pb collisions, see also \cite{Gordon:1994ut}). Figure adapted from \cite{Wilde:2012wc}.}
	\label{fig:ALICE_T_photons}
\end{figure}

At photon energies below 2-3 GeV, the measured photon spectra are approximately exponential and can be characterized by their inverse logarithmic slope, often called `effective temperature':
\begin{align}
\frac{d^2N}{p_{\text{T}} d p_{\text{T}} dy} \sim A \cdot \exp(-p_{\text{T}}/ T_{\text{eff}}).
\label{eq:photon_slope}
\end{align}
In \cite{Wilde:2012wc}, for example, an inverse slope parameter of $T_{\text{eff}}=304\pm 51^{\text{syst+stat}}$~MeV was extracted for Pb-Pb collisions at $\sqrt{s_{NN}}=2.76$~TeV from data taken by the ALICE experiment at the LHC, see also Fig.~\ref{fig:ALICE_T_photons}. In a similar analysis, an inverse slope parameter of $T_{\text{eff}}=221\pm 19^{\text{stat}}\pm 19^{\text{syst}}$~MeV was obtained for Au-Au collisions at $\sqrt{s_{NN}}=200$~GeV by the PHENIX collaboration at BNL \cite{PHENIX:2005yls,PHENIX:2008uif}.

\begin{figure}[t]
	\centering\includegraphics[width=0.9\textwidth]{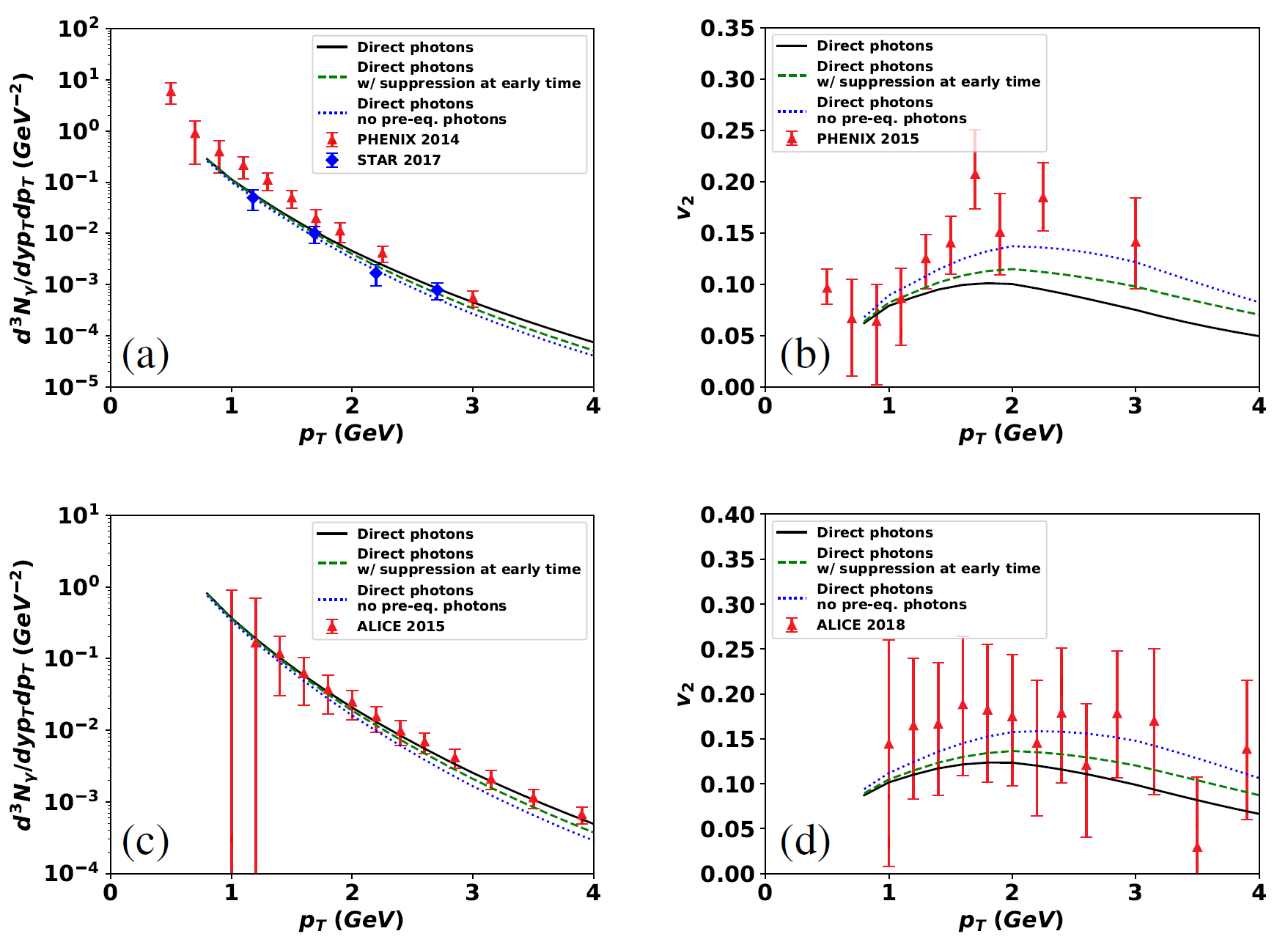}
	\caption{Top row: The spectrum (a) and $v_2$ (b) of direct photons in Au-Au collisions at $\sqrt{s}=200$~AGeV as measured by PHENIX \cite{PHENIX:2014nkk, PHENIX:2015igl} and STAR \cite{STAR:2016use} collaborations at RHIC, compared to theoretical computations obtained in \cite{Gale:2020xlg}. Bottom row: Same as top but for Pb-Pb collisions at $\sqrt{s}=2.76$~ATeV as measured by the ALICE collaboration at LHC \cite{ALICE:2018dti,ALICE:2015xmh}. Figure adapted from \cite{Gale:2020xlg}. For an update of the calculation in \cite{Gale:2020xlg} we refer to \cite{Gale:2021emg}, where most photon rates do have viscous corrections included.}
	\label{fig:Charles_QM2019}
\end{figure}

This effective temperature is, however, blue-shifted due to the transverse flow of the medium,
\begin{align}
T_{\text{eff}}=T\sqrt{\frac{1+\beta}{1-\beta}},
\label{eq:photon_shift}
\end{align}
with $\beta=v/c$ and $T$ the true temperature in the thermal rest frame, see also \cite{vanHees:2011vb} where the quantitative blueshift effect and the issue of the `true' temperature were first pointed out. By modeling the evolution of the radiating medium hydrodynamically, the relation between the effective temperature and the true temperature of the fireball has recently been investigated in \cite{Shen:2013vja}. It was found that at RHIC and LHC energies most photons are emitted from fireball regions with temperatures near the quark-hadron phase transition, but that their effective temperature is significantly enhanced by strong radial flow, see also \cite{vanHees:2011vb}. This finding, i.e.~that a large part of the photons comes from near $T_c$ and the hadronic phase, is an important step towards solving the so-called `$v_2$-puzzle' which refers to the difficulty to theoretically describe the large elliptic flow of photons measured in heavy-ion collisions, see for example \cite{Lohner:2012ct,PHENIX:2011oxq}. Recent comparisons between experimental data and theory, however, show an agreement within the uncertainties, see for example \cite{Gale:2020xlg,Gale:2021emg} and Fig.~\ref{fig:Charles_QM2019}.

\begin{figure}[t]
	\centering\includegraphics[width=0.9\textwidth]{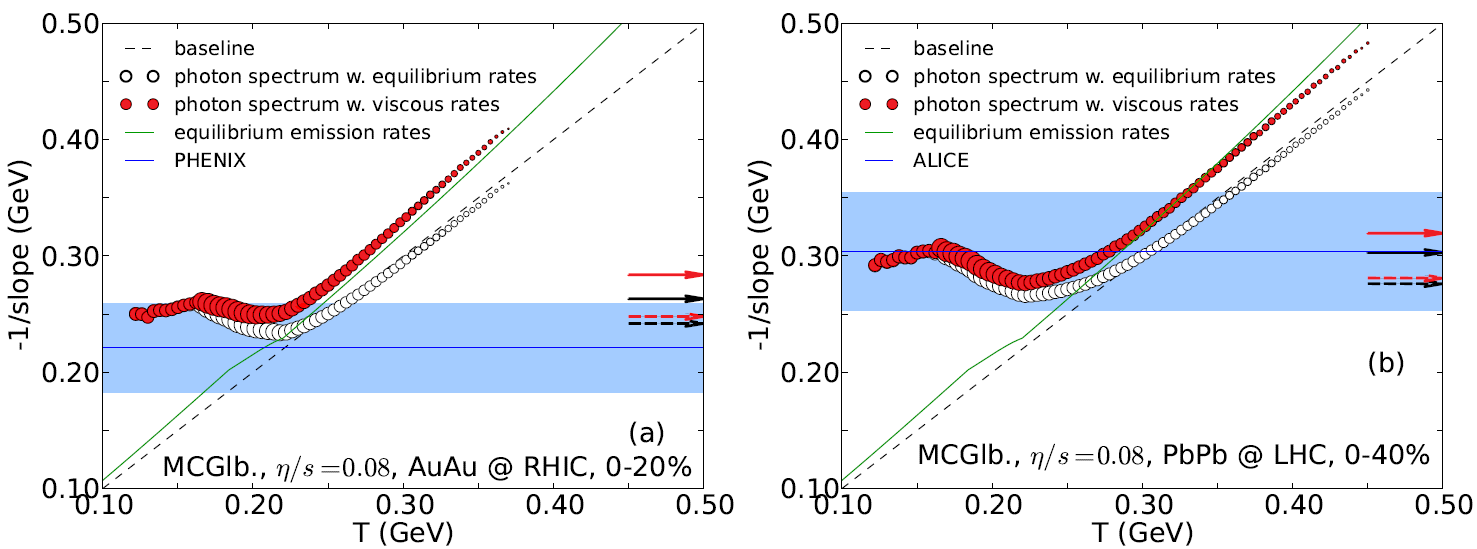}
	\caption{The inverse photon slope parameter $T_{\text{eff}}=-1/\text{slope}$ as a function of the local fluid cell temperature, from the equilibrium thermal emission rates (solid green lines) and from hydrodynamical simulations (open and filled circles), compared with the experimental values (horizontal lines and error bands), for (a) Au-Au collisions at RHIC as measured by the PHENIX collaboration \cite{PHENIX:2008uif} and for (b) Pb-Pb collisions at the LHC as measured by the ALICE collaboration \cite{Wilde:2012wc}. Arrows pointing to the right indicate the inverse slopes of the final space-time integrated hydrodynamic photon spectra: Solid black and red lines correspond to calculations assuming full chemical equilibrium from the beginning and using thermal equilibrium and viscously corrected photon emission rates, respectively. The dashed black and red arrows show the same for calculations with delayed chemical equilibration, see \cite{Shen:2013vja} for details. Figure adapted from \cite{Shen:2013vja}.}
	\label{fig:effective_temperature_hydro}
\end{figure}

In Fig.~\ref{fig:effective_temperature_hydro} the effective temperature as obtained from photon spectra measured by the PHENIX collaboration for Au-Au collisions \cite{PHENIX:2008uif} and by the ALICE collaboration for Pb-Pb collisions \cite{Wilde:2012wc}, respectively, is compared to the results obtained in \cite{Shen:2013vja} on the effective temperature $T_{\text{eff}}=-1/\text{slope}$ vs.~the true temperature $T$. The computed spectra include the thermal rates corrected for shear viscosity effects integrated over the viscous hydrodynamical space-time evolution, and also the prompt photons resulting from the very early interactions of the partons distributed inside the nucleons. The green lines in Fig.~\ref{fig:effective_temperature_hydro} show $T_{\text{eff}}$ vs. the true temperature $T$ for the equilibrium photon emission rates as extracted from an exponential fit. One sees that, due to the phase-space factors associated with the radiation process, the effective temperature of the emission rate is somewhat larger than the true temperature: at high $T$, the QGP emission rate goes roughly as $\exp(-E_\gamma/T) \log(E_\gamma/T)$ \cite{Kapusta:1991qp}, and the logarithmic factor is responsible for the somewhat harder emission spectrum.

The circles in Fig.~\ref{fig:effective_temperature_hydro} show the effective temperatures of photons emitted with equilibrium rates (open black circles) and with viscously corrected rates (filled red circles) from cells of a given temperature within the hydrodynamically evolving viscous medium. The area of the circles is proportional to the total photon yield emitted from all cells at that temperature. As the system cools, the effective photon temperature begins to deviate upward from the true temperature. This is caused by the strengthening radial flow: below $T\sim 200$~MeV, the radial boost effect on $T_{\text{eff}}$ overcompensates for the fireball cooling. We conclude that a robust understanding of the space-time evolution of the heavy-ion collision is necessary in order to extract reliable values for the true temperature.

Photons are also useful as a `viscometer', see e.g.~\cite{Shen:2013cca}, where viscous photon emission from nuclear collisions at RHIC and LHC was investigated by evolving fluctuating initial density profiles with event-by-event viscous hydrodynamics. Momentum spectra of thermal photons, radiated by these explosively expanding fireballs, and their $p_T$-differential anisotropic flow coefficients were computed, both with and without accounting for viscous corrections to the standard thermal emission rates. The overall effect of viscous corrections on the rates on the direct photon spectra was found to be small, which can be understood from the fact that viscous corrections are larger at higher $p_T$, where prompt photons dominate over thermal ones. The direct photon $v_2$, on the other hand, is suppressed at higher $p_T$ by both shear and bulk corrections to the photon rates, with the suppression being of the order of $20-30\%$ \cite{Paquet:2015lta}, see also \cite{Vujanovic:2019yih} for a study on the influence of bulk viscosity of QCD on dilepton tomography.

Traditionally, the vast majority of photon calculations account for the radiation from the very first nucleon-nucleon collisions and for that emitted throughout the fluid dynamical evolution \cite{Paquet:2015lta}. Closer attention is now being paid to the late stages \cite{Schafer:2020vvw}, and some results \cite{Churchill:2020yny, Greif:2016jeb} now also include the photons from early, pre-equilibrium, pre-hydro phases, see for example \cite{Gale:2020xlg}. Therein, a hybrid model which relies on QCD effective theory, K{\o}MP{\o}ST \cite{Kurkela:2018vqr}, to dynamically bridge the gap between IP-Glasma initial states \cite{Schenke:2012wb, McDonald:2020oyf} and viscous hydrodynamics. The hydrodynamical phase is then followed by dynamical freeze-out handled by UrQMD.  

The results of these photon calculations are shown in Fig.~\ref{fig:Charles_QM2019} for the photon spectrum and the photon elliptical flow. The dashed line shows the complete result on the direct photon signal with ``early time suppression'': the correction attributed to the gluon-dominated beginning of the K{\o}MP{\o}ST phase. The solid line is the direct photon signal without this correction and the dotted line is obtained by omitting the radiation from the effective kinetic theory period. The pre-hydro photons are found to contribute $\sim10-20\%$ to the net spectrum at higher transverse momenta. At RHIC, the calculated photon spectrum lies below the PHENIX data but is consistent with STAR measurements. We note, however, that the photon spectra and $v_2$ Fig.~\ref{fig:Charles_QM2019} do not yet account for viscous effects \cite{Paquet:2015lta}.

\subsection{Recent experimental results}
\label{sec:photons_experimental_results}
\vspace{2mm}

\begin{figure}[b!]
    \centering
    \includegraphics[width=.59\textwidth]{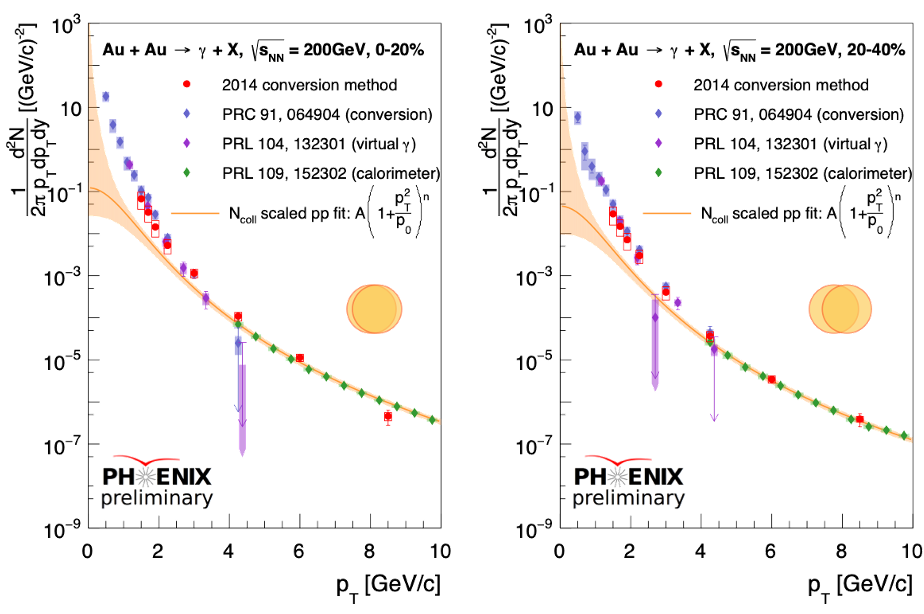}
    \includegraphics[width=.4\textwidth]{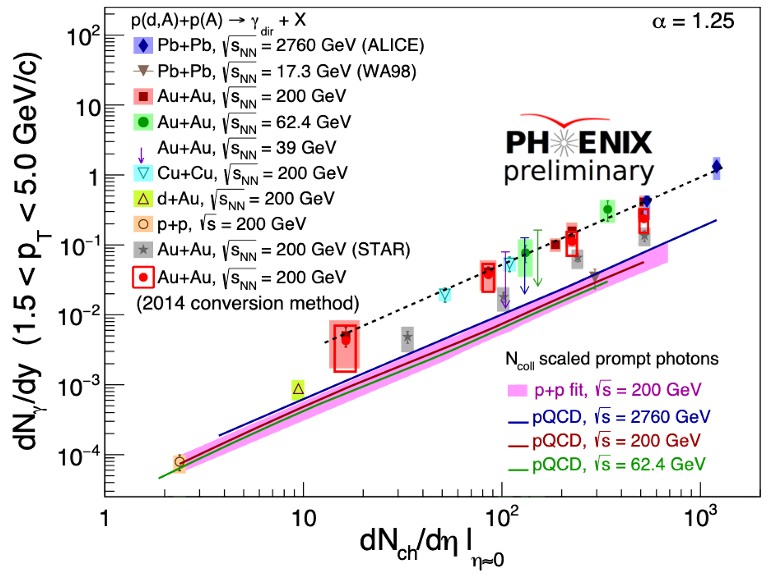}
    \caption{Direct photon yields in Au$+$Au collisions at $\sqrt{s_\mathrm{NN}}=200$~GeV \cite{Fan:2021rvi,PHENIX:2022rsx}.
    Left panel: for 0-20\% central collisions. Middle panel: for 20-40\% semi-peripheral collisions. Right panel: 
    universal scaling of low-$p_\mathrm{T}$ direct photon yields $dN_\gamma/dy$ with respect to the number of charged particles at midrapidity $dN_\mathrm{ch}/d\eta|_{\eta=0}$. Figures presented at Quark Matter 2019 \cite{Fan:2021rvi}.}
    \label{fig:phenixdirectgamma}
\end{figure}

The major experimental challenge in the measurement of direct photons is to disentangle contributions from the very large number of photons that stem from the decay of short-lived hadrons such as the $\pi^0$ and $\eta$ mesons. Consequently, direct photon measurements will have to be performed on a statistical basis as direct photons cannot uniquely be distinguished from decay photons. Electromagnetic decays from final state hadrons by far provide the largest contribution and as such form a substantial background to the measurement of direct photons. Early measurements of direct photon production in heavy-ion collisions at the SPS accelerator did not show significant results \cite{WA80:1990teq,CERES:1995rik,WA80:1995xza}. With the large area electromagnetic calorimeter, the WA98 collaboration was able to correct in a self-consistent manner the contributions from $\pi^0$ and $\eta$ mesons using the same detector and data set \cite{WA98:2000vxl}. In central Pb+Pb collisions at $\sqrt{s_{NN}}=17.2$~GeV it reported the first observation of direct photons with $p_T > 1.5$~GeV/$c$.  

Recently, the PHENIX collaboration reported on new measurements of the direct-photon spectrum from Au$+$Au collisions at $\sqrt{s_\mathrm{NN}}=$200~GeV \cite{Fan:2021rvi,PHENIX:2022rsx}. These new measurements, shown in Fig.~\ref{fig:phenixdirectgamma}, use the external-photon-conversion method and are based on the large RHIC Run-14 data sample. The results show a clear enhancement in the direct photon yields for $p_\mathrm{T} \le 3$~GeV/$c$, which continues to persist in the semi-peripheral data (middle panel). At high momenta, the results show consistency with $N_\mathrm{coll}$-scaled p$+$p results. In another important consistency check, PHENIX's new results show good agreement with previously published results based on different data sets \cite{Adare:2014fwh}, and/or different methods such as the virtual-$\gamma$ \cite{Adare:2008ab} and the calorimeter methods \cite{Adare:2011zr}. In the right panel of Fig.~\ref{fig:phenixdirectgamma}, the invariant yield of photons is plotted as a function of the charged hadron multiplicity $dN_\mathrm{ch}/d\eta$ at midrapidity. The new data from PHENIX are in line with the recently observed scaling \cite{Adare:2018wgc} of the photon yield with the charged particle yield $(dN_\mathrm{ch}/d\eta)^\alpha$, both at midrapidity. The value of $\alpha=1.25\pm0.02$ is based on a simultaneous fit of $N_\mathrm{coll}$ versus $dN_\mathrm{ch}/d\eta$ for a wide range of center-of-mass energies \cite{Adare:2018wgc}. In the same figure, data from the WA98 and STAR experiments are added. While the scaling appears to be similar, the rates are systematically lower. Data from the STAR Beam Energy Scan (BES) Phase-2 program should help resolve this apparent tension by adding several new data points at lower charged hadron multiplicities using a similar conversion technique \cite{STAR:2016use}.

\clearpage

\section{Dileptons in heavy-ion collisions}
\label{sec:dileptons_in_HICs}

\subsection{Classification and production channels}
\label{sec:dileptons_classification}
\vspace{2mm}

\begin{figure}[h!]
	\centering\includegraphics[width=0.8\textwidth]{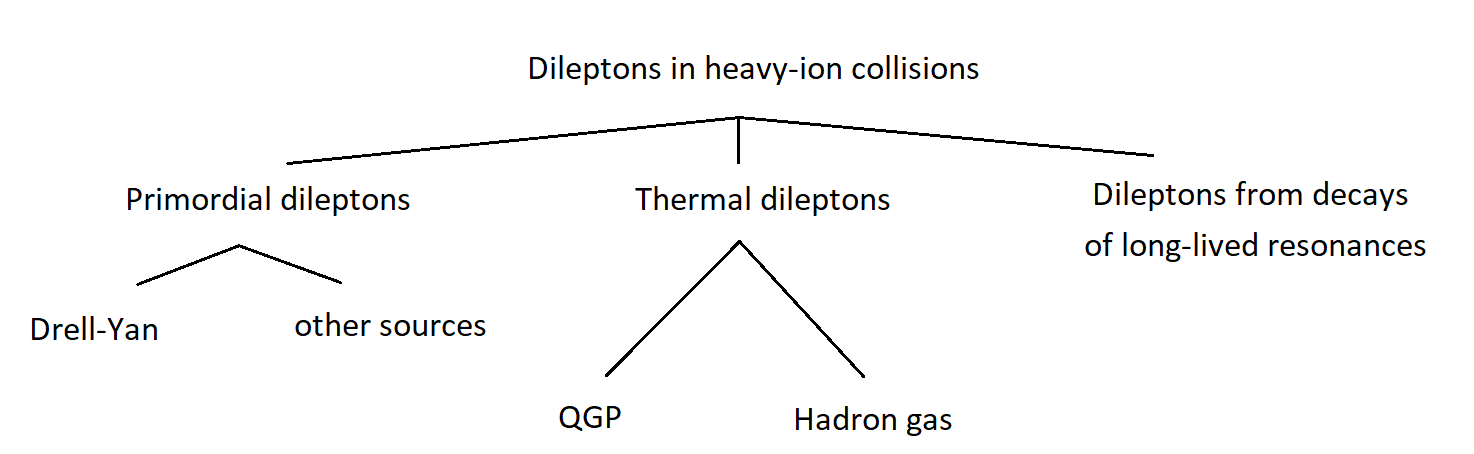}
	\caption{Classification of dileptons in heavy-ion collisions.}
	\label{fig:dilepton_processes}
\end{figure}

Similar to photons, dileptons, i.e., $e^+e^-$ or $\mu^+\mu^-$ pairs, are produced at all stages of the collision and can be classified as follows, see also Fig.~\ref{fig:dilepton_processes}:
\begin{itemize}
	\item `primordial' dileptons from $q\bar{q}$ annihilation, i.e.~from Drell-Yan processes like $NN\rightarrow e^+e^-X$
	\item thermal dileptons from the QGP as well as from the hot and dense hadron-gas phase originating from processes like $q\bar{q}\rightarrow e^+e^-$ and $\pi^+\pi^-\rightarrow e^+e^-$ as well as from multi-meson reactions (`$4\pi$') and decays of short-lived resonances like $\rho$, $\omega$, $a_1$, $\Delta$, $N^*$, ... 
	\item dileptons from decays of long-lived mesons and baryons (e.g. from $\pi^0$, $\eta$, $\phi$, $J/\Psi$, $\Psi$', $D\bar D$, ...) including the Dalitz decays for some of the light mesons (e.g. $\pi^0, \eta, \eta'\rightarrow\gamma e^+e^-$)
    \item especially at low center-of-mass energies, where baryons dominate over mesonic degrees of freedom, additional sources can contribute substantially such as Dalitz decays of resonances (e.g. $\Delta, N^*\rightarrow N e^+e^-$) and non-resonance, quasi-elastic NN bremsstrahlung (e.g. $NN\rightarrow NNe^+e^-$) \cite{HADES:2009cui}.
\end{itemize}
In the following, we will again focus on the thermal radiation from the QGP and the hadron-gas phase.

\subsection{Interpretation of dilepton spectra}
\label{sec:dileptons_interpretation}
\vspace{2mm}

\begin{figure}[t]
	\centering\includegraphics[width=0.55\textwidth]{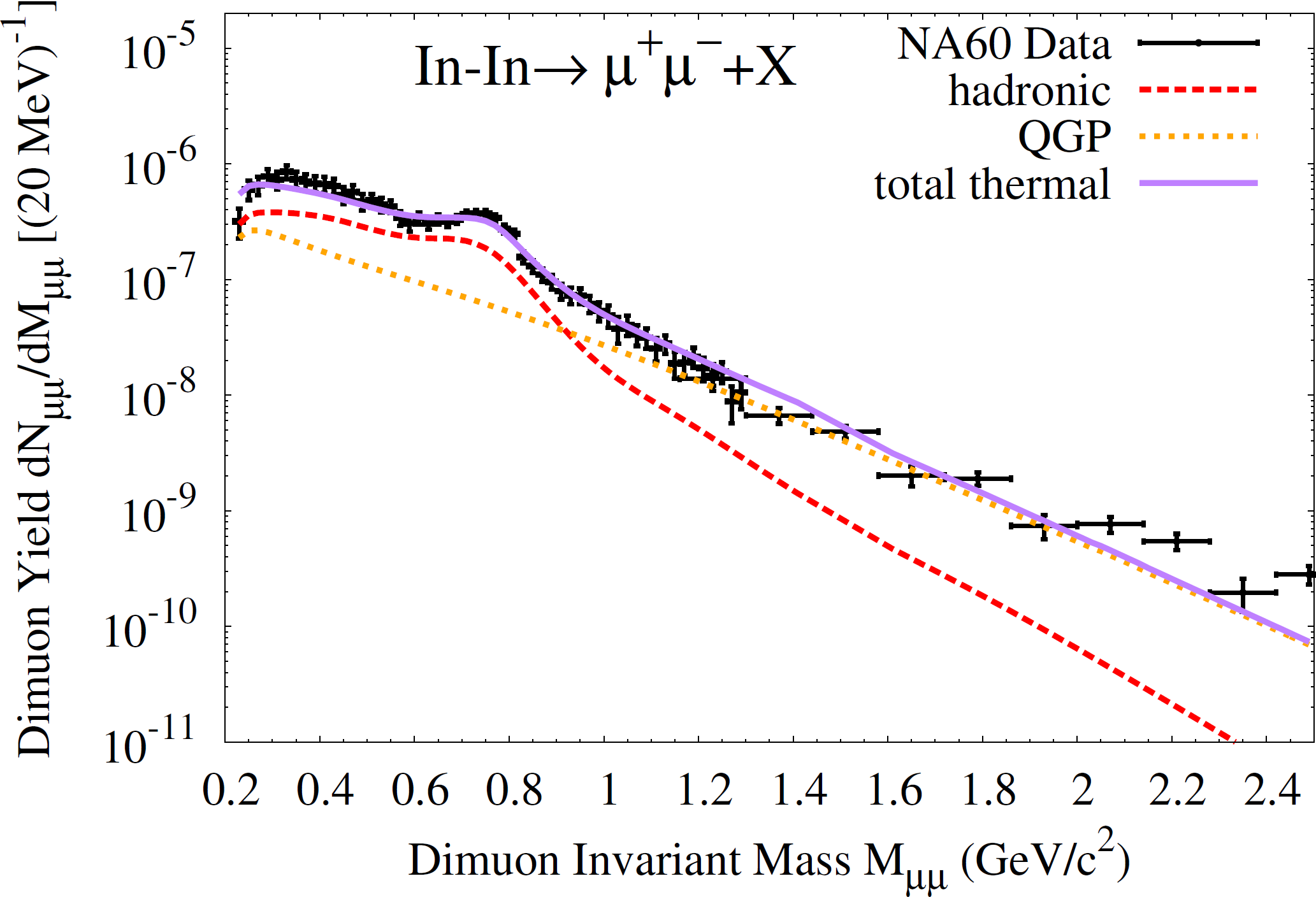}
	\caption{The excess dimuon invariant-mass spectrum in In-In collisions ($\sqrt{s_{NN}}=17.3$~GeV) as measured by NA60 at the SPS \cite{Specht:2010xu,NA60:2008ctj} is compared to a theoretical calculation \cite{Rapp:2014hha} composed of hadronic radiation (using in-medium $\rho$ and $\omega$ spectral functions and multi-pion annihilation with chiral mixing, dashed line) and QGP radiation (using a lattice-QCD inspired rate, dotted line).  Figure adapted from \cite{Rapp:2014hha}.}
	\label{fig:dilepton_spectrum_2}
\end{figure}

Dilepton invariant-mass spectra have long been recognized as the only observable which gives direct access to an in-medium spectral function of the QCD medium, most notably of the $\rho$ meson \cite{Brown:1991kk,Pisarski:1995xu,Rapp:1999ej,Harada:2003jx}. They also allow for a temperature measurement that is neither distorted by blue-shift effects nor limited by the hadron formation temperature \cite{Braun-Munzinger:2003htr}. In recent years, the agreement between experimentally measured dilepton spectra and theoretical predictions has reached an excellent quantitative agreement. In \cite{Rapp:2014hha}, for example, it was shown that the predictions of hadronic many-body theory for a melting $\rho$ meson, coupled with QGP emission utilizing a modern lattice-QCD based equation of state, yield a quantitative description of dilepton spectra in heavy-ion collisions at the SPS and the RHIC beam energy scan program. In Fig.~\ref{fig:dilepton_spectrum_2} the predictions from this approach are compared to the high-precision NA60 data on the excess dimuon invariant-mass spectrum as measured in In-In collisions ($\sqrt{s_{NN}}=17.3$~GeV) \cite{Specht:2010xu,NA60:2008ctj}. 

\subsubsection*{Dileptons as a barometer}
\label{sec:dileptons_barometer}
\vspace{2mm}

\begin{figure}[b!]
	\centering\includegraphics[width=0.55\textwidth]{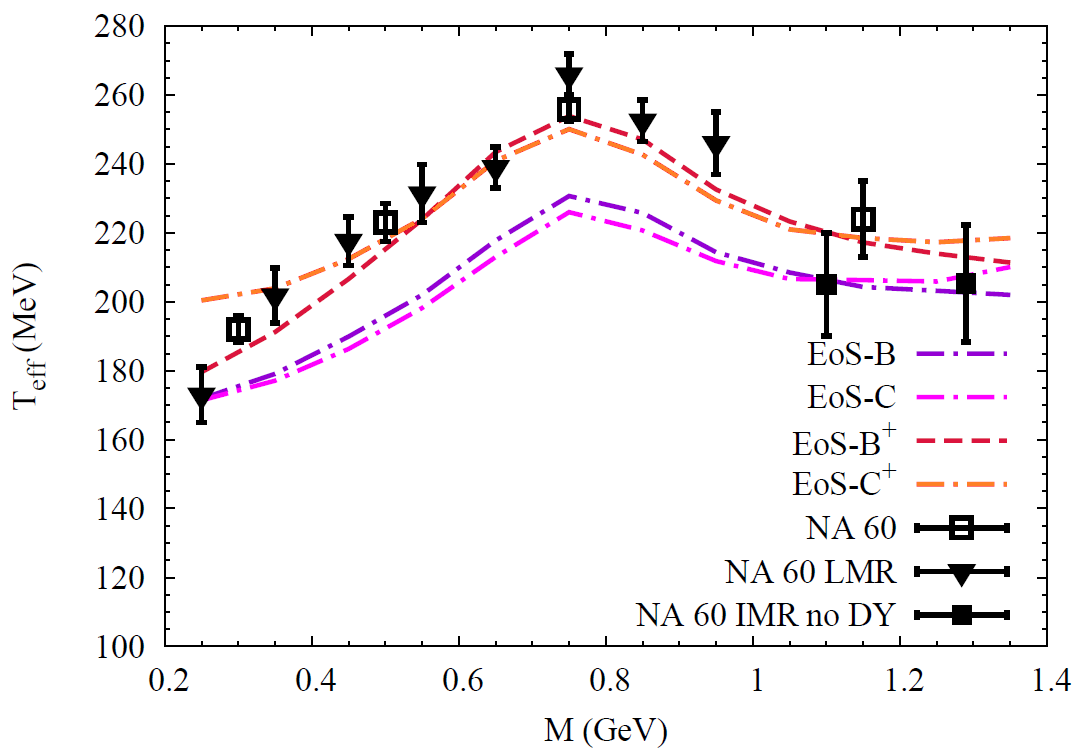}
	\caption{Effective slope parameter for excess dimuons as a function of their invariant mass as measured by NA60 in In-In collisions at $\sqrt{s}=17.3$~AGeV at SPS \cite{NA60:2008dcb}, compared to theory calculations \cite{vanHees:2007th}. Different equations of state and transverse fireball accelerations $a_t$ are used to ascertain the uncertainties of the result (where the latter simply describe the rate of expansion of the medium): EoS-B uses a scenario with a critical temperature of $T_c=160$~MeV while EoS-C uses $T_c=190$~MeV, and both use $a_t=0.085$~fm/c$^2$. EoS-B$^+$ and EoS-C$^+$ use the same equations of state as EoS-B and EoS-C, respectively,  but a transverse fireball acceleration of $a_t=0.1$~fm/c$^2$. Figure adapted from \cite{Rapp:2011is}.}
	\label{fig:dileptons_barometer}
\end{figure}

Similar to real photons, also the transverse-momentum spectra of dileptons can be used to learn about the effective temperature of the medium and its radial flow. The (non-relativistic) analogue of Eq.~(\ref{eq:photon_shift}) for massive particles, i.e.~virtual photons here, is given by
\begin{align}
    T_{\text{eff}}\simeq T+M \langle \beta \rangle^2,
    \label{eq:T_eff_dileptons}
\end{align}
see also \cite{Rapp:2011is}. An analysis of dilepton transverse-momentum spectra has for example been carried out by the NA60 collaboration at SPS for dimuons in In-In collisions at $\sqrt{s}=17.3$~AGeV, \cite{NA60:2008dcb}. Fig.~\ref{fig:dileptons_barometer} shows the effective temperatures extracted from excess dimuon spectra as a function of their invariant mass, compared to theory calculations \cite{vanHees:2007th}. Below the $\rho$ mass one finds a continuous rise of $T_{\text{eff}}$ with the invariant mass, as expected from Eq.~(\ref{eq:T_eff_dileptons}). Above the $\rho$ mass, however, the slope decreases again which can be interpreted in terms of a reduced Doppler shift since the higher-mass region is `biased' towards the earlier (possibly partonic) phase of the collision. The theoretical calculations give a reasonable description of the NA60 slopes and allow for independent confirmation of thermal emission from a QCD medium, see also \cite{Rapp:2011is}. The sensitivity to the large collective flow shows that dileptons can also serve as an accurate `barometer' of the fireball. As for the temperature determination, we note that invariant-mass (rather than transverse-momentum) dilepton spectra can, in principle, provide a cleaner measurement of the temperature since it does not suffer from a (Doppler) blue-shift effect.

\subsubsection*{Dileptons as a thermometer and chronometer}
\label{sec:dileptons_thermometer_chronometer}
\vspace{2mm}

The present robust understanding of all existing low-mass dilepton spectra allows to use this observable as a probe of fundamental fireball properties across the QCD phase diagram. In \cite{Rapp:2014hha} this has been put forward in two respects, i.e.~by extracting the total fireball lifetime from the excess yields in the low-mass region (LMR), and by extracting the early fireball temperatures from the invariant-mass slopes in the intermediate-mass region (IMR).

\begin{figure}[b!]
	\centering\includegraphics[width=0.42\textwidth]{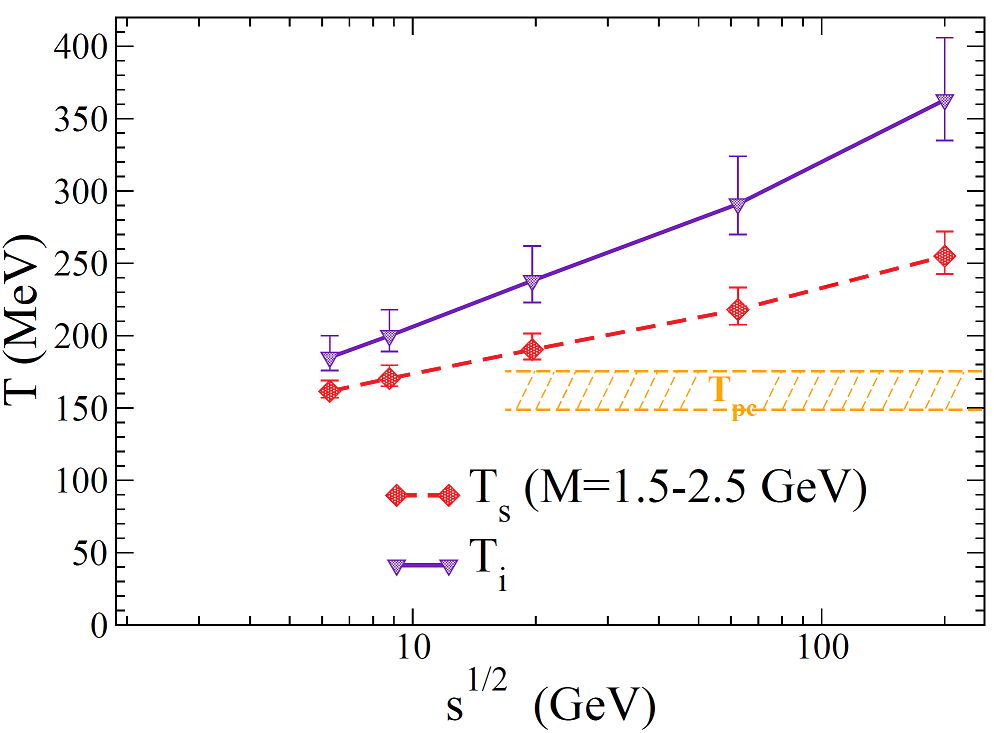}\hspace{0.1\textwidth}
	\centering\includegraphics[width=0.42\textwidth]{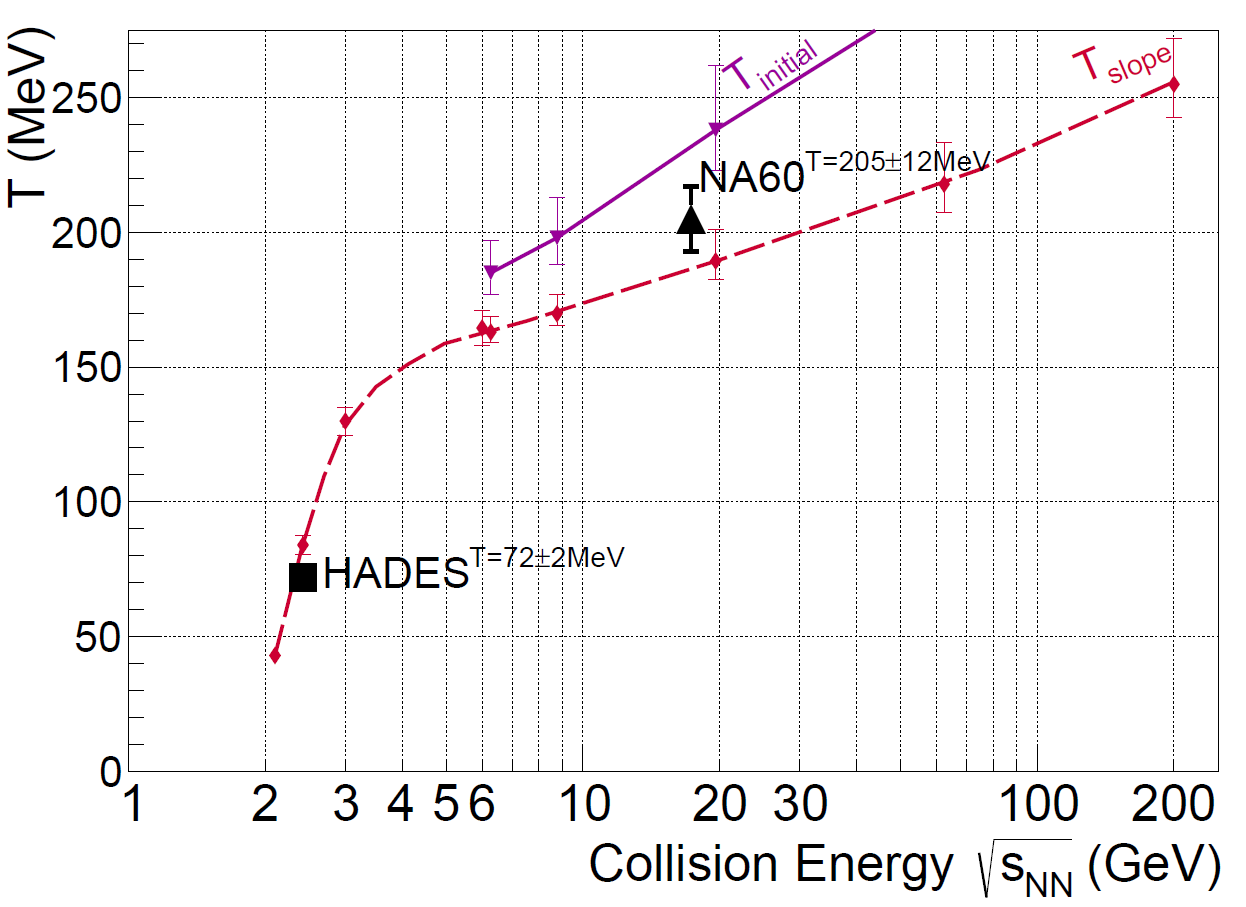}
	\caption{Left: Excitation function of the inverse slope parameter, $T_s$, from intermediate-mass dilepton spectra ($M=1.5-2.5$~GeV, diamonds connected with dashed line) and initial temperature (triangles connected with solid line) in central heavy-ion collisions  ($A\simeq 200$). The hatched area schematically indicates the pseudo-critical temperature regime at vanishing chemical potential. Right: The inverse slope parameters as measured by the HADES collaboration (black square), with $T_s=72\pm 2$~MeV \cite{HADES:2019auv}, and by the NA60 collaboration (black triangle), with $T_s=205\pm 12$~MeV \cite{Specht:2010xu,NA60:2008ctj}, are compared to the model predictions of \cite{Rapp:2014hha} (red and purple points and lines, as also shown on the left side). Figure adapted from \cite{Rapp:2014hha} (left) and \cite{Tetyana_private} (right).}
	\label{fig:dileptons_thermometer}
\end{figure} 

For the temperature determination, the IMR is used because here the medium effects on the EM spectral function are parametrically small, of order $T^2/M^2$, providing a stable thermometer. With $\text{Im}\Pi_{\text{EM}}\propto M^2$ one then obtains
\begin{align}
    \frac{dR_{ll}}{dM}\propto (MT)^{3/2}\exp{(-M/T)}
\end{align}
which is {\it independent} of the medium's collective flow. The observed spectra necessarily involve an average over the fireball evolution, but the choice of mass window, $1.5~\text{GeV}\leq M \leq 2.5\text{GeV}$, implies $T\ll M$ and thus enhances the sensitivity to the early high-$T$ phases of the evolution. Since primordial (and pre-equilibrium) contributions are not expected to be of exponential shape (e.g., power law for Drell-Yan), their `contamination' may be judged by the fit quality of the exponential ansatz.

The resulting inverse slopes, $T_s$, are shown in the left panel of Fig.~\ref{fig:dileptons_thermometer} for collision energies of $\sqrt{s_{NN}}=6-200$~GeV. Inverse slopes in the range from $T_s=160$~MeV to $260$~MeV are found based on certain assumptions, cf.~\cite{Rapp:2014hha}, which suggests that a thermalized QGP with temperatures well above the pseudo-critical one has been produced. Whether the produced medium is really thermalized remains however a difficult question to answer. The results furthermore quantify that the `measured' average temperature is about $30\%$ below the corresponding initial temperature, $T_i$. This gap significantly decreases when lowering the collision energy, to less than $15\%$ at $\sqrt{s_{NN}}=6$~GeV. This is in large part a consequence of the (pseudo-)latent heat in the transition which needs to be burned off in the expansion/cooling. The collision energy range below $\sqrt{s_{NN}}=10$~GeV thus appears to be well suited to map out this transition regime and possibly discover a plateau in the IMR dilepton slopes akin to a caloric curve. Such a transition may thus also be identified by combining available experimental results on the inverse slope parameter, $T_s$, as a function of collision energy, cf.~the right panel of Fig.~\ref{fig:dileptons_thermometer} where data points from HADES, with $T_s=72\pm 2$~MeV \cite{HADES:2019auv}, and from NA60, with $T_s=205\pm 12$~MeV \cite{Specht:2010xu,NA60:2008ctj}, are compared to the model predictions of \cite{Rapp:2014hha}. Of course, more experimental points from upcoming high-precision dilepton experiments at different beam energies are needed to identify such a transition based on deviations from the expected behavior for a scenario without phase transition shown in Fig.~\ref{fig:dileptons_thermometer}.

\begin{figure}[t]
	\centering\includegraphics[width=0.42\textwidth]{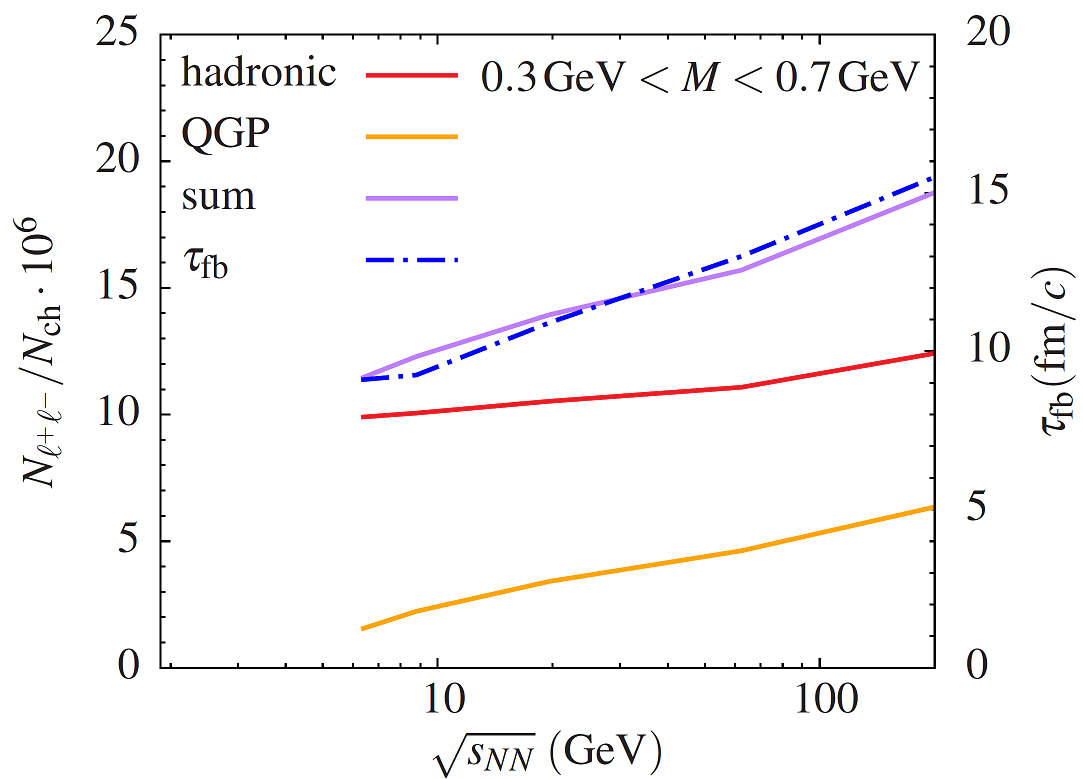}\hspace{0.1\textwidth}
	\centering\includegraphics[width=0.42\textwidth]{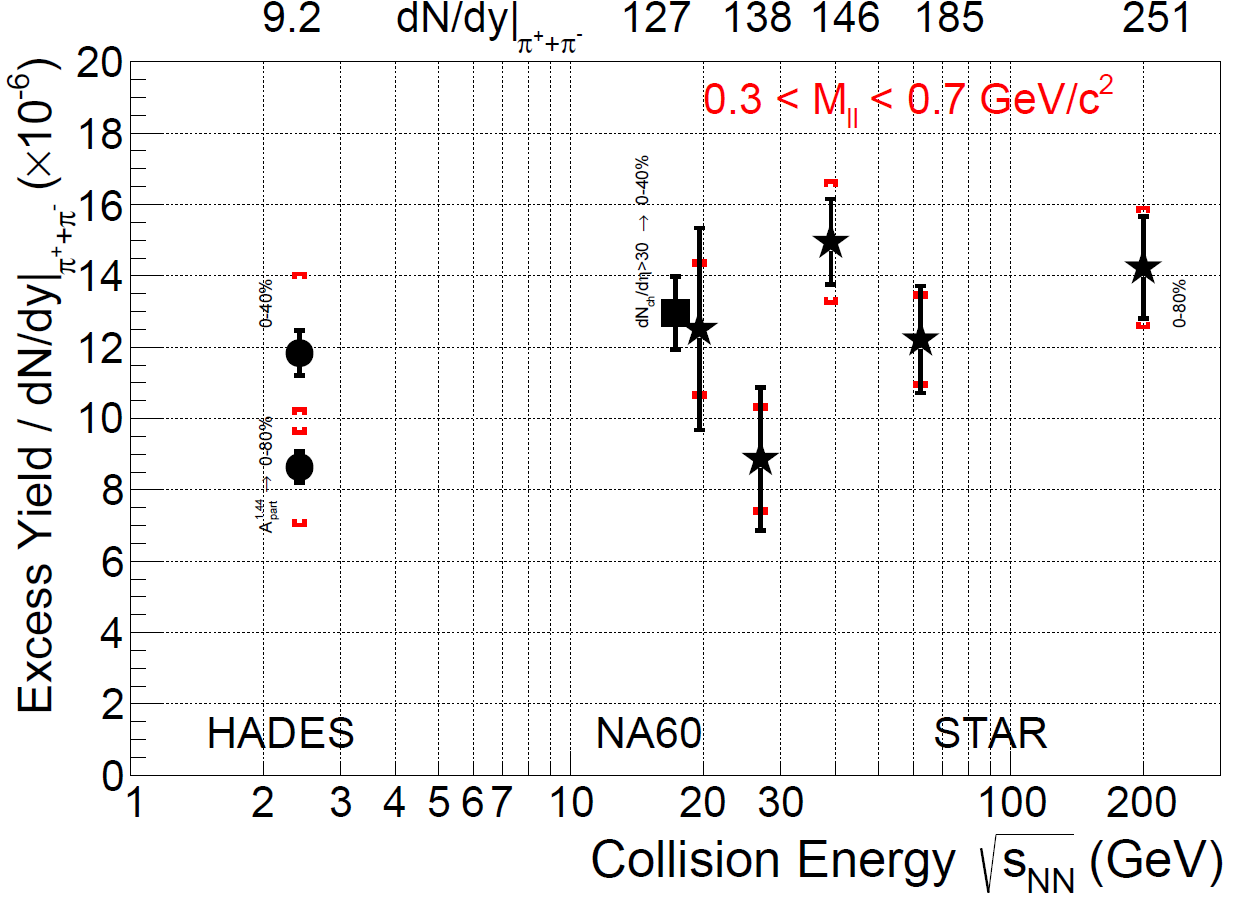}
	\caption{Left: Excitation function of low-mass thermal radiation (`excess spectra') integrated over the mass range $M=0.3-0.7$~GeV, as given by QGP (orange line) and in-medium hadronic (red line) contributions and their sum (purple line). The underlying fireball lifetime (dot-dashed line) is given by the right vertical scale. Right: Dilepton excess radiation as measured by the HADES collaboration \cite{HADES:2019auv}, the NA60 collaboration \cite{Specht:2010xu,NA60:2008ctj}, and the STAR collaboration \cite{STAR:2018xaj}. Figure adapted from \cite{Rapp:2014hha} (left) and \cite{Tetyana_private}  (right).}
	\label{fig:dileptons_chronometer}
\end{figure} 

For the determination of the fireball lifetime, it was shown in \cite{Rapp:2014hha} that low-mass dileptons can be utilized as a chronometer, see also \cite{Heinz:1990jw, Barz:1990mq}. In the left panel of Fig.~\ref{fig:dileptons_chronometer} the integrated LMR excess radiation is shown, i.e.~for a mass range $M=0.3-0.7$~GeV, below the free $\rho/\omega$ mass. It turns out that the integrated thermal excess radiation tracks the total fireball lifetime remarkably well, within less than $10\%$. An important reason for this is  that, despite the dominantly hadronic contribution, the QGP one is still significant. The latter would be relatively more suppressed when including the $\rho/\omega$ peak region. Likewise, the hadronic medium effects are essential to provide sufficient yield in the low-mass region. With such accuracy, low-mass dileptons are an excellent tool to detect any `anomalous' variations in the fireball lifetime. The right panel of Fig.~\ref{fig:dileptons_chronometer} shows a compilation of experimental results on the integrated LMR dilepton excess radiation for different beam energies. These results were obtained by the HADES collaboration \cite{HADES:2019auv}, the NA60 collaboration \cite{Specht:2010xu,NA60:2008ctj}, and the STAR collaboration \cite{STAR:2018xaj}. 
All shown data points are acceptance-corrected and represent the excess radiation as obtained by subtracting cocktail contributions excluding the $\rho$-meson (since its contributions are expected to be strongly modified by the medium) from the dielectron yields. For more details on the individual measurements and the employed methodology, we refer to the corresponding citations given above.

The general behavior agrees with the theoretical expectations shown in the left panel of Fig.~\ref{fig:dileptons_chronometer} but additional data points with smaller uncertainties are needed to make quantitative statements and possibly identify anomalous behavior such as a peak which may indicate the existence of a critical point in whose vicinity the system lives longer and thus produces extra radiation. We conclude that dilepton radiation is well suited to provide direct information on the QCD phase boundary in particular in a region where a critical point and an onset of first-order transitions are conjectured.


\subsubsection*{Dileptons as a polarimeter}
\label{sec:dileptons_polarimeter}
\vspace{2mm}

We now turn to another application of dileptons, namely as a polarimeter. Recently, it was proposed that the polarization of real and virtual photons can be used to study the momentum anisotropy of the distributions of quarks and gluons \cite{Ipp:2007ng, Baym:2014qfa, Baym:2017qxy}. In a first measurement of the dilepton angular anisotropy, the NA60 collaboration found that the anisotropy coefficients in 158~$A$GeV In-In collisions are consistent with zero \cite{NA60:2008iqj}, while the HADES collaboration finds a substantial transverse polarization in Ar-KCl at 1.76~$A$GeV \cite{HADES:2011nqx}. The invariant mass spectrum and $q_T$ dependence of low-mass dileptons ($M< 1$ GeV) produced in heavy-ion collisions are consistent with an equilibrated, collectively expanding source \cite{NA60:2008ctj,NA60:2007lzy}. Moreover, the lack of dilepton anisotropy found in \cite{NA60:2008iqj} has been interpreted as evidence for a thermalized medium. However, as noted in \cite{Hoyer:1986pp}, also a fully thermalized medium, in general, emits polarized photons. 

In general, the polarization state of a virtual photon is reflected in anisotropies of the angular distribution of the lepton pair. Thus, different photon production mechanisms give rise to characteristic shapes for the dilepton angular distribution \cite{Gottfried:1964nx, NA10:1986fgk, HERA-B:2009iab,Speranza:2016tcg}. The angular distribution of the leptons originating from the decay of a virtual photon, expressed in the photon rest frame, is of the form \cite{Gottfried:1964nx, NA10:1986fgk,Faccioli:2011pn,Faccioli:2010kd}
\begin{align}\label{general_ang_distr0}
\frac{d\Gamma}{d^4q d\Omega_\ell}={}\, \mathcal{N} \Big(1+\lambda_\theta\cos^2\theta_\ell 
&+\lambda_\phi \sin^2\theta_\ell\cos2\phi_\ell+\lambda_{\theta\phi}\sin2\theta_\ell\cos\phi_\ell \nonumber\\
&+\lambda^{\bot}_\phi \sin^2\theta_\ell\sin2\phi_\ell+\lambda^{\bot}_{\theta\phi}\sin2\theta_\ell\sin\phi_\ell \Big),
\end{align}
where $\Gamma\equiv\frac{d N}{d^4 x}$ is the dilepton production rate per unit volume, $q^\mu$ the virtual photon momentum while  $\theta_\ell$ and $\phi_\ell$ are the polar and azimuthal angles of, e.g., the negative lepton in the rest frame of the photon and $d\Omega_\ell=d\cos \theta_\ell\, d\phi_\ell$. The normalization $\mathcal{N}$ is independent of the lepton angles. The coefficients $\lambda_\theta$, $\lambda_{\phi}$, $\lambda_{\theta\phi}$, $\lambda^{\bot}_\phi$ and $\lambda^{\bot}_{\theta\phi}$ are the anisotropy coefficients, $\lambda^{\bot}_\phi$ and $\lambda^{\bot}_{\theta\phi}$ being non-zero only for processes that are not symmetric with respect to reflections in the production plane.  We note that the anisotropy coefficients depend on the choice of the quantization axis and therefore on the chosen frame of reference. Typical frames are for example given by the helicity frame, where the quantization axis is along the photon momentum, and the Collins-Soper frame, where it is the bisector of the angle formed by the beam and target momenta in the photon rest frame, see for example \cite{Faccioli:2010kd, Collins:1977iv} for more details.

\begin{figure}[t]
	\centering\includegraphics[width=0.85\textwidth]{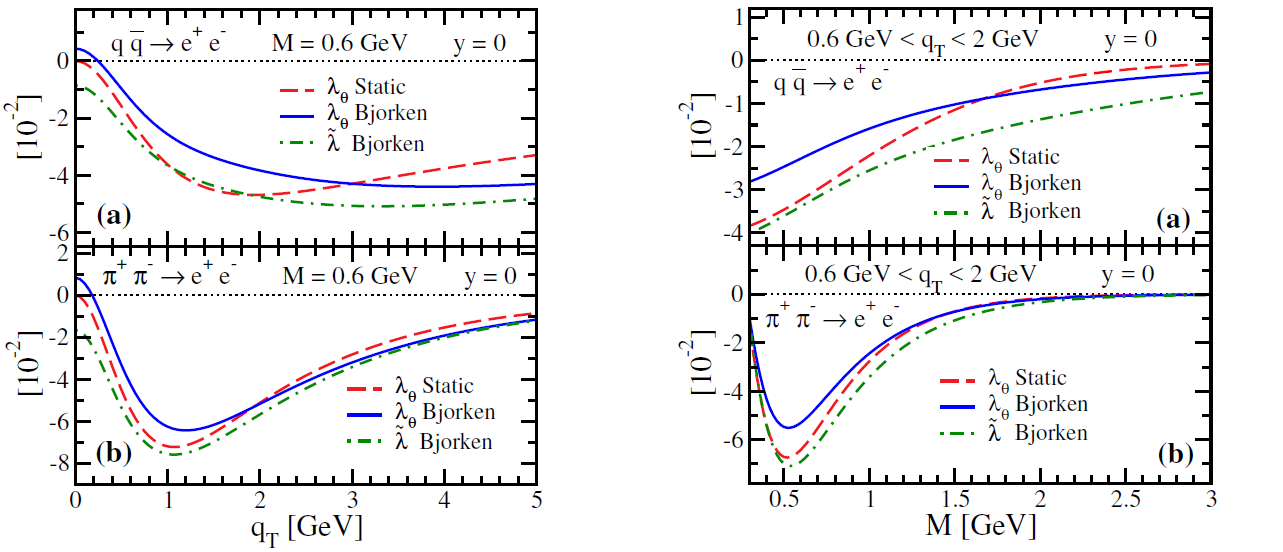}
	\caption{Left: Anisotropy coefficients as functions of the virtual photon transverse momentum at an invariant mass $M=0.6$~GeV for (a) the Drell-Yan process, and (b) pion annihilation. Right: Anisotropy coefficients integrated over transverse momentum in the range between $0.6$~GeV and $2$~GeV as functions of the invariant mass $M$.	The red dashed lines correspond to a static medium, while the blue and green lines correspond to a longitudinal Bjorken expansion. Figure adapted from \cite{Speranza:2018osi}.}
	\label{fig:polarization}
\end{figure}

In \cite{Speranza:2018osi} a general framework for studying photon polarization and the associated angular anisotropies of dileptons produced at high collision energies was presented. In particular, it was shown how the velocity and temperature profiles describing the evolution of the medium are reflected in the anisotropy coefficients. In  Fig.~\ref{fig:polarization} results on the anisotropy coefficients for dileptons emitted from a thermalized static medium as well as from a longitudinally expanding medium (Bjorken) are shown for the Drell-Yan and pion annihilation processes. Therein, also the frame-invariant combination 
\begin{align}
\tilde{\lambda}\equiv \frac{\lambda_\theta+3\lambda_\phi}{1-\lambda_\phi}
\end{align}
is shown. One observes that, in the static case, the anisotropy coefficient tends to zero for small values of the photon transverse momentum and vanishes at $q_T=0$~GeV, for both processes. For large values of $q_T$ the anisotropy coefficients again approach zero, because the momentum distributions are well approximated by the Boltzmann distribution, leading to unpolarized photons. In the case of the Bjorken expansion, the anisotropy coefficients do not vanish in the limit $q_T\rightarrow 0$. This is a consequence of the fact that a photon with vanishing momentum in the center of mass (c.m.) frame has a non-zero momentum in the local fluid rest frame if emitted with a fluid element with flow.

In the right panel of Fig.~\ref{fig:polarization} the anisotropy coefficients, integrated over $q_T$ between 0.6 and 2~GeV, are shown as functions of the photon invariant mass $M$. Here, the Boltzmann limit, with vanishing anisotropy, is approached for large $M$. Interestingly, the two processes considered yield rather similar anisotropy patterns, although the photon polarizations in the corresponding elementary reactions are distinctly different. In the Drell-Yan process, the photons are purely transverse ($\lambda_\theta=1$), while in the pion-annihilation process they are purely longitudinal ($\lambda_\theta=-1$) in a frame where the $z$-axis is along the `beam' axis, defined by the momenta of the incident particles in the c.m.~frame, see \cite{Speranza:2018osi} for more details.

Integrated over $M$, $q_T$ and $y$, the coefficient $\lambda_\theta$ is of the order of $1\%$ or smaller and thus consistent with the finding of the NA60 collaboration \cite{NA60:2008iqj} that the anisotropy coefficients are small, and within experimental error, compatible with zero. We also note that the large transverse polarization obtained by the HADES collaboration in Ar-KCl at 1.76~$A$GeV \cite{HADES:2011nqx} is not consistent with the annihilation processes in local thermal equilibrium considered here. The observed anisotropy may be due to non-equilibrium effects or dominated by another process, as for example $\Delta$ Dalitz decay, as pointed out in \cite{Speranza:2018osi}.

We conclude that future experiments with higher statistics could provide an unambiguous signal of virtual photon polarization effects in heavy-ion collisions. Since the anisotropy coefficients depend on the underlying elementary reaction they can also be used to extract information on the production mechanism of dileptons and thus to distinguish different phases and degrees of freedom.

\subsubsection*{Dileptons as a multimeter: electrical conductivity}
\label{sec:dileptons_conductivity}
\vspace{2mm}

Transport coefficients are an important tool for characterizing hot and dense nuclear matter. Their computation, however, is challenging and recent results on, e.g., the electrical conductivity $\sigma_{\text{el}}$ of hot hadronic matter vary considerably \cite{Caron-Huot:2006pee,Fernandez-Fraile:2005bew,Finazzo:2013efa,Aarts:2014nba, Amato:2013naa}. The significance of the electrical conductivity can be exemplified as follows. First, the electrical conductivity reflects the sensitivity of the charge-dependent directed flow of final state hadrons on the early stage charge asymmetry \cite{Hirono:2012rt,Voronyuk:2014rna}. Second, the EM response plays a crucial role in determining the thermal photon and dilepton emission rates entering through the current-current correlator, such that the transverse momentum spectra and elliptic flow are also sensitive to the temperature dependence of $\sigma_{\text{el}}$. In addition, $\sigma_{\text{el}}$ also appears as input in the (hydrodynamical) evolution equations of the fireball in the presence of EM fields \cite{Gursoy:2014dca,Gursoy:2014aka,Tuchin:2013ie}.

The electrical conductivity is defined as the proportionality constant between an external electric field $E^\mu$ and the induced current density $J^\mu$
\begin{align}
    J^\mu=\sigma_{\text{el}} E^\mu.
\end{align}
The electrical conductivity can also be extracted from the EM current correlator $\Pi_{\text{EM}}$ in the zero-momentum, low-energy limit,
\begin{align}
    \sigma_{\text{el}}(T)=-e^2 \lim_{q_0\rightarrow 0}\frac{\text{Im}\Pi_{\text{EM}}(q_0,\vec{q}=0,T)}{q_0}.
    \label{eq:conductivity}
\end{align}
The electrical conductivity can, therefore, at least in principle, also be extracted from dilepton spectra at very low energies. At the moment, however, such an endeavor does not seem to be feasible.

Within the Vector Meson Dominance model, one can further relate $\Pi_{\text{EM}}$ to the $\rho$ propagator $D_\rho$ via
\begin{align}
    \Pi_{\text{EM}}(q,T)=\frac{m_\rho^4}{g_\rho^2}D_\rho(q, T)=\frac{m_\rho^4}{g_\rho^2}\frac{1}{q^2-m_\rho^2-\Sigma_\rho(q,T)},
\end{align}
where $\Sigma_\rho(q,T)$ is the $\rho$ self energy. Such a computation was performed in \cite{Atchison:2017bpl} using hadronic many-body theory for the calculating of the $\rho$ self energy in a pion gas, with the inclusion of vertex corrections to maintain gauge invariance. 

\begin{figure}[t!]
	\centering\includegraphics[width=0.85\textwidth]{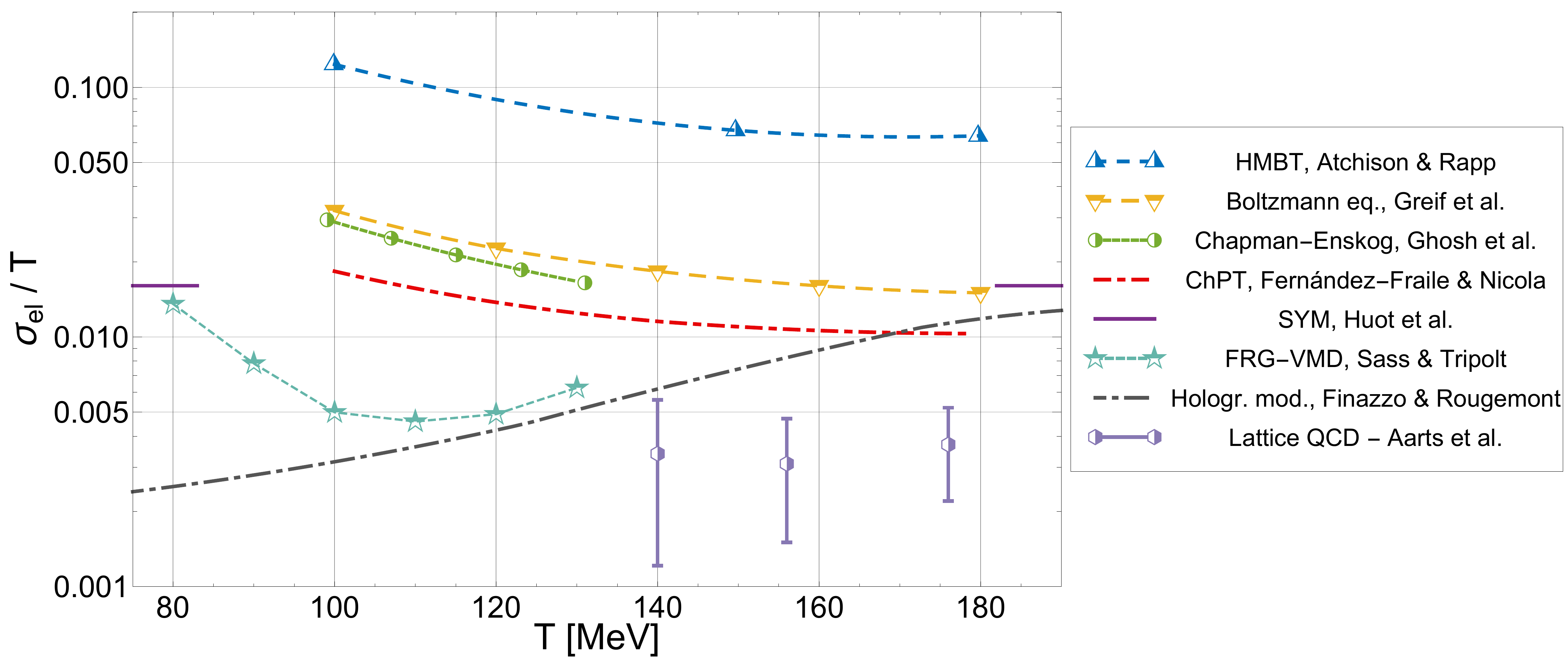}
	\caption{Compilation of results on the electrical conductivity: Hadronic many-body theory \cite{Atchison:2017bpl}, linearized Boltzmann equation \cite{Greif:2016skc}, relativistic transport with Chapman-Enskog technique \cite{Ghosh:2017njg}, chiral perturbation theory \cite{Fernandez-Fraile:2005bew}, $N=4$ super-Yang-Mills theory \cite{Caron-Huot:2006pee}, FRG with Vector Meson Dominance \cite{Sass2021, Tripolt:2014wra,Tripolt:2016cey}, non-conformal holographic model \cite{Finazzo:2015xwa}, 2+1 flavour anisotropic lattice data \cite{Amato:2013naa,Aarts:2014nba}.}
	\label{fig:conductivity}
\end{figure}

The obtained electrical conductivity is shown in Fig.~\ref{fig:conductivity} together with results from other approaches. The result in \cite{Ghosh:2017njg} was obtained by solving the relativistic transport equation in presence of a finite electric field employing the Chapman-Enskog technique where the collision term has been treated in the relaxation time approximation. Therein, the scattering amplitudes of charged pions modeled by $\rho$ and $\sigma$ meson exchange using an effective Lagrangian have been obtained at finite temperature by introducing self-energy corrections in the thermal propagators in the real-time formalism. The results in \cite{Greif:2016skc} were obtained from a kinetic theory approach involving isotropic cross sections using the Boltzmann equation. In \cite{Fernandez-Fraile:2005bew} the electrical conductivity of a pion gas was studied at low temperatures in the framework of linear response and chiral perturbation theory. Therein, the standard ChPT power counting was modified to include pion propagators with a nonzero thermal width in order to properly account for collision effects typical of kinetic theory. The lattice data in \cite{Amato:2013naa,Aarts:2014nba} was obtained for 2+1 flavour anisotropic configurations. Therein, the maximum entropy method was used to construct spectral functions from correlators of the conserved current which were then used in the Kubo formula for the electrical conductivity. The holographic estimate for the electrical conductivity from \cite{Finazzo:2015xwa} for a strongly coupled quark-gluon plasma was obtained using a bottom-up Einstein-Maxwell-Dilaton (EMD) holographic model. The result from \cite{Caron-Huot:2006pee} was obtained for $\mathcal{N}=4$ super-Yang-Mills theory and is given by
\begin{align}
    \sigma_{\text{el}}(T)=e^2\frac{N_c^2T}{16\pi}\approx 0.016 \,T
\end{align}
in the strong coupling limit. Finally, the FRG results obtained in \cite{Sass2021, Tripolt:2014wra,Tripolt:2016cey} were obtained using Eq.~(\ref{eq:conductivity}) where the $\rho$ self energy was computed using the full pion propagator at finite temperature as obtained in \cite{Tripolt:2014wra,Tripolt:2016cey}. We note that most approaches seem to converge to similar results on $\sigma_{\text{el}}$ near the crossover temperature $T_c$. However, in particular, at lower temperatures, as relevant for the hadronic regime in heavy-ion collisions, future work will be necessary to arrive at quantitatively comparable results.

\subsection{Recent experimental results}
\label{sec:dileptons_experimental_results}

\vspace{2mm}

Dilepton invariant-mass spectra bring a plethora of physics channels
from different stages of the evolution of the medium that can be `tuned in' by selecting the relevant mass window and thus not only include the leptonic decay channels of various light, strange, and charm mesons but also allow
for the measurement of virtual direct photons from similar sources. 

\begin{figure}[b!]
    \centering
    \includegraphics[width=0.98\textwidth]{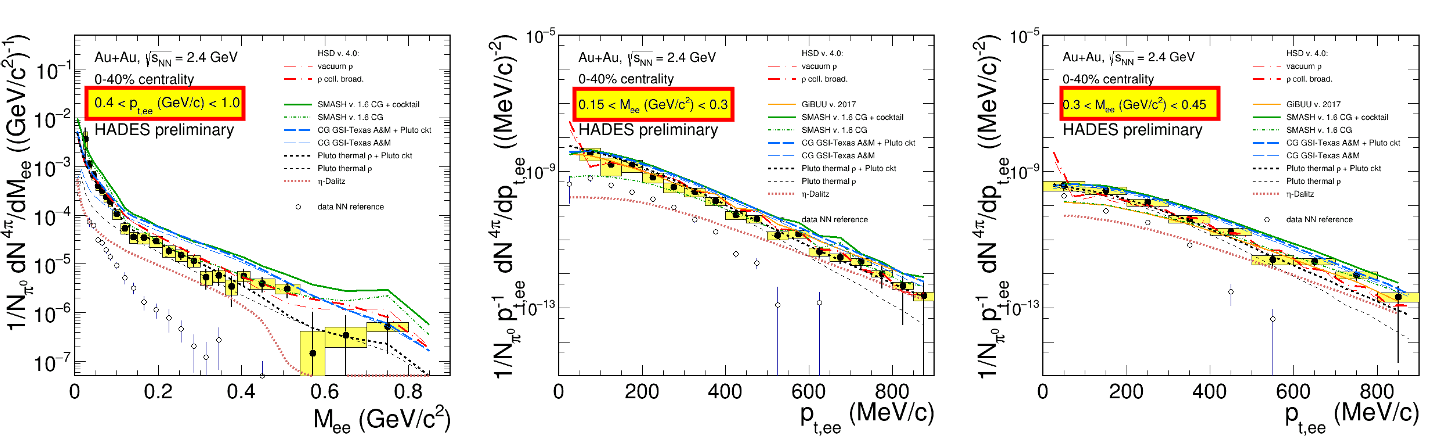}
    \caption{Thermal dielectron measurements from Au$+$Au collisions at $\sqrt{s_\mathrm{NN}}=2.42$~GeV by the HADES collaboration \cite{Harabasz:2021gmb}, figures presented at Quark Matter 2019. Left panel: invariant-mass spectra for low-$p_\mathrm{T}$ dielectrons. Middle and right panels: transverse momentum distributions for $0.15 < M_{ee} < 0.3$~GeV/$c^2$ and $0.3 < M_{ee} < 0.45$~GeV/$c^2$, respectively.}
\label{fig:HADESee}
\end{figure}

Results from measurements of dielectron production in Au$+$Au collisions at $\sqrt{s_{NN}}=2.42$~GeV by the HADES collaboration  \cite{Adamczewski-Musch:2019byl} confirm at this energy the strong in-medium modification of the $\rho$ meson, first reported at SPS energies by the NA60 collaboration \cite{Arnaldi:2006jq}. After careful removal of the hadronic contributions to the invariant mass spectrum, an average temperature of the radiating fireball of $71.8\pm2.1$~MeV was extracted based on a black-body spectral function fit. Recently, the collaboration reported on its multi-differential measurements of the dielectron invariant-mass and $p_\mathrm{T}$ spectra in both the Au$+$Au at $\sqrt{s_{NN}}=2.42$~GeV and Ag$+$Ag at $\sqrt{s_{NN}}=2.42$~and~$2.55$~GeV systems \cite{Harabasz:2021gmb, qm2022:Hades:Harabasz}. In Fig.~\ref{fig:HADESee}, a comparison of the invariant-mass yield (left panel) and momentum spectra in two mass windows (middle and right panels) are compared with several model descriptions. The precision of these preliminary data already demonstrates sufficient sensitivity to the details of these model descriptions. The Pluto framework \cite{Frohlich:2007bi} provides for the calculation of the hadron cocktail (see Sect.~\ref{sec:em-spectroscopy}. It allows to include contributions from thermal $\rho$ meson emissions as was also shown in \cite{HADES:2019auv} for Au$+$Au collisions at $\sqrt{s_{}NN}=2.42$~GeV; here, too, it describes these data well. The GSI-Texas model uses  a coarse-graining procedure for the underlying transport evolution to compute the time-dependent emission of dielectrons \cite{GalatyukHohlerRappEtAl2016}. On top of the previously mentioned hadronic cocktail, these simulations appear to overpredict the data in the lower mass range but do describe the $\rho$-mass reasonably well within the experimental data precision. The same applies to HSD model \cite{BratkovskayaAichelinThomereEtAl2013} in which an off-shell microscopic hadron-string-dynamics transport approach is used on the case of free, vacuum $\rho$ and that of collisional broadening scenario. In the $\rho$ meson mass range, the preliminary data appears to disfavor the former albeit with relatively large uncertainties. The SMASH hadronic transport model relies on dilepton rates from resonance interactions with vacuum properties. It has seen good agreement with experimental $pp$ data at SIS energies \cite{StaudenmaierWeilSteinbergEtAl2018}. Combined with a coarse-graining approach the model appears to generally overpredict the Au$+$Au data, shown in Fig.~\ref{fig:HADESee}.


Dilepton-based measurements of the azimuthal anisotropy $v_2$ as a function of $p_\mathrm{T}$ in different invariant mass regions have been long been proposed as an alternative way to study medium at the different stages \cite{Chatterjee:2007xk}. However, measuring the dielectron $v_2$ is a statistics-hungry challenge, see e.g.~\cite{Adamczyk:2014lpa}. Preliminary results from the HADES collaboration based on Au$+$Au at $\sqrt{s_\mathrm{NN}}=2.42$~GeV and Ag$+$Ag at $\sqrt{s_\mathrm{NN}}=2.55$~GeV show a consistency in the comparison of the $v_2$ of dielectrons in the $\pi^0$ Dalitz mass range compared to that of charged pions \cite{Harabasz:2021gmb,qm2022:Hades:Harabasz}. Importantly, for higher dielectron masses where the yield is predominantly from the medium, the measured $v_2$ is consistent with zero which agrees with the general picture of dielectrons as deep penetrating probes of the hot and dense medium.

\begin{figure}[t!]
    \centering
    \includegraphics[width=0.35\textwidth]{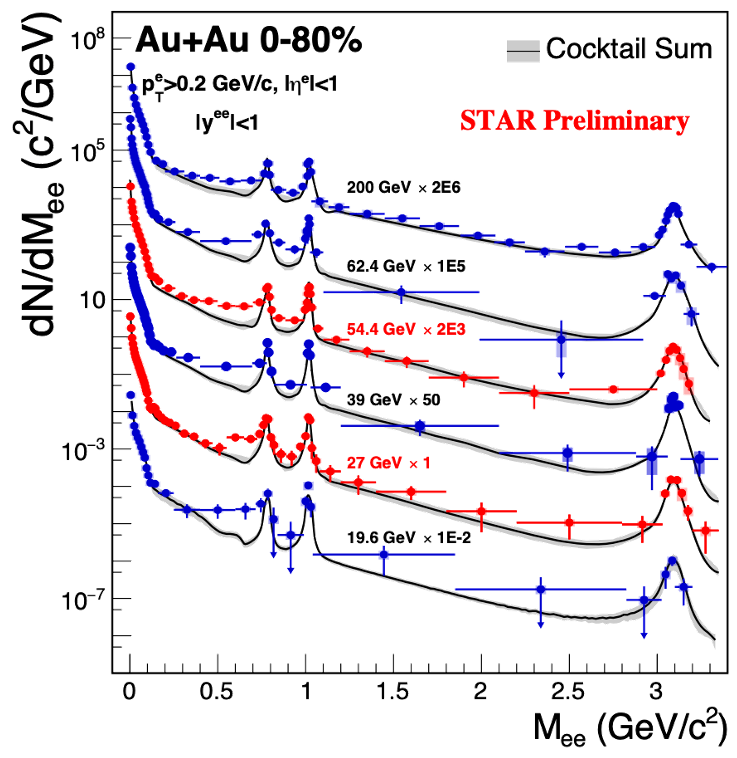}\hspace{14mm}
    \includegraphics[width=0.48\textwidth]{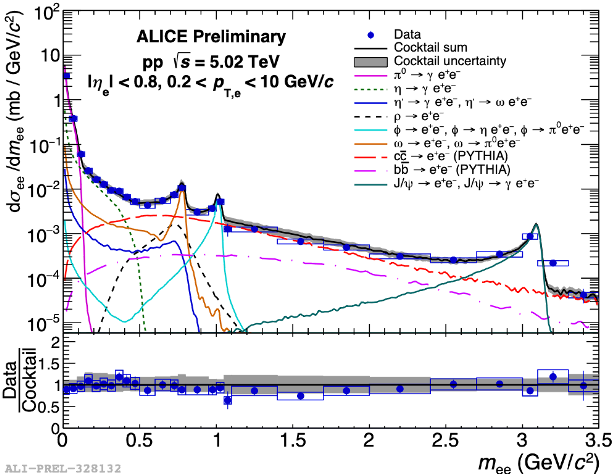}
    \caption{Dielectron invariant mass spectra in Au$+$Au and  $p$$+$$p$ collisions from STAR and ALICE, respectively. Left panel: high-statistics measurements at $\sqrt{s_\mathrm{NN}}=27$ and 54.4~GeV (red symbols) by STAR \cite{Seck:2021mti}. Right panel: ALICE results for p$+$p in collisions at  $\sqrt{s}=5.02$~TeV, \cite{Scheid:2020hmm}. Figures presented at Quark Matter 2019.}
    \label{fig:dielectronSTAR_ALICE}
\end{figure}

A systematic beam-energy scan study (BES) by the STAR collaboration of the production of dielectrons in the low-mass range \cite{Adam:2018qev} has recently been augmented by two high-statistics data samples at  $\sqrt{s_\mathrm{NN}}=27$~GeV and 54.4~GeV \cite{Seck:2021mti,qm2022:STAR:Ye}. As shown in the left panel of Fig.~\ref{fig:dielectronSTAR_ALICE}, these new data sets involve a ten-fold increase in event statistics compared to earlier BES data. Such an increase is expected to better constrain the cocktail by  direct measurements of the $\omega$ and $\phi$ mesons, and will allow for virtual direct photon measurements. The uncertainties in these preliminary results are considered good indicators of the expected precision for the highly anticipated BES Phase-2 energies between $\sqrt{s_\mathrm{NN}}=7.7$ and 19.6~GeV. The new data sets will also allow for temperature extraction from the intermediate mass range \cite{qm2022:STAR:Ye}. As mentioned earlier in this paper, such temperature measurements do not suffer from a radial-flow driven blue-shift as is the case for the effective temperatures extracted from dilepton momentum spectra. However, care is still needed in the subtraction of the correlated charm contributions, especially in the intermediate mass range. De-correlation effects on the decay electrons due to the medium are estimated by randomly assigning angular coordinates and accounted for in the systematic uncertainties of the results.

In the right panel of Fig.~\ref{fig:dielectronSTAR_ALICE}, preliminary dielectron invariant-mass results from the ALICE collaboration are shown for p$+$p collisions at $\sqrt{s}=5.02$~TeV. These results form the vacuum baseline for Pb$+$Pb studies and are found to be well described by the expectations from the hadronic cocktail. The distinct shape of the charm and beauty contributions in the intermediate mass range ($1.1\le M_\mathrm{ee} \le 2.7~$GeV/$c^2$) is used to extract the charm and beauty cross sections which are found to be consistent with independent heavy-flavor measurements \cite{Scheid:2020hmm}. Recently, the collaboration released its first preliminary results from high-multiplicity p$+$p collisions at $\sqrt{s}=13$~TeV which -within uncertainties- did not reveal any signs of thermal radiation \cite{qm2022:ALICE:Jung}. Additionally, preliminary results from central Pb$+$Pb collisions are consistent with cocktail descriptions at low invariant mass. The nuclear modification factor $R_{AA}$ as a function of $(m_{ee}$ does also not show any enhancement within uncertainties.

The ALICE collaboration used the p$+$Pb invariant mass spectra to verify initial state nuclear modification,
\begin{align}
 R_\mathrm{pPb}=\frac{1}{\langle N_\mathrm{coll}\rangle}\frac{dN/dM_\mathrm{ee}|_\mathrm{pPb}}{dN/dM_\mathrm{ee}|_\mathrm{pp}}
\end{align}
at $\sqrt{s_\mathrm{NN}}=5.02$~TeV. The results in the intermediate mass range do not show significant modifications, in agreement with previous D-meson measurements from the ALICE collaboration. However, in the low mass range ($M_\mathrm{ee} \le 1$~GeV/$c^2$ a deviation from unity is observed. This deviation is expected as light-flavor production at low $p_\mathrm{T}$ does not scale with $N_\mathrm{coll}$ and is also observed when comparing to cocktail ratios that include scaling of light flavor in p$+$Pb \cite{Scheid:2020hmm}.

\begin{figure}[h]
    \centering
    \includegraphics[width=.3\textwidth]{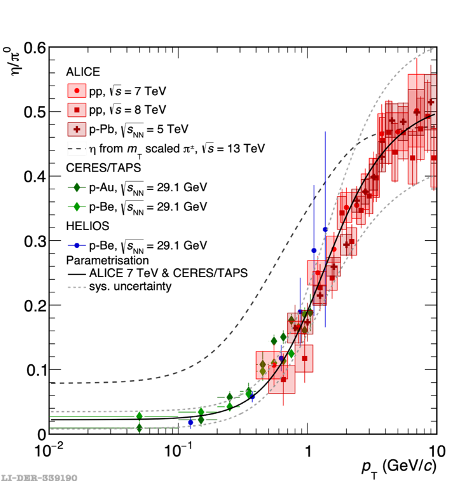}
    \includegraphics[width=.34\textwidth]{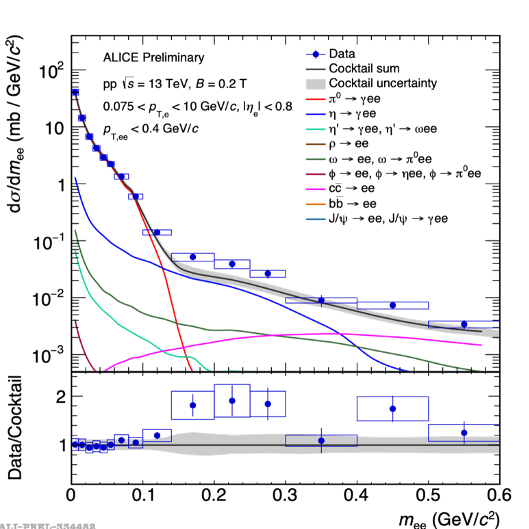}
    \includegraphics[width=.34\textwidth]{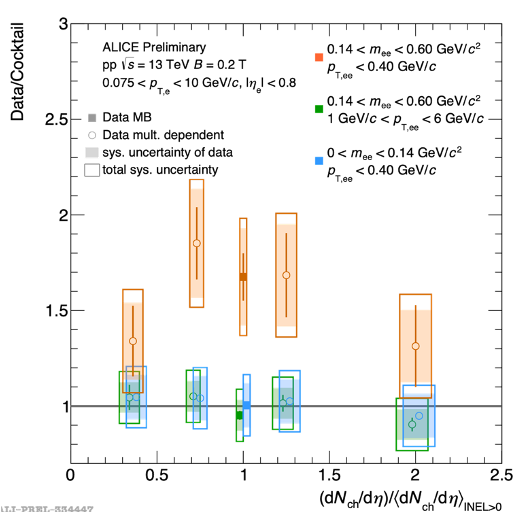}
    \caption{Soft dielectron production in p$+$p collisions at $\sqrt{s}=13$~TeV measured by 
    ALICE \cite{Scheid:2020hmm}. Left panel: new parametrization of the $\eta/\pi$ ratio. Middle panel: low dielectron invariant-mass spectrum in p$+$p collisions at $\sqrt{s}=13$~TeV. Right panel: data-over-cocktail ratio for low-$p_\mathrm{T}$ $\eta$ mesons compared to $\pi^0$ mesons in the same low-$p_\mathrm{T}$ range, $\eta$ mesons in a  higher $p_\mathrm{T}$ range. Figures presented at Quark Matter 2019.}
    \label{fig:softdielectrons}
\end{figure}

Recently, the ALICE collaboration reported on its potential for studying in p$+$p collisions at $\sqrt{s}=13$~TeV a soft dielectron enhancement \cite{Bailhache:2019wyf} that was first reported
at the ISR by the Axial Field Spectrometer collaboration for p$+$p at $\sqrt{s}=63$~GeV \cite{Hedberg:1987yq}. At that time, large uncertainties on the contribution of the $\eta$ meson to the hadronic cocktail and limited statistics did not allow for a quantitative conclusion. A special run in which the field of the ALICE solenoid magnet was lowered to $B=0.2$~T allowed the low-$p_\mathrm{T}$ reach for electrons to drop to 75~MeV/$c$ \cite{Scheid:2020hmm}. A reevaluation of the $\eta$ contribution is shown in the left panel of Fig.~\ref{fig:softdielectrons}. Combined with the new low $B$-field run, these improvements now show a significant enhancement over the cocktail for $p_{T,\mathrm{ee}} < 0.4$~GeV/$c$ in the $\eta$ mass range as can be seen in the middle panel. Interestingly, and shown in the right panel of Fig.~\ref{fig:softdielectrons}, is a comparison of dielectron yields in the $\pi^0$ and $\eta$-meson ranges versus a normalized charged-particle multiplicity. The differently colored symbols indicate different combinations of the two invariant mass ranges ($M_{ee} <0.14$~GeV/$c^2$ and $0.14 < M_{ee} <0.60$ GeV/$c^2$). These two mass ranges are dominated by contributions from $\pi^0$ and $\eta$ decays, respectively.  The plot also shows different dielectron momentum ranges ($p_T^{ee}<0.4$~GeV/$c$ and $1 <p_T^{ee} < 6$~GeV/$c$). For the high-momentum $\eta$-meson and the low-momentum $\pi$ ranges the data are well described by the cocktail for the minimum bias data (filled symbols) and do not seem to show any multiplicity dependence. However, for the low-momentum $\eta$ range, the cocktail significantly underestimates the data for minimum-bias data and seems to indicate a slight multiplicity dependence. In the absence of a deconfined medium, the physical mechanism for this enhancement of the data compared to the expectations from the hadronic cocktail is yet to be understood.

\clearpage 
\section{Conclusions and Outlook}
\label{sec:conclusions}

Electromagnetic probes, i.e.~photons and dileptons, enjoy a unique status in heavy-ion collisions since they do not interact strongly with the created fireball and can provide us with a wide range of insights on the properties of hot and dense QCD matter. Dileptons are especially useful since they have an additional `degree of freedom', i.e., the invariant mass. They can therefore provide basic kinematical information such as the fireball temperature, the degree of collectivity, and the lifetime, but also dynamical information on in-medium spectral functions encoding changes in degrees of freedom and chiral symmetry restoration as well as on transport coefficients like the electrical conductivity. 

In recent years, the melting of the $\rho$ meson in a strongly-interacting hadronic medium was confirmed by various experiments and theoretical calculations, indicating a transition from hadronic degrees of freedom towards a quark-antiquark continuum that is consistent with chiral restoration. There is also emerging consensus that chiral partners become degenerate at the ground state mass in a way that the chiral mass splitting burns off but the ground-state mass, which is then likely generated by another mechanism based on QCD interactions, remains.

New theoretical developments, e.g.~from the Functional Renormalization Group or lattice QCD, are expected to provide chirally and thermodynamically consistent vector-meson spectral functions that will allow for a phenomenologically successful description of experimentally measured dilepton spectra while at the same time being well-founded in theory. Together with high-precision measurements expected from running and upcoming experiments, this will allow to establish a clear connection to chiral symmetry restoration and eventually identify QCD phase transitions such as the chiral first-order transition or the critical endpoint.

In particular the regime of the QCD phase diagram at high baryon chemical potential and low to moderate temperatures has received increased attention in recent years since one expects important landmarks and possibly new phases in this region. For this, measurements of thermal dilepton radiation need to be performed with high statistics and an excellent discrimination of background, in particular at lower beam energies ranging from $\sqrt{s_{\text{NN}}}=2$~GeV to $\sqrt{s_{\text{NN}}}=20$~GeV. Several such projects are planned and discussed below. 

 The high-$\mu_B$ region of the QCD phase diagram will in particular be explored by new upcoming experiments such as the CBM experiment at FAIR, where HADES is currently operational, the STAR BES-2 run, and NICA at the Joint Institute of Nuclear Research, see also \cite{Galatyuk:2019lcf} for a more detailed overview. With these detectors becoming operational, high statistics data on thermal dilepton radiation down to SIS18 energies will become available and, when combined with a robust theoretical understanding, allow for an unprecedentedly clear glimpse at the phase structure and the properties of strong interaction matter in the high-$\mu_B$ region of the QCD phase diagram.
 
 The STAR experiment has recently concluded its data collection for its dilepton program as part of the Beam Energy Scan (BES) and will provide dielectron spectra with good statistics in the low-invariant-mass range at lower collider energies between 7.7 and 19.6~GeV. These data, combined with the results from the first BES campaign between 19.6 and 62.4~GeV will provide for a unique and broad data set that can be used to systematically tie future high-precision measurements at LHC energies, to existing results at SPS and FAIR, and ultimately to future measurements at low-energy facilities such as SPS, FAIR, NICA, and J-PARC \cite{Scomparin:2022bsb, Durante:2019hzd, Senger:2022bzm, Sako:2019hzh}. Furthermore, STAR's BES dielectron measurements at lower collider energies should be used to provide additional data points that may help clarify the tension in the experimental results from PHENIX and STAR. In that context, high-statistics dielectron measurements scheduled for 2023-2025 and which will complete RHIC's mission should also be used to provide new insights.
 
 At SPS energies, the next generation of the NA60 experiment, i.e.~NA60+, is proposed to take data \cite{NA60:2018ckg} starting 2029. It can deliver the precision needed to identify $\rho-a_1$ modifications, which would be a clear indication of chiral symmetry restoration. Additionally, a high-precision thermal dimuon measurement would also bring a unique measurement to the caloric curve. At lower energies, the CBM detector will play a crucial role. Its rate capabilities are expected to top that of most other experiments by several orders of magnitude. Moreover, the detector can be configured to measure either dielectrons or dimuons. 
 
The ALICE experiment will continue dilepton spectroscopy with a focus on thermal radiation in run 3. These studies are expected to reach much higher precision than the current measurements, thanks to the upgrade of the ALICE detector \cite{Citron:2018lsq}. The main goal will be to measure dileptons in the low-mass and intermediate-mass regions in order to access the in-medium $\rho$ meson spectral function and the thermal radiation in the region above the light vector mesons. Fast-forward to Run 5 and the ALICE3 detector will set an all-new stage using an ultra-thin tracker with an unprecedented low material budget \cite{Adamova:2019vkf}. High precision tracking, combined with a very high rate capability, should position the ALICE3 detector in an exceptional position to perform high-precision dilepton measurements down to very low mass and $p_T$ values. At the top LHC energies, this would allow for a direct measurement in the chiral symmetry restoration through the earlier mentioned $\rho-a_1$ mixing.

In conclusion, the physics of electromagnetic probes in heavy-ion collisions is complex - both in experiment and theory. The wealth of experimental data and theoretical developments over the past many years has been nothing but impressive. But, it is only the beginning of a new era in which new theoretical insights are providing guidance to the next generation of dedicated, highly sophisticated experiments that are being readied to deliver high precision and reach. The excitement in the community is palpable and we are thrilled to be part of this.

\section*{Acknowledgments}
\label{sec:acknowledgments}

R.-A.~T.~would like to thank Jochen Wambach for a critical reading of the manuscript. In addition, R.-A.~T.~would like to thank Tetyana Galatyuk, Charles Gale, Hendrik van Hees, Ralf Rapp, Dirk Rischke and Lorenz von Smekal for valuable discussions. R.-A.~T.~is supported by the Deutsche Forschungsgemeinschaft (DFG, German Research Foundation) through the Collaborative Research Center CRC-TR 211 ``Strong-interaction matter under extreme conditions'' -- Project No. 315477589-TRR 211. F.G.~is supported in part by the U.S.~Department of Energy Office of Science under grant No.~DE-SC0005131.

\clearpage
\bibliography{QCD,references}



\end{document}